\documentclass{aa}
\usepackage{graphicx}
\usepackage{amsmath}
\usepackage{txfonts}
\usepackage{url}
\usepackage{acronym}
\usepackage{natbib}
\usepackage{footnote}
\usepackage{color}
\usepackage{caption}
\usepackage{subcaption}
\usepackage{multirow}
\usepackage{float}
\usepackage{pdflscape}
\usepackage{xcolor}
\usepackage[scientific-notation=true]{siunitx}
\usepackage[version=3]{mhchem}
\newcommand{\htu}{H{\sc ii}~}
\newcommand{\msun}{M$_{\odot}$}
\newcommand{\micron}{$\mu$m~}
\newcommand{\lsun}{L$_{\odot}$~}
\begin{document} 
%------------------acronym-------------------------------------------------------------------------- 
\acrodef{HMCs}{Hot molecular cores}
\acrodef{COMs}{complex organic molecules}
\acrodef{GMCs}{Giant Molecular Clouds}
\acrodef{IRDCs}{Infrared dark clouds}
%
%------------------------------------------ --------------------------------------------------------
\title{Evolution of complex organic molecules in hot molecular cores}
\subtitle{Synthetic spectra at (sub-)mm wavebands}
\author{R. Choudhury\thanks{rchoudhury@mpe.mpg.de} \inst{1,2}
          \and
          P. Schilke \inst{1}
          \and
          G. St\'{e}phan \inst{1,3}
          \and
          E. Bergin \inst{4}
          \and
	  T. M\"{o}ller \inst{1}
          \and
          A. Schmiedeke \inst{1}
          \and
          A. Zernickel \inst{1}	  
          }
   \institute{I. Physikalisches Institut, Universit\"{a}t zu K\"{o}ln, Z\"{u}lpicher Strasse 77, 50937, K\"{o}ln, Germany
         \and
	    Max-Planck-Institut f\"{u}r extraterrestrische Physik, Giessenbachstrasse 1, 85748, Garching, Germany
	 \and   
            LERMA, Observatoire de Paris, 61 Av. de l'Observatoire, 75014, Paris, France
         \and
	    Department of Astronomy, University of Michigan, Ann Arbor, MI 48109,  USA
             }
\date{Received/Accepted}
\titlerunning{Synthetic spectra of Hot Molecular Cores}  
\authorrunning{R. Choudhury et. al.}

%---------------------------------------------------------------------------------------------------
\abstract
% context heading (optional)
{Hot molecular cores (HMCs) are  intermediate stages of high-mass star
  formation  and are also  known for  their rich  chemical  reservoirs and
  emission line spectra at (sub-)mm wavebands. Complex organic molecules
  (COMs)  such   as  methanol  (\ce{CH3OH}),   ethanol  (\ce{C2H5OH}),
  dimethyl ether (\ce{CH3OCH3}) and methyl formate (\ce{HCOOCH3}) produce
  most of these observed lines.  The observed spectral feature of HMCs
  such as total number  of emission lines and associated line intensities
  are also found to vary with evolutionary stages.}
% aims
{We aim to investigate the spectral evolution of these COMs to explore
  the  initial  evolutionary  stages   of  high-mass  star  formation
  including HMCs.}
% method
{We  developed various  3D models  for HMCs  guided by  the evolutionary
  scenarios proposed by recent empirical and modeling studies. We then
  investigated  the   spatio-temporal  variation  of   temperature  and
  molecular  abundances  in HMCs  by  consistently coupling  gas-grain
  chemical evolution with  radiative transfer calculations. We explored
  the effects  of varying physical conditions  on molecular abundances
  including density distribution and  luminosity evolution of the central
  protostar(s) among  other parameters.  Finally,  we simulated the
  synthetic  spectra  for   these  models  at  different  evolutionary
  timescales to compare with observations.}
%Results
{Temperature has  a profound effect  on the formation of  COMs through
  the  depletion and  diffusion  on grain  surface  to desorption  and
  further  gas-phase  processing.    The  time-dependent  temperature
  structure of  the hot core  models provides a realistic  framework for
  investigating  the spatial variation  of ice  mantle evaporation  as a
  function of evolutionary timescales.  We find that a slightly higher
  value  (15K)  than  the  canonical dark  cloud  temperature  (10K)
  provides a  more productive environment for COM  formation on grain
  surface.  With  increasing  protostellar  luminosity, the  water  ice
  evaporation font ($\sim$100K)  expands and  the spatial
  distribution of gas phase abundances  of these COMs also spreads out.
  We calculated the temporal variation  of the radial profiles of these
  COMs for  different hot core  models. These profiles   resemble
  the so-called jump profiles with relative abundances higher
  than 10$^{-9}$ within the evaporation font will furthermore be useful to
  model the observed  spectra of hot cores.  We  present the simulated
  spectra  of these  COMs for  different  hot core  models at  various
  evolutionary timescales. A qualitative comparison of the simulated and
  observed spectra suggests that these self-consistent hot core models
  can  reproduce the  notable trends  in hot  core  spectral variation
  within  the typical hot  core timescales  of 10$^{5}$  year. These
  models  predict that  the spatial  distribution of  various emission
  line maps will also  expand with evolutionary time; this  feature can be
  used to constrain the  relative desorption energies of the molecules
  that mainly form on the grain surface and return to the gas phase via
  thermal desorption. The detailed  modeling of the thermal structure of
  hot cores with similar masses along with the characterization of
  the  desorption  energies of  different  molecules  can  be used  to
  constrain the  luminosity evolution  of the central  protostars. The
  model predictions can be compared with high resolution observation
  that can probe scales of a few  thousand AU in high-mass star forming
  regions  such as from  Atacama Large  Millimeter/submillimeter Array
  (ALMA). We used a spectral fitting method to analyze the
  simulated spectra and find  that it significantly underestimates some
  of  the physical parameters  such as  temperature.  The  coupling of
  chemical   evolution  with   radiative  transfer   models   will  be
  particularly useful  to decipher the physical structure  of hot cores
  and  also to constrain  the initial evolutionary  stages of high-mass star formation.}
%Conclusion
{}
%---------------------------------------------------------------------------------------------------
\maketitle
%-------------------------------------------------------------------------------------------------------------
\section{Introduction}
\label{sec:intro}
\ac{HMCs},  also known  as hot  cores, are  associated with  the early
evolutionary  stages  of  high-mass  star  formation.   Our  current
understanding  of  high-mass  star  formation  is  incomplete,  but  a
plausible   evolutionary   scenario,    emerging   from   the   recent
observational  studies, is  as follows:  \ac{IRDCs}  $\rightarrow$ hot
molecular  cores   $\rightarrow$  hyper/ultra-compact   \htu  regions
$\rightarrow$  \htu region  surrounding the  ionizing  high-mass stars
\citep[see][]{Beuther2007a}.    The   intermediate    phases   between
\ac{IRDCs} and hypercompact  \htu regions are known as  hot cores; at
this phase  the central protostars  heat up their  surrounding gaseous
and dusty materials to high temperatures ($\ge$ 100~K) without ionizing
it  (due  to  lack  of  UV  photons).  \cite{Kurtz2000}  characterized
HMCs as  a relatively compact  phase of high-mass  star formation
($\le$ 0.1~pc) with a typical  density and temperature in the range of
$\ge$ 10$^{7}$~cm$^{-3}$ and $\ge$ 100~K. The evaporation
of water and organic rich ice  mantles leads to the detection of a wide
variety of  \ac{COMs} such as  \ce{CH3OH}, \ce{CH3OCH3}, \ce{HCOOCH3} and
\ce{C2H5OH} in a number of \ac{HMCs} \citep{Herbst2009}.

The  chemical   richness  and   diversity  of  \ac{HMCs}   has received
significant attention in  recent times and has also triggered  a number of
studies that  modeled the relevant chemical  processes responsible for
building    up   the   molecular    reservoir   in    \ac{HMCs}   (see
\citealt{Herbst2009} for  a detailed  review of existing  models). Among
these,  gas-grain models  (which  include both  gas phase  and
grain surface  chemical reactions) emerged as the  most promising ones
to    explain   the    abundance   of    \ac{COMs}   in    hot   cores
\citep{Charnley1992}. These  studies suggested that  the chemical richness
of  \ac{HMCs} originates  from the  evaporation and  subsequent  gas phase
processing of the molecules that  became locked on icy mantles during the
initial collapse  phase of high-mass protostar formation.  Recent hot
core models  \citep{Garrod2008,Garrod2013} successfully reproduced the
abundances of a number of COMs inferred from the observational studies
of hot  cores. One of  the crucial parameters  of these models  is the
gradual warm-up  of hot cores  to a temperature  of 100~K and higher
to facilitate  the formation of  COMs on grain surface  and eventually
their evaporation to the  gas phase. Recent observations indicate that
\ac{COMs}  are  also  present  in  cold dark  clouds  with  a  typical
temperature                          of                          10~K
\citep{Cernicharo2012,Bacmann2012}.   \cite{Vasyunin2013a}   suggested
some  alternative  mechanisms that  do  not  require high  temperature
(e.g.,   reactive    desorption)   for   COM    formation   in   these
environments. However, these processes  are probably less important in
the hot core phase.

The salient  observational feature of \ac{HMCs}  are numerous emission
lines at  (sub-)mm wavebands,  mostly originating from  the rotational
transitions       of      COMs.      Dedicated       line      surveys
(e.g., \citealt{Schilke1997,Schilke2001,Zernickel2012,Crockett2014,Neill2014},
see also  Table. 2 of \citealt{Herbst2009}) are very  useful to explore
the inventory  of COMs in  hot cores. Most  of these studies  used 
local thermodynamic  equilibrium (LTE) fitting of  spectra, for example, rotational
diagram  analysis \citep{Goldsmith1999},  XCLASS \citep{Zernickel2012},
or similar techniques for estimating the physical parameters and chemical
abundances.  However,  these techniques, while useful  to estimate the
average physio-chemical scenario of the hot cores, are not suitable to
explore the detailed structures \citep{Herbst2009}. Investigations of
the chemical evolution  of selected molecules  using a number of  hot core
sources    are    also     carried    out    in    selected    studies
\citep{Bisschop2007,Beuther2009,Gerner2014}.  It has  been  found that
the intensity as well as the spatial distribution of emission lines of
various molecules also varies with  evolutionary stages and from one hot
core  to another.  These features  can  be attributed  to the  varying
physio-chemical  condition during the  initial evolutionary  stages of
high-mass star formation.  Although the  results are  promising, 
high  angular resolution  spectral  line observations  of the  initial
stages of high-mass star formation are still rare \citep {Brouillet2013}.

With the availability  of the Atacama Large Millimeter/submillimeter Array
(ALMA)\footnote{\url{http://www.almaobservatory.org/}},   it   is  now
possible to explore the chemical segregation of high-mass star forming
regions down to a spatial  resolution of a few hundred AU in relatively
short  time.  Observations  of many  sources  at various
evolutionary stages  will also provide the opportunity  to explore the
chemical diversity.  In short, these  observed chemical probes  can be
used  as powerful  tools to  decipher  the internal  structure and  to
constrain  the evolutionary  stages  of  hot cores.  So  far, only  few
astrochemical   studies  \citep[e.g.,][]{Doty2002,Oberg2013,Gerner2014}
incorporated a radial variation of density and temperature. However, a
self-consistent approach to explore the temporal evolution of physical
parameters  associated  with  high-mass star  formation  is  not
considered in these models.  Moreover, there are
no dedicated studies that explore  the trends of temporal variation in
simulated spectra which arise due to the varying physical and chemical conditions
of evolving hot cores. To bridge this gap, we investigated the effects
of the spatio-temporal  variation of  physical conditions on  the chemical
evolution  of hot cores  in a  self-consistent way.  In this  work, we
investigate   the  chemical   evolution  of   selected  oxygen-bearing
\ac{COMs} such as, \ce{CH3OH},  \ce{C2H5OH}, \ce{HCOOCH3} and \ce{CH3OCH3}
and simulate the spectral data cubes at (sub-)mm wavebands for various
hot core models at different evolutionary stages.

We describe the model set-up, initial physical conditions and chemical
modeling   in  Sect.~\ref{sec:method}.  Spatio-temporal   evolution  of
molecular   abundances   of    selected   COMs   are   summarized   in
Sect.~\ref{sec:results}.  The  simulated spectra and a qualitative comparison with the
observed   spectra   are presented  in   Sect.~\ref{sec:anal}.   In
Sect.~\ref{sec:dis},   we  summarize   this  study   and   discuss  the
possibilities  of  using  these   simulated  spectra  along  with  the
observations  to explore  the evolution  of hot  cores,  and  finally we
present our conclusion in Sect.~\ref{sec:conc}.
%----------------------------------------------------------------------------------------------------------------------------------
%-------------------------------------------------------------------------------------------------------------------------------------
\section{Method}
\label{sec:method}
\begin{figure}
\includegraphics[scale=0.5]{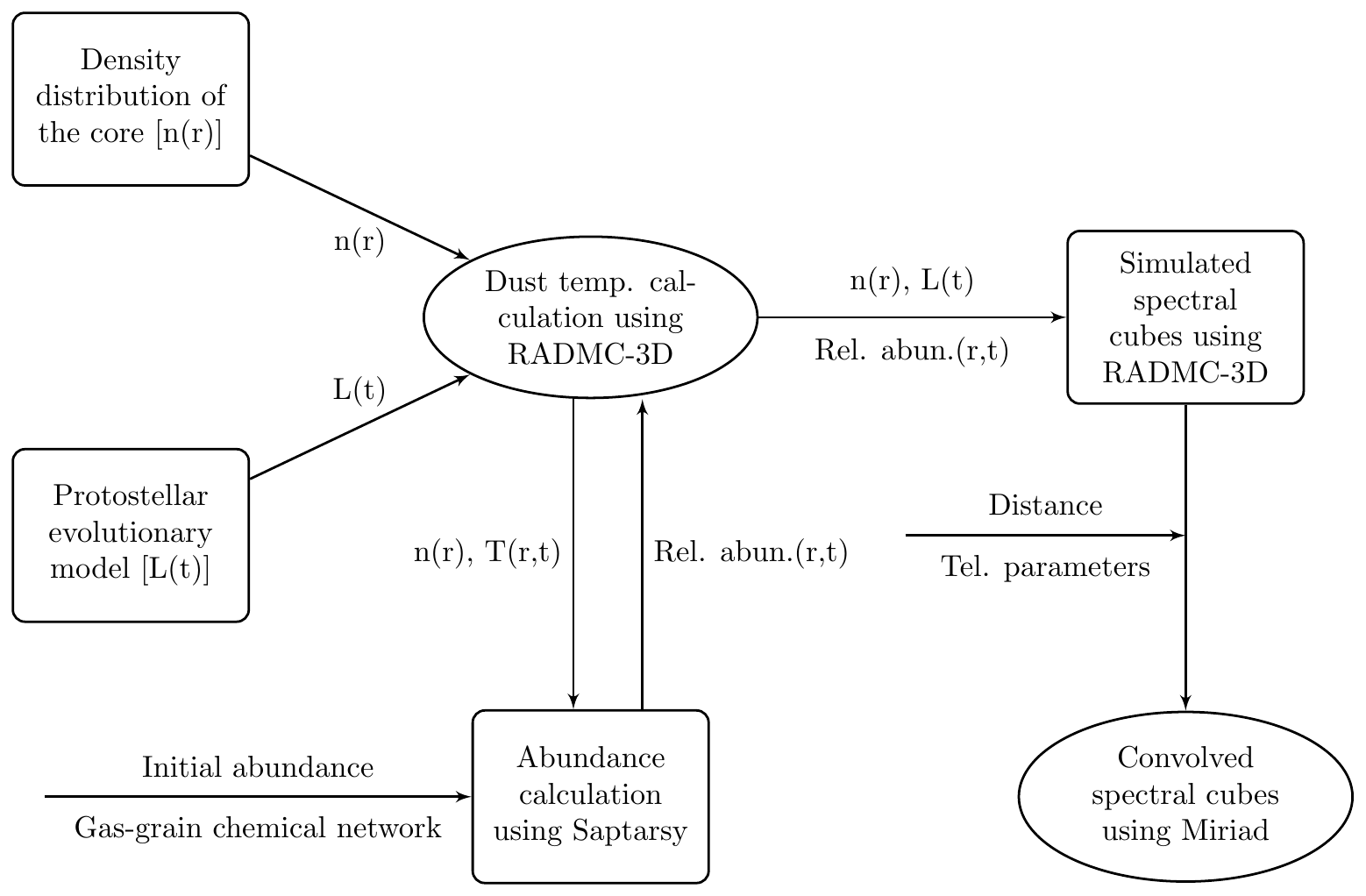}
\caption{Flowchart of the modeling framework.} 
\label{fig:flow_path}
\end{figure}
Most of  the high-mass star forming  regions are distant  and heavily
obscured.   The   physical   structure,  evolutionary   sequence   and
timescales  of  high-mass  star  formation  are  thus only poorly
understood.   Molecular  emission lines  provide  good diagnostics  to
understand    the   internal    structure   of    these   environments
\citep{Rolffs2011}. In  the case  of low-mass star  formation, several
studies  incorporated  the results  of  hydrodynamic simulations  into
chemical  models   to  investigate  the  role   of  evolving  physical
conditions    on   the    chemical    composition   \citep{Aikawa2008,
  Furuya2012}. Similarly,  the ideal way  to model hot  core chemistry
would  be   to  couple  the  chemical   modeling  with  hydrodynamical
simulations of high-mass star formation.  However, models of high-mass
protostellar evolution  along with  the surrounding gaseous  and dusty
materials  are  rare,  with  only  a  handful  of  available  solutions
\citep{Kuiper2013}. In  order to investigate the  coupling of physical
conditions and molecular abundances, we set-up a simple physical model
of hot  cores guided  by the current  understanding of high-mass star
formation and used this set-up to examine the spatio-temporal variation
of  chemical evolution  of hot  cores and  finally  generated synthetic
spectra  following the  flowchart  shown in  Fig.~\ref{fig:flow_path}.
The flowchart  is executed using a  Python script (Schmiedeke
et al., 2014, in preparation). We describe the physical parameters and
chemical models in detail below.
%------------------------------------------------------------------------------------------------------------------------------------------------
%
\subsection{Chemical model}
\label{subsec:cmodel}
We   developed   a   rate-equation-based   1D   astrochemical   code
(\emph{Saptarsy})  in Fortran90;  it calculates  the spatio-temporal
evolution  of various  molecular  species using  gas phase  reactions,
gas-grain interactions, and surface  chemistry (Choudhury et al., 2014,
in  preparation). The code  has its  origin in  the \emph{Astrochem}
code of  \cite{Bergin1995}, and  after several major  modifications, it
has become an  improved and new astrochemical code  that can be
easily       integrated      with      other       external      codes
(e.g., \emph{RADMC-3D}\footnote{\url{http://www.ita.uni-heidelberg.de/~dullemond/software/radmc-3d/}}).
The  ordinary   differential  equation  solver   from  Netlib  library
\emph{dvodpk}\footnote{\url{http://www.netlib.org/ode/}} and MA28\footnote{\url{http://www.hsl.rl.ac.uk/}}, the advance solver
of sparse systems of linear equations from the HSL library, are used to  compute  the  chemical  evolution. Gas-grain  interaction  via
accretion  and  desorption   and  surface  reactions  are  implemented
following the recipe  of \cite{Hasegawa1992} and \cite{Hasegawa1993a}.
The temporal evolution of  grains such as grain coagulation and formation of
the  grain  mantles  etc.  can  affect  the  grain  surface  chemistry
\citep{Taquet2012, Garrod2013a},  but these processes are  beyond the scope
of this study and are, therefore, not considered in this work.

\emph{Saptarsy}   is   suitable    to   explore   the   effects   of
spatio-temporal variation of physical parameters (e.g., temperature) on
chemical  evolution.  It  has  been  benchmarked  against  the  models
described    in    \cite{Semenov2010}.    The    gas-grain    chemical
network\footnote{\url{http://www.physics.ohio-state.edu/~eric/research.html}}
from   the  Ohio  State   University  Astrophysical   Chemistry  Group
($\sim$7840  reactions   and  $\sim$840  species)  is   used  in  this
study.  The  detailed  description  of  the network  and  reaction  rate
calculations  of the  associated chemical  reactions can  be  found in
\cite{Garrod2008} and \cite{Semenov2010}.  We adopted standard grain
parameters  such as radius of  0.1 \micron,  density of 3  g cm$^{-3}$  and a
dust-to-gas mass ratio of 0.01. The initial abundance and
desorption  energy   of  the   molecular  species  were   adopted  from
\cite{Garrod2008}  and \cite{Garrod2013}.  The ratio  of
diffusion-to-desorption  energy  is  assumed  to  be  0.5.  We  used
\emph{Saptarsy} to obtain the spatio-temporal abundance evolution by
solving the  rate equations based  on the overall  molecular formation
and destruction  terms. Chemical evolutionary models  are simulated up
to a time-scale of  \num{1e5} year using discrete time-steps separated
by few hundreds to thousands year. Conservative values of relative and
absolute  tolerances of 10$^{-6}$ and  10$^{-20}$,  respectively, were
used   throughout  the   calculation.  To   avoid   complexity,  direct
photo-dissociation  and -ionization  reactions are  not  considered in
this work  since hot  core phases were  not dominated by  UV radiation.
However,  the effects  of UV  photons  become prominent  at the  final
phases  of  hot  core  evolution  that is during  the  transformation  to
ultracompact  \htu  regions  which  require  a detailed  modeling  of  the
radiation  field inside hot  cores; this  is the  focus of  a separate
study (St\'{e}phan et al., 2014, in preparation).

%--------------------------------------------------------------------------------------------------------------------------------------------------------------------
\subsection{Density distribution}
\label{subsec:pmodel}
\begin{table}
\caption{List of abbreviations used in the nomenclature of various models }              
\label{tab:m_name}     
\centering                                     
\begin{tabular}{l l l }          
\hline\hline   
Description/parameter & Value  & Abbrv. \\
\hline  
Cold core models          & --     & cc  \\
Hot molecular core models & --     & hmc \\
\hline
Plummer radius        & 3000~AU & r3 \\
                      & 3500~AU & r3.5\\
                      & 4000~AU & r4 \\
                      & 4500~AU & r4.5\\
\hline                     
Constant temperature   & 10~K     & t10 \\
at cold core phase    & 15~K     & t15 \\
\hline
Luminosity evolution  & L$^{2}$ & \emph{l}2 \\
during warm-up phase  & L$^{4}$ & \emph{l}4 \\
                      & L$^{7.5}$ & \emph{l}7 \\
\hline                      
Cosmic-ray ionization & \num{1.37e-17}~\si{s^{-1}} & cr17 \\
rate                  & \num{1.37e-16}~\si{s^{-1}} & cr16 \\
\hline                                            
\end{tabular}
\end{table}
In  a complex  environment such as  a hot  core, the  physical conditions
(e.g., density, temperature) not only vary with time, but also with
the   distance   from   the   central  protostars.   Recent   (sub-)mm
interferometric studies also indicated that internal rotational motion,
outflows,  etc. can  be associated  with  \ac{HMCs} \citep{Beltran2011,
  Furuya2011}. Several  studies investigated the  density distribution
of     these     objects     using     dust     continuum     emission
\citep{vanderTak2000,Beuther2002}   and  found  a   power-law  density
distribution  on larger scales;  recent studies  using interferometric
observations suggested  a Plummer-type distribution  at smaller scales
\citep{Rolffs2011}. Plummer-type profiles  can be represented as n(r)=
n$_{c}$(1+c(r/r$_{p}$)$^{2}$)$^{\gamma}$,  where c  is  a constant  and
n$_{c}$ and  r$_{p}$ represent the central density  and Plummer radius
(this parameter  sets the size of  the core at which  n$_{c}$ drops to
half of its  value) and the exponent  ($\gamma$) sets the
density distribution at radii larger than the Plummer radius.  We set up a
spherical hot  core model that contains  a high-mass  protostar at the
center embedded in a dense core  of gas and dust. We adopted a Plummer-type  distribution for  the hot  core models  with a  peak  density of
$2\times10^{7}$~cm$^{-3}$, c=0.31951 and $\gamma$=-5/2 throughout this
work.  The density  distribution profile  does not  change  with time,
but we  have considered  various  types  of  density profiles  by
varying the Plummer radius (see Table.~\ref{tab:m_name}). Assuming the
standard dust-to-gas mass  ratio, the mass of the model  core with Plummer
radius $\sim$3000~AU is estimated  to be around 45\msun, indicating that
it   barely   has  the   potential   to   form   a  high-mass   star
($\textgreater$~10\msun).
%

%----------------------------------------------------------------------------------------------------------------------------------------------------------------------
\subsection{Protostellar evolutionary models}
\label{subsec:proto_evol}
The  central protostar  is the  only source  of heating  in  our model
set-up and thus the temperature  of hot cores depend on the luminosity
of  the  protostar.  We   used  the protostellar  evolutionary  models  of
\cite{Hosokawa2009}  (spherical   accretion  model  with   a  constant
accretion  rate of  10$^{-3}$~\msun  yr$^{-1}$ \citep{Rolffs2011})  to
investigate the  spatio-temporal evolution  of the temperature in  our hot
core  models.  The  temperature  evolution  from IRDC  phases
(10--15~K) to the formation of protostars is not explicitly covered in
this   model.  \cite{Garrod2008}  explored   the  effect   of  warm-up
timescales (the required time to reach the characteristic temperature
of 100~K) and found that  with a longer warm-up timescale the chemical
complexity also increases. To imitate the luminosity evolution during
the warm-up  phase, we interpolated  the luminosity over a  fixed time
interval  (\num{5e4} to  \num{7.5e4} year)  using different  power-law
indices of 2, 4, and 7.5 (see Fig.~\ref{fig:lum_and_temp}(a)). The
starting  point of  the  luminosity interpolation  (\num{5e4}
year) was  chosen in such a  way that the temperature  at the immediate
vicinity of  the protostars is around 15~K; the  luminosity at
the end point is adopted from the models of \citeauthor{Hosokawa2009} so that
the temperature reaches up to 100~K  at the center. We then used these
luminosity profiles to calculate  the temperature. The power-law index
7.5 was chosen to reproduce  a similar temperature evolution as used in
\cite{Garrod2008}.   We  restricted the  highest  value of  the protostellar
luminosity   to  \num{1e5}~\lsun  following   the  result   of  recent
observational studies of hot cores \citep{Beuther2009}.

%-----------------------------------------------------------------------------------------------------------------------------------------------------------------------
%
\subsection{Nomenclature of the models}
\label{subsec:m_name}
We  developed  various models  by  varying  the  physical parameters  to
explore    the    spatio-temporal    variation   of    the    chemical
evolution. Instead  of commonly used tags  (e.g., model:a), we
use  different  combinations of  abbreviations  to  refer to a  particular
model.   We   summarize  the   physical   parameters  and   associated
abbreviations  in  Table~\ref{tab:m_name}. For  example,  a hot  core
model tagged  as \emph{cr16-r3-l2} refers to  a model where
the Plummer  radius is 3000~AU  (r3), the power-law index  of the protostellar
luminosity evolution is 2  (\emph{l}2), and the cosmic-ray ionization
rate  is \num{1.37e-16}~\si{s^{-1}}.  Similarly, the  \emph{cr-17 models}
refers to all the hot core models with a cosmic-ray ionization rate of
\num{1.37e-17}~\si{s^{-1}}.
%
%---------------------------------------------------------------------------------------------------------------------------------
\begin{figure*}
\includegraphics[width=17cm]{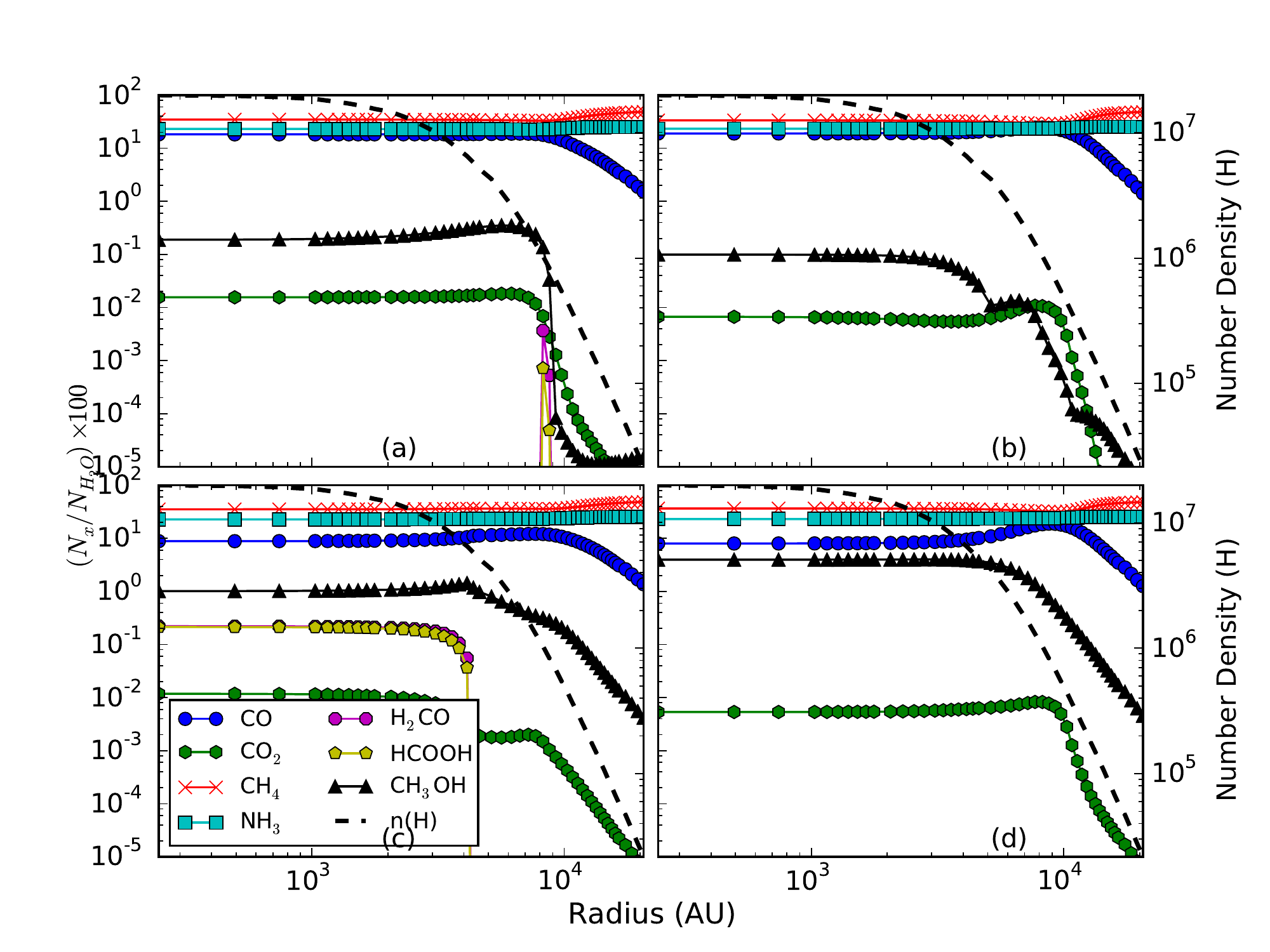}
\caption{%
Evolution of molecular abundance on grains with respect to \ce{s-H2O} in percent at 5x10$^{4}$ year for different cold core models: (a) \emph{cc-cr17-t10},
(b) \emph{cc-cr16-t10}, (c) \emph{cc-cr17-t15}, and (d) \emph{cc-cr16-t15}.} 
\label{fig:ice_abun}
\end{figure*}
%-----------------------------------------------------------------------------------------------------------------------------
%
\subsection{Simulation of spectral data cubes}
\label{subsec:data_cube}

The radiative transfer code  RADMC-3D was used to simulate the
spectral  data  cubes  of  different  COMs at  various  wavebands  and
evolutionary stages;  the   radius adopted for the  hot core  models was
0.1~pc ($\sim$20600~AU).  We started with  a density distribution  and a
luminosity  evolution  profile  of   the  central  protostar  and  used
RADMC-3D  to  calculate   the  spatio-temporal  evolution  of the
temperature.  In  the  next  step,  we  extracted  the  1D  density  and
temperature distribution  profiles from the 3D hot  core models. These
density-weighted profiles along with the initial abundance of selected
molecules  served  as  the  input   for  \emph{Saptarsy}, which
calculates the  spatio-temporal chemical evolution in  hot cores.  The
1D  abundance distribution  profiles of  respective molecules  were then
used  as  input  abundance  profiles  in  RADMC-3D.  Finally,
simulated  spectra   spanning  over  a  frequency   range  at  desired
evolutionary  timescales were  generated  assuming local  thermodynamic
equilibrium (LTE). As suggested by \cite{Herbst2009}, the ideal way to
generate simulated  spectra would be  to use full  non-LTE statistical
equilibrium  excitation of  the molecules.  However,  collisional rate
coefficients required  for non-LTE modelings  are available only for  a few
molecules. Moreover,  the high  densities and radiative  coupling with
far-infrared     radiation     fields      make     LTE     a     good
approximation.  Therefore,  for  this   study  we  only  considered  LTE
excitation  to  generate comparable  simulated  spectra for  different
COMs.  To facilitate the comparison with observations, simulated spectra were
convolved    with    different     telescope    beam    sizes    using
\emph{MIRIAD}\footnote{\url{http://bima.astro.umd.edu/miriad/}},
also  taking into account  the distance  to the  source. The  hot core
models  were placed  at a  fiducial distance  of 2~kpc.  Molecular data
files  of  various  COMs  were  retrieved  from  the Cologne  Database  for
Molecular                                                  Spectroscopy
(CDMS)\footnote{\url{http://cdms.ph1.uni-koeln.de/cdms/portal/}}
\citep{cdms2,cdms1}, the  JPL Catalog  \citep{jpl_cat}, and  the VAMDC database
\citep{Rixon2011}.
%---------------------------------------------------------------------------------------------------------------------------------------------------------------
\section{Results}
\label{sec:results}

\begin{table*}
\caption{Molecular abundances on grains with respect to \ce{s-H2O} in percent: observed and simulated values}              
\label{table:ice_abun}     
\centering                                     
\begin{tabular}{l c c c c c c c c c}          
\hline\hline                        
Molecules & W33A$^{1}$ & NGC~7538        & Sgr A*$^{2}$ &\citeauthor{Garrod2008} & \citeauthor{Oberg2011} & \citeauthor{Garrod2013} & Model        & Model\\   
          &            &  IRS~9$^{2}$    &              &  (2008)                &   (2011)               &    (2013)               & \emph{cc-cr17-t15} & \emph{cc-cr16-t15} \\
\hline                                  
\ce{CO}    & 8   & 16 & <12   & 19      &13    & 57     & 8.75& 7.9 \\     
\ce{CO2}   & 13  & 22 & 14    & 4.1(-3) &13    & 18     & 0.01& 0.005 \\
\ce{CH4}   & 1.5 & 2  & 2     & 22      & 2    & 1.9    & 35.2& 36.6 \\
\ce{NH3}   & 15  & 13 & 20-30 & 25      & 5    & 18     & 22.7& 23.1  \\
\ce{H2CO}  & 6   & 4  & <3    & 4.3     & $\leq$2& 1.6  & 0.22& 9(-10) \\
\ce{HCOOH} & 7   & 3  & 3     & 3.2(-6) & --   &  --    & 0.21& 1.7(-10)\\
\ce{CH3OH} & 18  & 5  & <4    & 15      &  4   &  6.9   & 1.03&  4.0 \\
\hline                                            
\end{tabular}
\tablebib{
(1)~\citet{Gibb2000}; (2)~ see \citet{Gibb2000} for original references.  a(b) refers to a$\times$10$^{b}$  throughout the manuscript.
}
\end{table*}
%----------------------------------------------------------------------------------------------------------------------------------
 \begin{figure*}
 \centering
   \begin{tabular}{@{}ccc@{}}
     \includegraphics[width=.3\textwidth]{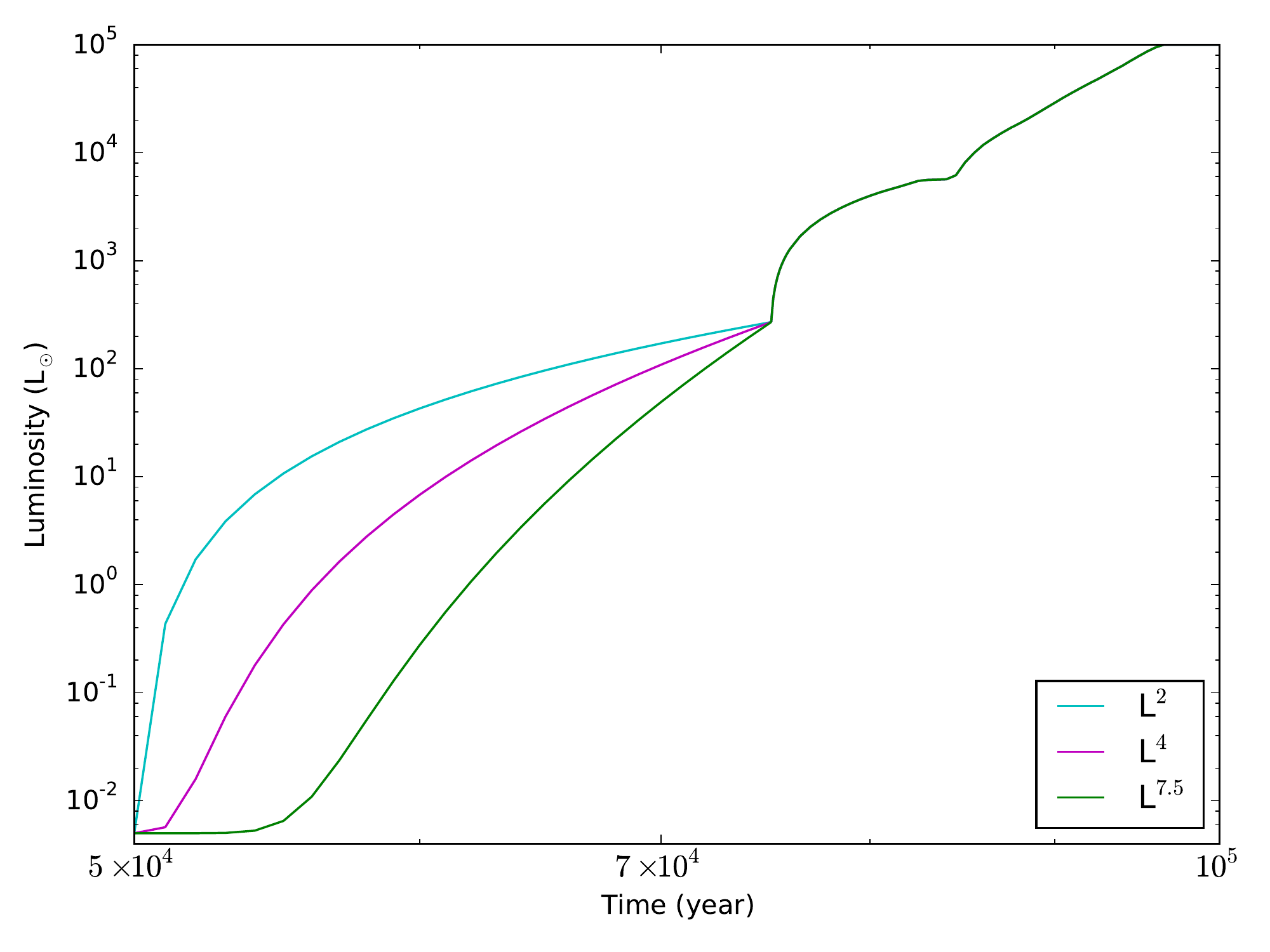} &
     \includegraphics[width=.31\textwidth]{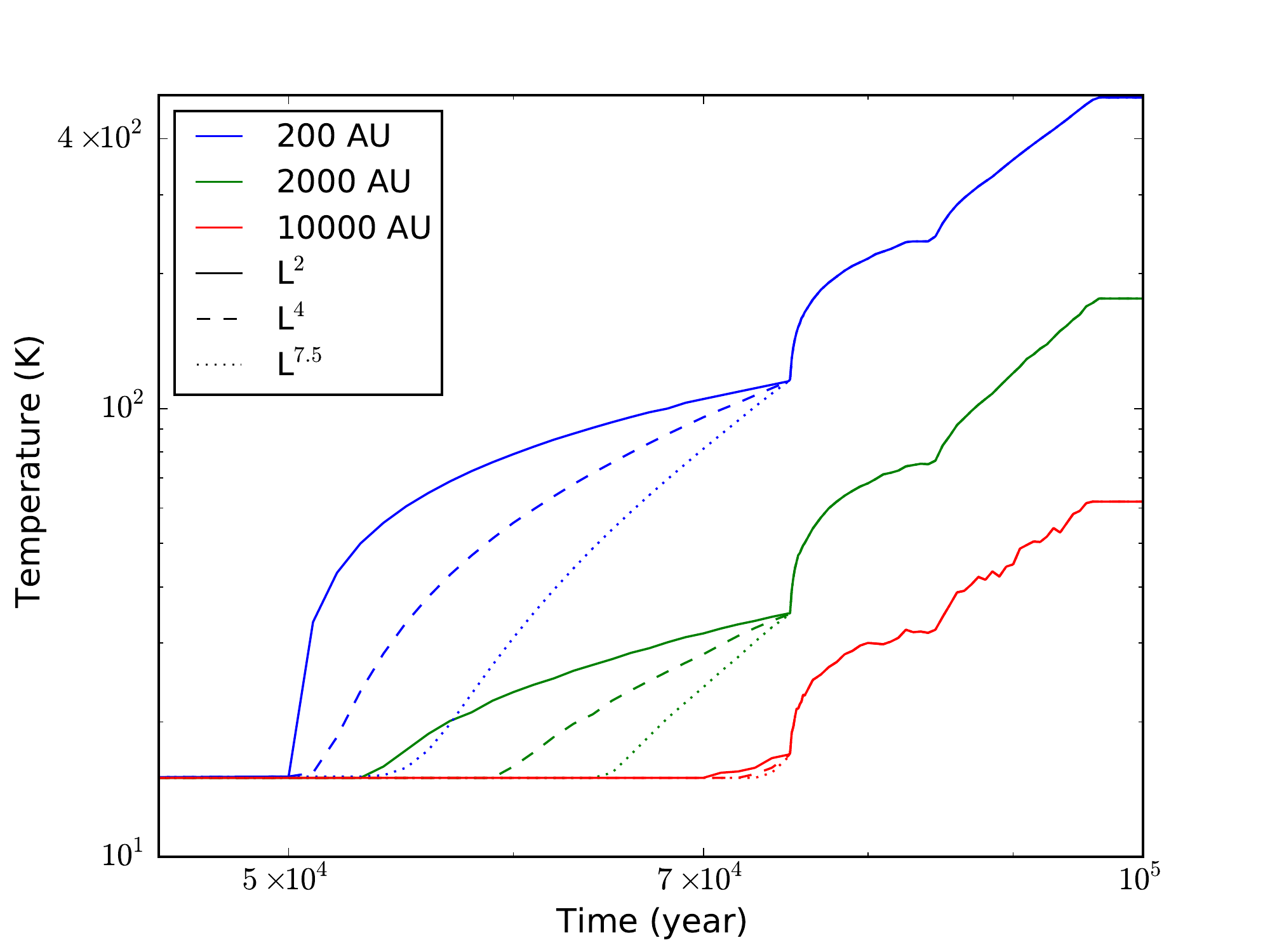} &
     \includegraphics[width=.31\textwidth]{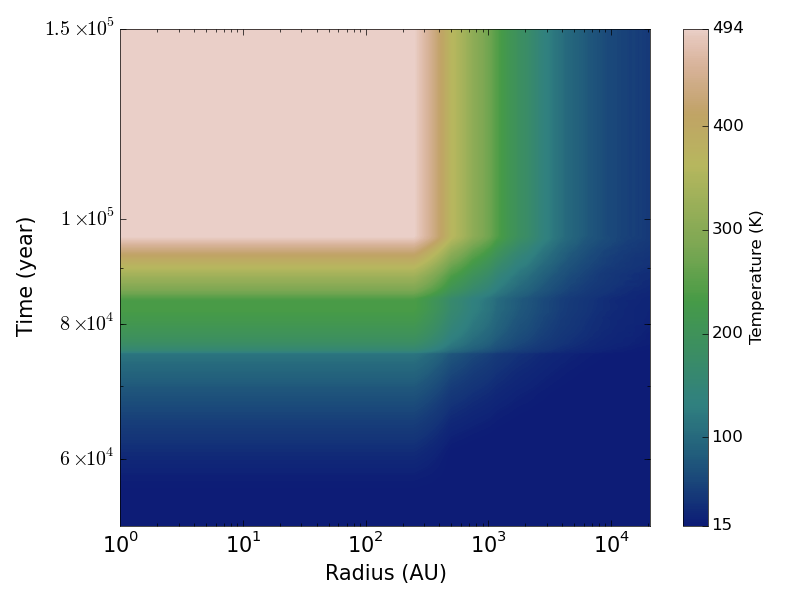} \\
     (a) & (b) & (c) \\
   \end{tabular}
   \caption{(a) Luminosity evolution profiles for different power-law indices. (b) Temperature evolution in the hot core model \emph{hmc-r3} at different radii:  $\sim$200~AU (\emph{blue}), $\sim$2000~AU
 (\emph{green}) and $\sim$10000~AU (\emph{red}). Temperature distribution obtained using different protostellar luminosity evolutions are shown with continuous
 (\emph{l2}), dashed (\emph{l4}), and dotted (\emph{l7}) lines. (c) Spatio-temporal evolution of the temperature in the hot core model \emph{hmc-r3}.}
 \label{fig:lum_and_temp}
 \end{figure*}

%--------------------------------------------------------------------------------------------------------------------------------
\subsection{Molecular abundance on grains}
\label{subsec:ice_abun}
The  abundance of various  species on  grains that is  the \emph{ice}
abundances,  are  important  parameters  that  influence  the  chemical
evolution  of  hot cores.  At  low  temperature (10--15~K),  molecules
accrete onto  grains and  thus become locked  on the grain  surface; this is also
known as freeze-out or depletion of molecules. We represent the grain
surface species with \emph{s-}  meaning that  \ce{s-H2O} represents water on
grain surface.  A reduced abundance of some molecules at the center of
the cores, that is  in the high density regions, have  been observed in a
number of  starless cores indicating  the clear presence  of molecular
ice formation at this  stage \citep{Bergin2007}. The accretion rate of
molecules onto  grain surfaces depends on the  sticking coefficient of
molecules to  the grain, the grain  surface area, the  velocity of the
molecules in gas phase, and the number density of dust grains. Another
critical factor that regulates  the \emph{ice} formation is the dust
temperature  at the  early pre-stellar  stages. Most  of the  hot core
models  so far considered  the initial  temperature as  10~K. However,
recent large scale surveys of  high-mass star forming regions indicate
an    average    temperature    of    15~K    for    starless    cores
\citep{Wienen2012,Hoq2013}. The slight increase of temperature from 10
to  15~K increases  the  mobility of  molecules  on grains  (reactive
radicals such as \ce{OH}, \ce{CH3}) and thus provides a more productive
environment to form complex molecules.

We therefore explored models  with  temperatures of 10~K and 15~K during
the prestellar phase. Another factor that influences the
abundance  of molecules frozen  to grain  surfaces is  the interaction
with cosmic rays that can dissociate and even evaporate ices. There is
a debate  about the cosmic-ray rate in  the literature based  on the
observations  of  \ce{H3+}  \citep{Indriolo2012}. Motivated  by  these
studies, we  also explored  the effect of   cosmic-ray  ionization rate
with   two  distinct   values   of.  \num{1.37e-16}~\si{s^{-1}}   and
\num{1.37e-17}~\si{s^{-1}}   \citep{Padovani2009}.     We   ran   four
different     models    to     calculate     the    ice     abundances,
 \emph{cc-cr17-t10}, \emph{cc-cr16-t10}, \emph{cc-cr17-t15},
and  \emph{cc-cr16-t15}.  The  timescale of  the starless  phase is
uncertain,  but we  have some  guidance  from large  scale surveys  of
high-mass star forming regions \citep{Tackenberg2012}; thus we adopted
a  timescale   of  \num{5e4}  year  for  the   pre-stellar  phase.  In
Fig.~\ref{fig:ice_abun}  we  present the  grain  surface abundance  of
selected molecules at the pre-stellar stage with respect to \ce{s-H2O}
in  percent.  All  these   models  show  a similar  \ce{s-H2O}  relative
abundance  \num{2e-4} with respect to the total hydrogen number
density  (n(\ce{H}+\ce{2H2})) at  the  center. We  also tabulated  the
\emph{ice}  abundance  of these  molecules  from  the literature  in
Table~\ref{table:ice_abun}.  Comparing the  observed and our simulated
values,  it appears  that  cold core  models \emph{cc-cr17-t15}  and
\emph{cc-cr16-t15}  produce  \ce{s-CH3OH}  abundances close  to  the
observed values among the  other molecules. Since \ce{s-CH3OH} acts as
the parent  molecule of other COMs,  we have used these  two models for
the subsequent calculations.
\subsection{Spatio-temporal variation of temperature}
\label{subsec:e-temp}
Temperature plays an important role  in chemical evolution as the rate
of most  of the  chemical reactions strongly  depend on  it.  Previous
studies  \citep[e.g.,][]{Viti1999, Garrod2006} investigated  the effect
of  temperature  evolution  assuming  a power-law  distribution  in  a
constant density set-up. However  no self-consistent treatment is used
in  these studies to  calculate the  spatial structure  of temperature
using any protostellar luminosity function.  A significant improvement
for  temperature  evolution  is  achieved by  using a  proper  radiative
transfer  calculation  that  is   tied  to  an  evolving  protostellar
luminosity  model.  As required  by  the  chemical evolutionary  model
\emph{Saptarsy}, temperature  evolution of  the hot core  models for
different  luminosity  evolution   are  calculated  at  some  discrete
time-steps  using \emph{RADMC-3D}  (Fig.\ref{fig:lum_and_temp}).  We
estimated  the   heating  and  cooling  timescales   of  dust  grains
explicitly to  check whether dust temperature  calculation at discrete
time-steps   would   yield   a   realistic  estimate   of   temperature
evolution.  To  estimate  the  cooling  timescales,  we  compared  the
absorption and emission processes of dust grains. A typical dust grain
of radius  0.1~\micron at 20~K  would emit 1.14  $\times$ 10$^{-12}$
erg s$^{-1}$  assuming an efficiency factor  of 10$^{-4}$ (considering
the  opacities  at  (sub-)mm  wavelengths). Similarly,  considering  our
assumed density of grains (3~\si{g cm^{-3}}), the thermal energy of the
dust grain  would be \num{2.5e-6}~erg. Therefore, the  time-scale for
thermalization of dust  grains would be about the fraction of a
year (private  communication with C.  P. Dullemond). As a  result, any
changes  in   temperature  of  the   hot  core  model   become  nearly
instantaneous compared to the changes in luminosity and thus ensure a
nearly  steady-state  condition in  the  hot  core  models. Dust  mass
opacity coefficients were  retrieved from \cite{Ossenkopf1994}. We also
assumed that gas  temperature is same as the  dust temperature, which is
 a reasonable assumption considering the high density of hot cores.
\cite{Doty2002} made a detailed calculation of the gas temperature
for their hot core model and  found that the values are similar to the
ones  obtained  using   the  assumption  of  T$_{gas}$=T$_{dust}$.  In
Figs. ~\ref{fig:lum_and_temp}(b)  and  \ref{fig:lum_and_temp}(c), the
spatio-temporal   evolution  of  temperature   for  hot   core  models
\emph{hmc-r3}  are  shown. At  the  center  of  the hot  cores,  the
temperature remains  constant (15~K) up  to \num{5e4} year  (cold core
phase), then it rises to  the typical hot core temperature (100~K) during
the next \num{2.5e4} year (warm-up phase), and after that, it reaches up
to 500~K at the end of the evolutionary time-scale that is \num{1e5} year
(hot core phase). It is evident  from the figures that the duration of
these phases varies as a function of distance from the central star.
\subsection{Spatio-temporal abundance variation}
\label{subsec:com_abun}

\begin{figure*}
\centering
\includegraphics[width=\textwidth,height=0.8\textwidth]{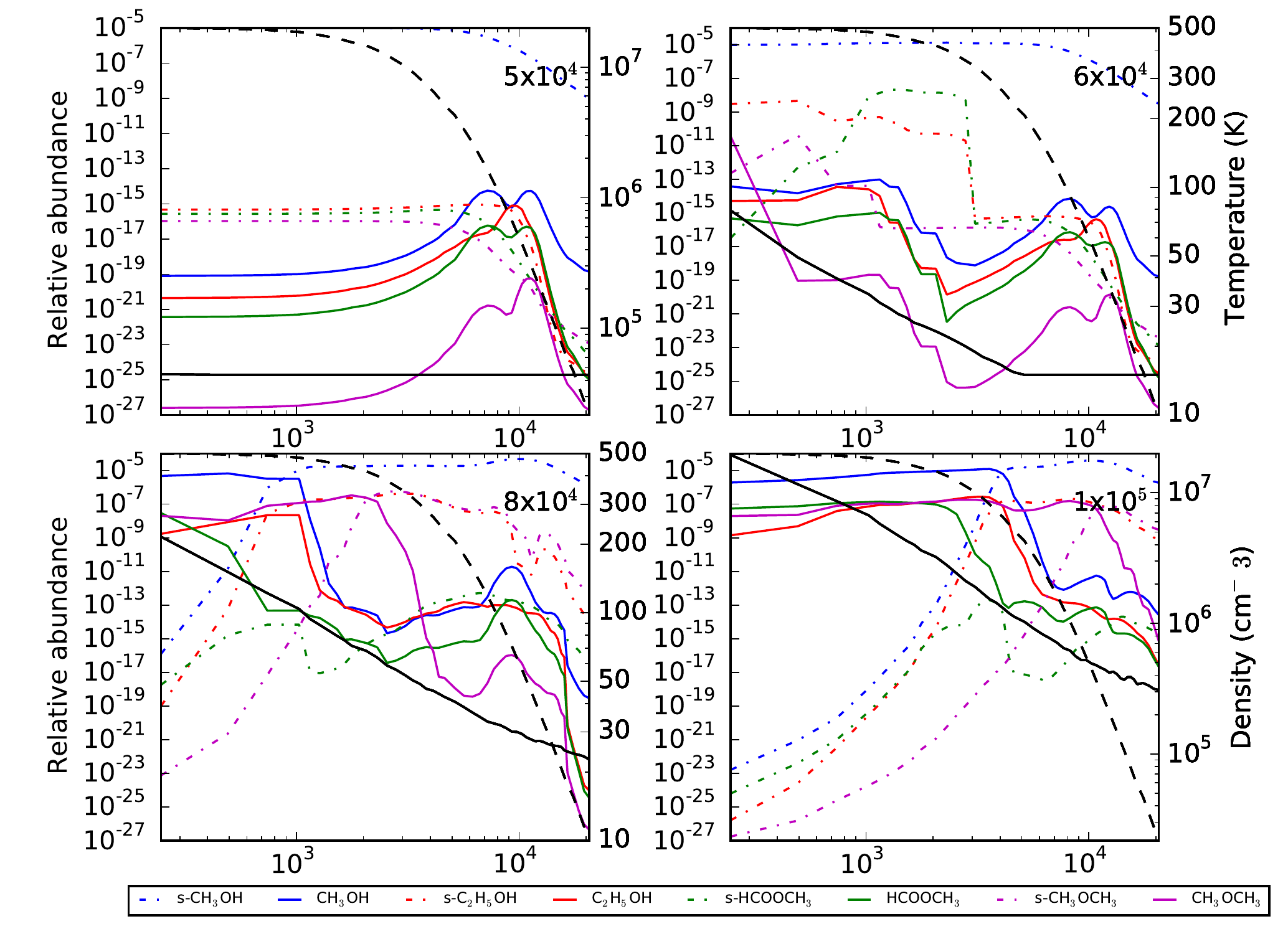}
\caption{Spatio-temporal abundance variation of selected COMs on the grain (\emph{dot-dashed line}) and in gas phases (\emph{continuous line}) are shown
for the hot core model \emph{cr16-r3-\emph{l}2}. \emph{Black dashed} and \emph{continuous} lines show the radial variation of the density and the temperature
respectively. \emph{Time-stamps} (in year) for different temporal snapshots  are annotated in the upper right corner of the respective plots. Note the
alternative labeling of the \emph{density} and \emph{temperature} axes on the right side of the respective plots.}
\label{fig:com_evol_a}
\end{figure*}

The  most  abundant molecule  that  dominates  the  spectra at  (sub-)mm
wavebands   is   \ce{CH3OH}.     Following   \ce{CH3OH},   the   other
oxygen-bearing \ac{COMs} that contribute most of the lines in (sub-)mm
wavebands   (480--1907~GHz)   are   \ce{CH3OCH3},   \ce{HCOOCH3},   and
\ce{C2H5OH}  \citep{Zernickel2012}.  These COMs  are also
referred  as  \emph{first-generation} molecules  \citep{Bisschop2007,
  Garrod2008,  Herbst2009};  these molecules  are  mainly produced  by
reactions  between primary  \emph{ice}  species (also  known as  the
\emph{zeroth-generation}  molecules).  To investigate  the favorable
physical conditions for the  formation of \emph{first-generation} COMs
during  the warm-up  phase, we  ran further  models using  the initial
physical conditions discussed in Sect.~\ref{subsec:ice_abun}. We ran
24  hot core  models  in total  up  to \num{1e5}  year with  different
physical  parameters  such  as  with  four  different  static  density
distributions    (\emph{r3},    \emph{r3.5},    \emph{r4},    and
\emph{r4.5})  along  with   three  different  luminosity  evolutions
(\emph{l2},  \emph{l4}, and  \emph{l7})  and two  different
cosmic-ray  ionization rates (\emph{cr-16}  and \emph{cr-17}) (see
Table~\ref{tab:m_name}).   The dust  temperature is  one of  the most
dominant   factors   that  control   the   gas   and  grain   surface
abundances. Thanks to the spatio-temporal evolution of the temperature
structure, our model set-up  is particularly useful to investigate the
transition from grain to gas phase abundance of various molecules as a
function  of  evolutionary  stages  and  distances  from  the  central
star. The  spatio-temporal variations of grain  and gas-phase abundance
of  selected   COMs  for  various   hot  core  models  are   shown  in
Figs.~\ref{fig:com_evol_a},
\ref{fig:com_evol_b}--\ref{fig:com_evol_e}. At the initial stages, the
\emph{ice} abundance of COMs dominate over the gas phase abundances but  with increased temperature  \emph{ices} evaporate  from the
grain  surfaces and  thus  boost  the  gas  phase abundances.  With
temporal  evolution, the  radius of  the evaporation  font  (where the
temperature  is  equal  to   the  water  ice  evaporation  temperature
of  100~K)  also   increases  (see  Fig.~\ref{fig:100k_radius}),  and
consequently,  the spatial  distribution of  gas phase  abundances also
spreads   out.    However,   at   the    outer   region   (\textgreater
\num{4.5e3}~AU),   where  the temperature   remains   lower  than   100~K,
\emph{ice} abundance of COMs dominate the gas phase abundances.
One of  the important  features of this  work is the  derived temporal
variation  of the  spatial distribution  of  COMs for  solid  and
gaseous phases.  These profiles will  be particularly useful  for modeling
the   observed   high  resolution   emission   line   maps  of   these
molecules.  Gas phase  abundance of  selected COMs  from  the existing
observational    and    modeling     studies    are    collected    in
Table~\ref{table:obs_mol_abun};  a similar compilation  of  the observed
value of  these molecules for  a number of  hot cores can be  found in
\cite{Bisschop2007}. The gas phase  abundance of different COMs at the
center   of    various   hot    core   models   are    summarized   in
Tables~\ref{table:cr16_mol_abun}  and  \ref{table:cr17_mol_abun}. These
abundances are comparable with  the observed abundance of various COMs
at  least  for  the \emph{cr16}  models. However,  while  comparing  the
observed and  simulated values, one should  also keep in  mind that the
quoted  observed values are  mostly conversions  of the  derived column
densities with respect to \ce{H2}  or \ce{CO} column densities using a
constant  temperature,  which  introduces  additional  uncertainty.  We
discuss     this     aspect    in     detail     in
Sect.~\ref{subsec:myxclass_fit}.

%----------------------------------------------------------------------------------------------------------------------------------------
%-------------------------------------------------------------------------------------------------------------------------------------------------------%
%---------------------------------------------------------------------------------------------------------------------------------------------------------------
\begin{table*}
\caption{Abundance of COMs in Hot molecular cores: observed and simulated values}              
\label{table:obs_mol_abun}     
\centering                                     
\begin{tabular}{l c c c c c c c }          
\hline\hline                        
Molecules    & IRAS            & Orion Hot   & NGC        & G29.96$^{4}$  & G327.3-0.6$^{5}$& \citeauthor{Garrod2008} &  \citeauthor{Garrod2013}\\   
             & 16293-2422$^{1}$&  Core$^{2}$ & 6334I$^{3}$&               &                 &  (2008)                 &    (2013)               \\
\hline                                                                                         
\ce{CH3OH}   &3.0(-7)          &  2.2(-6)       & 4.7(-6)    & 6.6(-8)        & 5.3(-8)        &  3.2(-6)                & 1.1(-5)                 \\
\ce{C2H5OH}  & --              &  --            & 4.7(-8)    & 1.0(-8)        & 2.6(-9)        &  1.5(-9)                & 5.9(-8)                 \\
\ce{CH3OCH3} &2.4(-7)          &  6.8(-8)       & 1.0(-6)    & 3.3(-8)        & 3.4(-7)        &  3.0(-9)                & 4.8(-8)                 \\
\ce{HCOOCH3} &2.0(-7)          &  --            & 2.4(-7)    & 1.3(-8)        & 5.0(-8)        &  7.8(-11)               & 9.2(-8)                  \\
\hline                                            
\end{tabular}
\tablebib{
(1)~\citet{Cazaux2003}; (2)~\citet{Crockett2014}; (3)~\citet{Zernickel2012}; (4)~\citet{Beuther2009}; (5)~\citet{Gibb2000a};.
}
\end{table*}

%-----------------------------------------------------------------------------------------------------------------------------------
\begin{table*}
\small
\caption{Simulated values of COMs at the center of various hot core models with the cosmic ray ionization rate of \num{1.37e-16}}              
\label{table:cr16_mol_abun}     
\centering                                     
\begin{tabular}{l c c c c c c c c c c c c}          
\hline\hline                        
Molecules & r3-\emph{l}2 & r3-\emph{l}4  & r3-\emph{l}7 & r3.5-\emph{l}2 & r3.5-\emph{l}4& r3.5-\emph{l}7 & r4-\emph{l}2  & r4-\emph{l}4 & r4-\emph{l}7  & r4.5-\emph{l}2 & r4.5-\emph{l}4 & r4.5-\emph{l}7 \\
\hline                                  
\ce{CH3OH}    & 4.5(-7) & 1.3(-6) & 2.0(-6) & 1.0(-6)   & 1.7(-6) & 1.7(-6) & 1.3(-6) & 9.8(-7) & 1.9(-6) & 6.1(-7) & 1.3(-6) & 2.4(-6)\\ 
\ce{C2H5OH}   & 4.3(-10)& 2.3(-9) & 9.5(-9) & 9.0(-10)  & 3.8(-9) & 7.9(-9) & 1.1(-9) & 2.3(-9) & 8.3(-9) & 5.8(-10)& 3.0(-9) & 1.0(-8)\\   
\ce{CH3OCH3}  & 6.6(-9) & 1.7(-8) & 3.5(-8) & 1.3(-8)   & 2.0(-8) & 3.1(-8) & 1.6(-8) & 1.4(-8) & 3.2(-8) & 8.9(-9) & 1.7(-8) & 3.4(-8)\\ 
\ce{HCOOCH3}  & 3.4(-8) & 1.1(-7) & 1.4(-7) & 4.6(-8)   & 1.1(-7) & 1.4(-7) & 4.7(-8) & 1.1(-7) & 1.4(-7) & 3.7(-8) & 1.1(-7) & 1.4(-7)\\ 
\hline                                            
\end{tabular}
\end{table*}
%---------------------------------------------------------------------------------------------------------------------------------------------
\begin{table*}
\small
\caption{Simulated values of COMs at the center of  various hot core models with the cosmic ray ionization rate of \num{1.37e-17}}              
\label{table:cr17_mol_abun}     
\centering  
\resizebox{\textwidth}{!}{%
\begin{tabular}{l c c c c c c c c c c c c}          
\hline\hline                        
Molecules & r3-\emph{l}2 & r3-\emph{l}4  & r3-\emph{l}7 & r3.5-\emph{l}2 & r3.5-\emph{l}4& r3.5-\emph{l}7 & r4-\emph{l}2 & r4-\emph{l}4 & r4-\emph{l}7 & r4.5-\emph{l}2 & r4.5-\emph{l}4 & r4.5-\emph{l}7 \\
\hline                                  
\ce{CH3OH}    & 2.1(-6) & 2.2(-6) & 2.3(-6) & 2.2(-6) & 2.2(-6) & 2.3(-6) & 2.1(-6) & 2.2(-6) & 2.7(-6) & 2.1(-6) & 2.6(-6) & 2.3(-6)\\ 
\ce{C2H5OH}   & 9.3(-11)& 1.2(-10)& 1.5(-10)& 9.3(-11)& 1.2(-10)& 1.5(-10)& 9.2(-11)& 1.2(-10)& 1.5(-10)& 9.4(-11)& 1.2(-10)& 1.5(-10)\\
\ce{CH3OCH3}  & 6.8(-9) & 7.0(-9) & 8.1(-9) & 6.4(-9) & 7.1(-9) & 8.1(-9) & 6.9(-9) & 7.0(-9) & 6.3(-9) & 6.9(-9) & 4.7(-9) & 8.1(-9) \\
\ce{HCOOCH3}  & 3.4(-8) & 3.3(-8) & 3.7(-8) & 3.2(-8) & 3.3(-8) & 3.7(-8) & 3.5(-8) & 3.3(-8) & 2.6(-8) & 3.5(-8) & 1.8(-8) & 3.7(-8) \\ 
\hline                                            
\end{tabular}%
}
\end{table*}

%---------------------------------------------------------------------------------------------------------------------------------------------------------

\cite{Garrod2008} suggested that  cosmic-ray induced photodissociation
of   molecules  (e.g.  \ce{CH3OH})   produce  heavier   radicals  that
effectively  diffuse  over  grains  at relatively  higher  temperature
($\textgreater$20~K) and  produce other COMs such as  \ce{C2H5OH}. We
found   that  a   higher   value  of   the cosmic-ray  ionization   rate
of \num{1.37e-16}  supports the formation  of more COMs  over grains
and eventually  also in the gas phase. At  the initial stages  of the evolution,
the \emph{cr-16} models show nearly similar grain surface abundances for
\ce{C2H5OH},  \ce{CH3OCH3}, and \ce{HCOOCH3}  over larger  radii;  but for
the \emph{cr-17} models,  these abundances not  only differ by  orders of
magnitude but also  vary with radii. The relative  abundance of these
COMs  with respect  to  \ce{CH3OH} also  differ significantly  between the
\emph{cr-17}       and        \emph{cr-16}       models       (see
Figs.~\ref{fig:com_evol_a},
\ref{fig:com_evol_b}--\ref{fig:com_evol_e}). During the warm-up phase,
relative gas-phase abundances of COMs varies noticeably depending on the
physical  conditions.  Moreover, even  with a  constant
warm-up time  scale, a slower  rate of temperature  increment provides
favorable        platforms       to        form        more       COMs
(Tables~\ref{table:cr16_mol_abun} and \ref{table:cr17_mol_abun});
a similar  behavior was  also found by  \cite{Garrod2006}. In  both 
models,  the grain  surface  abundance  of  \ce{C2H5OH}  are  higher  than
those of \ce{CH3OCH3} and  \ce{HCOOCH3} at the center,  but the final
gas  phase  abundance  of  \ce{CH3OCH3} and  \ce{HCOOCH3}  are  higher
than that of \ce{C2H5OH},  suggesting that \ce{CH3OCH3} and \ce{HCOOCH3}
also  have effective  gas-phase formation  routes in  addition  to the
grain surface formation routes.  The grain surface abundance of these COMs
reach  their highest values  at  the  outer regions  where the  highest
temperature reaches only  $\sim$50~K.  Interestingly, a similar pattern
for grain surface abundances of \ce{HCOOCH3} $\textless$ \ce{C2H5OH}
$\textless$ \ce{CH3OCH3}  $\textless$ \ce{CH3OH} has  been observed at
the outer  regions of  both the  \emph{cr-16} and  \emph{cr-17} models.
During  the  warm-up  phase,    the gas  phase  abundance  of
\ce{CH3OCH3} initially dominates other  COMs; the final highest
gas-phase  \ce{CH3OCH3}  abundance  extends  more than  that of \ce{CH3OH}  in
the \emph{cr-16}  models,   irrespective  of  density   distribution  and
temperature evolution.  In the \emph{cr-17} models  the spatial distribution
of \ce{CH3OCH3}  shows two  different peaks; the  extent of  the first
peak is always narrower than the \ce{CH3OH} distribution, but the extension
and magnitude  of the second  peak depends on the  density distribution
(see Figs.~\ref{fig:com_evol_d} and \ref{fig:com_evol_e}).

\subsection{Spatio-temporal variation of formation routes of COMs}
\label{subsec:com_form}
As    stated   earlier,   we used the    OSU   gas-grain    chemical   network
\citep{Garrod2008} and the  associated chemical
pathways  are   described  extensively  in   the  corresponding  paper
\citep[see also][for a general summary]{Herbst2009}. For completeness,
we briefly summarize the important grain surface reactions and outline
the variation  of formation routes  of COMs depending on  the physical
conditions.  At the  initial  stages, the  ice  mantles  primarily
consist   of  \ce{s-H2O},   \ce{s-CH4},  \ce{s-H2CO},   \ce{s-NH3},  or
\ce{s-CH3OH} etc.  \ce{s-H2O},  \ce{s-NH3} and \ce{s-CH4}
are  mainly formed by  successive hydrogenation  of O,  N, and  C atoms
\citep{Tielens1982}  and \ce{s-H2CO}  and  \ce{s-CH3OH} by  successive
hydrogenation   of  CO   \citep{Charnley2009}.   Cosmic-ray  induced
photodissociation  of these  molecules,  specifically \ce{s-H2CO}  and
\ce{s-CH3OH}, produce other  important radicals (chemical species that
actively  react  with   other  species)  such as  \ce{s-OH},  \ce{s-HCO},
\ce{s-CH3},    \ce{s-CH3O}.    During   the    warm-up    phase
($\textless$100~K), these radicals effectively diffuse over grains and
produce  highly  abundant   oxygen-bearing  COMs, for example,  \ce{s-C2H5OH},
\ce{s-HCOOCH3},  and \ce{s-CH3OCH3};  the following
reactions act as the main  formation routes of these COMs depending on
the physical conditions:
\begin{small}
\begin{equation}
\label{eq:meth1}
 \ce{s-H + s-CH3O -> s-CH3OH }
\end{equation}
\vspace{-0.75cm}
\begin{equation}
\label{eq:meth2}
 \ce{s-H + s-HCOOCH3 -> s-CH3OH + s-HCO }
 \end{equation}
 \vspace{-0.75cm}
 \begin{equation}
\label{eq:meth3} 
 \ce{s-OH + s-CH3 -> s-CH3OH }
 \end{equation}
 \vspace{-0.75cm}
\begin{equation}
\label{eq:meth4}
 \ce{s-H2CO + s-CH2OH -> s-CH3OH + s-HCO }
\end{equation}
\vspace{-0.75cm}
\begin{flalign}
\label{eq:eth1}
&\ce{s-CH2OH + s-CO -> s-CH2OHCO}& \nonumber \\
&\ce{s-CH2OHCO + s-H -> s-CH2OHCHO}& \\
&\ce{s-CH3 + s-CH2OHCHO -> s-C2H5OH + s-HCO}& \nonumber 
\end{flalign}
\vspace{-0.5cm}
\begin{align}
\label{eq:eth2}
\ce{s-CH3 + s-CH2OH -> s-C2H5OH}
\end{align}
\vspace{-0.7cm}
\begin{equation}
\label{eq:di1}
\ce{s-CH3 + s-CH3O -> s-CH3OCH3}
\end{equation}
\vspace{-0.5cm}
\begin{flalign}
\label{eq:for1}
&\ce{s-CH3O + s-CO -> s-CH3OCO}& \nonumber \\[-1em]
& & \\
&\ce{s-H + s-CH3OCO -> s-HCOOCH3}& \nonumber 
\end{flalign} 
\vspace{-0.7cm}
\begin{equation}
\label{eq:for2}
\ce{s-CH3O + s-H2CO -> s-HCOOCH3 + s-H}
\end{equation} 
\vspace{-0.8cm}
\begin{equation}
\label{eq:for3}
\ce{s-CH3O + s-HCO -> s-HCOOCH3}
\end{equation} 
\end{small}
The mobility of the radicals  strongly depend on the temperature, which
shows a significant spatio-temporal variation and thus also affect the
formation pathways of COMs. To facilitate the discussion, we divided the hot
core models  into three different  subregions, central,  middle, and
outer regions.  The typical temperatures of  these zones at  the end of
the evolution  are $\textgreater$100~K, 60--100~K,  and $\textless$60~K.  In  Table~\ref{table:form_routes},  we  summarize  the
important reactions at different  regions that mainly produce the COMs
described above.

%-------------------------------------------------
\begin{table*}
\caption{Spatial variation of formation routes of selected COMs}              
\label{table:form_routes}     
\centering   
\begin{tabular*}{\textwidth}{@{\extracolsep{\fill}}llll}
\hline 
\multirow{2}{*}{Molecules}& Central region      & Middle region    & Outer region \\
                          & Radius:$\textless$ 2000~AU & 2000--4500~AU    & $\textgreater$ 4500~AU\\
\hline        
\ce{s-CH3OH}  &  Reac. 1, 2 \& 3               &  Reac. 1, 2, 3 \& 4             &  Reac. 1  \& 3 \\
\ce{s-C2H5OH} &  Reac. 5 and 6 ($\geq$30~K)    &  Reac. 5 and 6 ($\geq$25~K)     &  Reac. 5 and 6 ($\geq$20~K) \\
\ce{s-CH3OCH3}&  Reac. 7                       &  Reac. 7                        &  Reac. 7                  \\
\ce{s-HCOOCH3}&  Reac. 8, 10 \& 9 ($\geq$60~K) &  Reac. 8, 10 \& 9 ($\geq$60~K)  &  Reac. 8 \& 10       \\
\hline
\end{tabular*}
\end{table*}
%-------------------------------------------------------------------------------------------------------------------

At  the initial  stages  of evolution,  that is  at constant  temperature
(15~K), Reaction~\ref{eq:meth1}  acts as the primary
pathway   of  \ce{s-CH3OH}   formation   for  all   the  models.   For
the \emph{cr-16} models, up  to 30~K, Reaction~\ref{eq:meth2} dominates and
beyond   that   until   the   onset  of   evaporation   (i.e.,   100~K),
Reaction~\ref{eq:meth3}  emerges  as  the  main channel  of  \ce{s-CH3OH}
formation.  Reaction~\ref{eq:meth4} also occasionally becomes  the main  contributor of
\ce{s-CH3OH} formation  in the middle region. However, in
the outer region, Reaction~\ref{eq:meth1} dominates up to 30~K, and beyond
that,  Reaction~\ref{eq:meth3} produces  most of  the  \ce{s-CH3OH}. These
trends do  not vary significantly with  different luminosity evolution
models.  However,  for  the \emph{cr-17}  models, some  differences  are
observed: although  Reaction~\ref{eq:meth1} initially dominates,
Reaction~\ref{eq:meth2} begins  to contribute  soon and remains  the main
contributory  channel  for  temperatures  up to  60--70~K.  At  higher
temperatures Reaction~\ref{eq:meth3} acts as the main formation route for
\ce{s-CH3OH} until it finally evaporates in the gas phase. But in the outer
region Reaction~\ref{eq:meth2}  begins to strongly  dominate at relatively
higher  temperatures ($\textgreater$ 15--40~K)  also depending  on the
density. However, for temperatures around  60~K and above, the trend is
similar to the \emph{cr-16} models that is  Reaction~\ref{eq:meth3} acts as
the main formation pathway.

Formation  of  other   COMs,  for instance,  \ce{s-C2H5OH},  \ce{s-CH3OCH3},  and
\ce{s-HCOOCH3},  also  strongly depend  on  the  temperature since  the
diffusion of  relatively heavy radicals  such as \ce{s-CH3}, \ce{s-CH3O},
and \ce{s-CH2OH}  is involved.  For  the \emph{cr-16}  models  the
formation of \ce{s-C2H5OH}  is dominated by Reaction~\ref{eq:eth1} around
20~K, and above  that temperature until the desorption  to the gas phase, 
Reaction~\ref{eq:eth2} becomes the main  contributing channel. For the
\emph{cr-17}  models,  Reaction~\ref{eq:eth1}
dominates until  the middle  region   up  to 60~K and  above that, Reaction~\ref{eq:eth2} acts  as the
main  formation route;  but in  the outer  region  both of  these
reactions   contribute  to  the \ce{s-C2H5OH}  formation.  The formation  of
\ce{s-CH3OCH3}  is   always  dominated  by   Reaction~\ref{eq:di1}.   The
formation of \ce{s-HCOOCH3} in  the \emph{cr-16} models is controlled by
Reaction~\ref{eq:for1} and  \ref{eq:for3} for temperatures  lower than 60~K, 
and  above  that  Reaction~\ref{eq:for2}   acts  as  the  main  formation
route. For  the \emph{cr-17} models the behavior is  similar, except that
Reaction~\ref{eq:for3} does not contribute significantly to the formation
of  \ce{s-HCOOCH3}.  A more  detailed  discussion  of  the  formation  and
destruction pathways  of various molecular species  including the COMs
was reported in \cite{Garrod2008}.

%------------------------------------------------------------------------------------------------------------------------------------------------------------
\section{Analysis}
\label{sec:anal}
\subsection{Qualitative comparison of simulated and observed spectra}
\label{subsec:sim_compare}
\begin{figure}
\centering
\includegraphics[width=8cm]{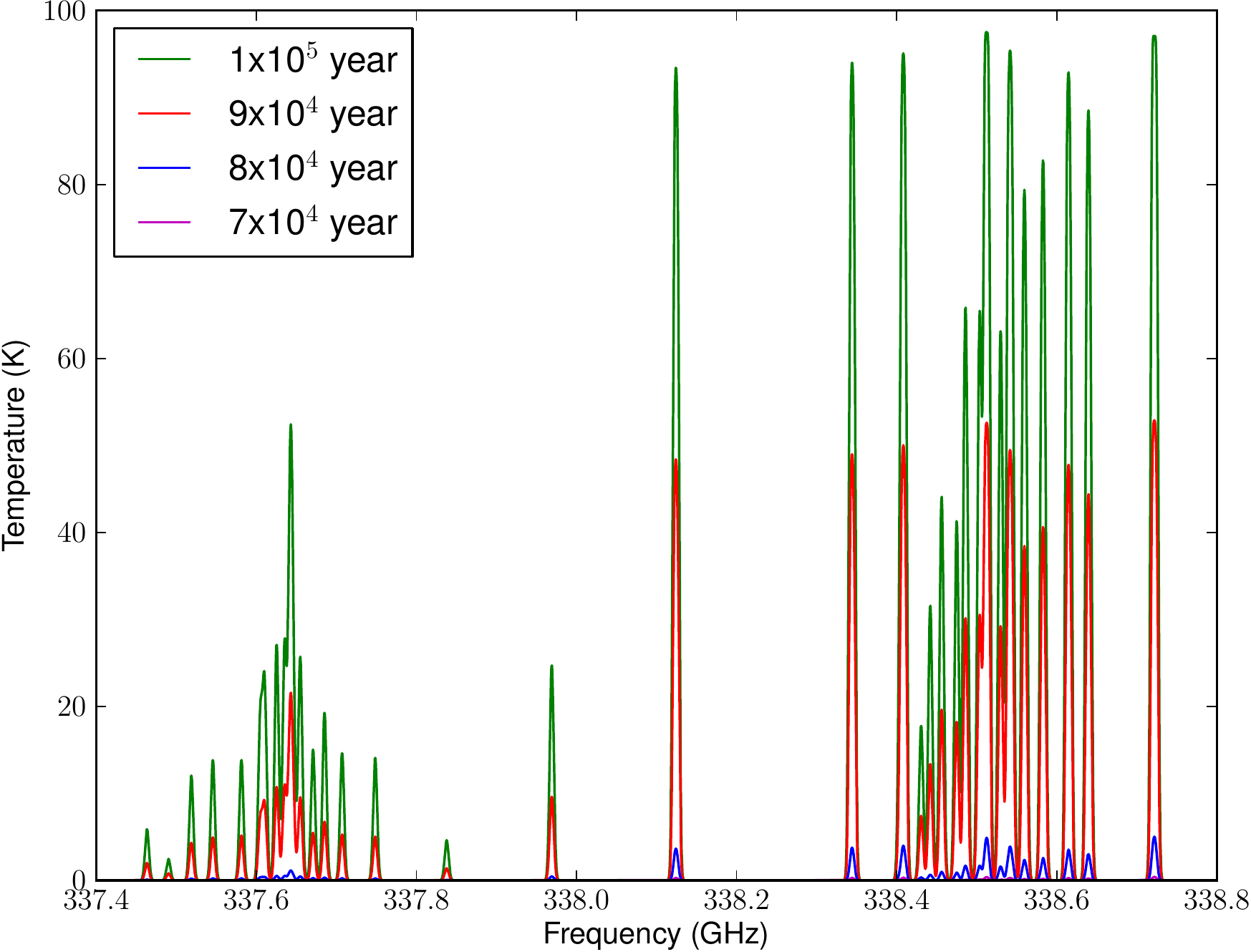}
\caption{Simulated spectra of \ce{CH3OH} for the  hot core model \emph{cr16-r3-l2} at different evolutionary timescales. The adopted distance to the
source and the telescope beam size is 2~kpc and 2.2\arcsec.}
\label{fig:spec_four_evol}
\end{figure}

\begin{figure}
\includegraphics[width=8cm]{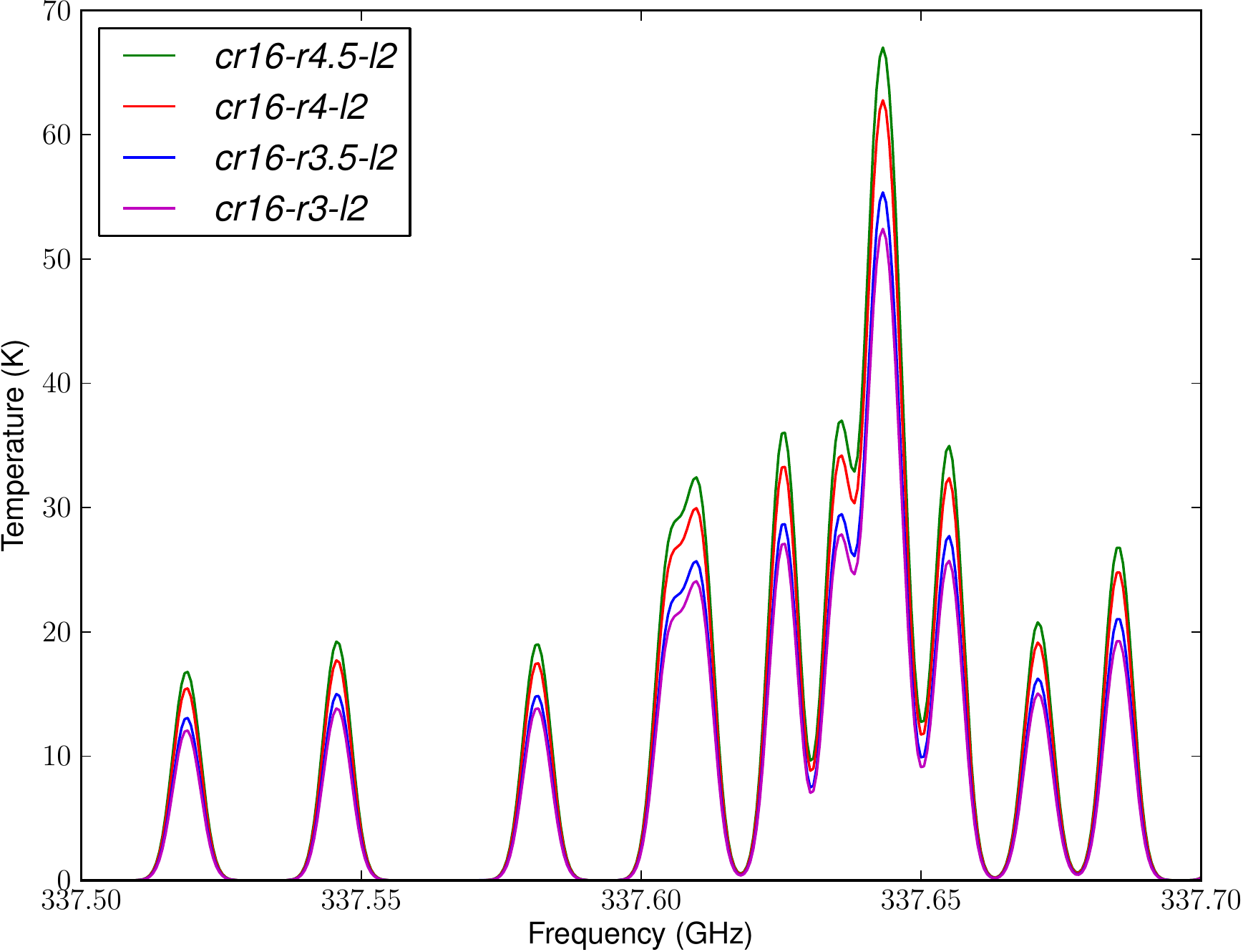}
\caption{Simulated spectra of \ce{CH3OH} for the hot core models \emph{cr16-r3-\emph{l}2}, \emph{cr16-r3.5-\emph{l}2}, \emph{cr16-r4-\emph{l}2},  and
\emph{cr16-r4.5-\emph{l}2} at \num{1.0e5} year. The adopted distance to the
source and the telescope beam size is 2~kpc and 2.2\arcsec.}
\label{fig:spec_four_model}
\end{figure}

\begin{figure*}
\centering
\begin{tabular}{@{}cc@{}}
\includegraphics[width=8.5cm]{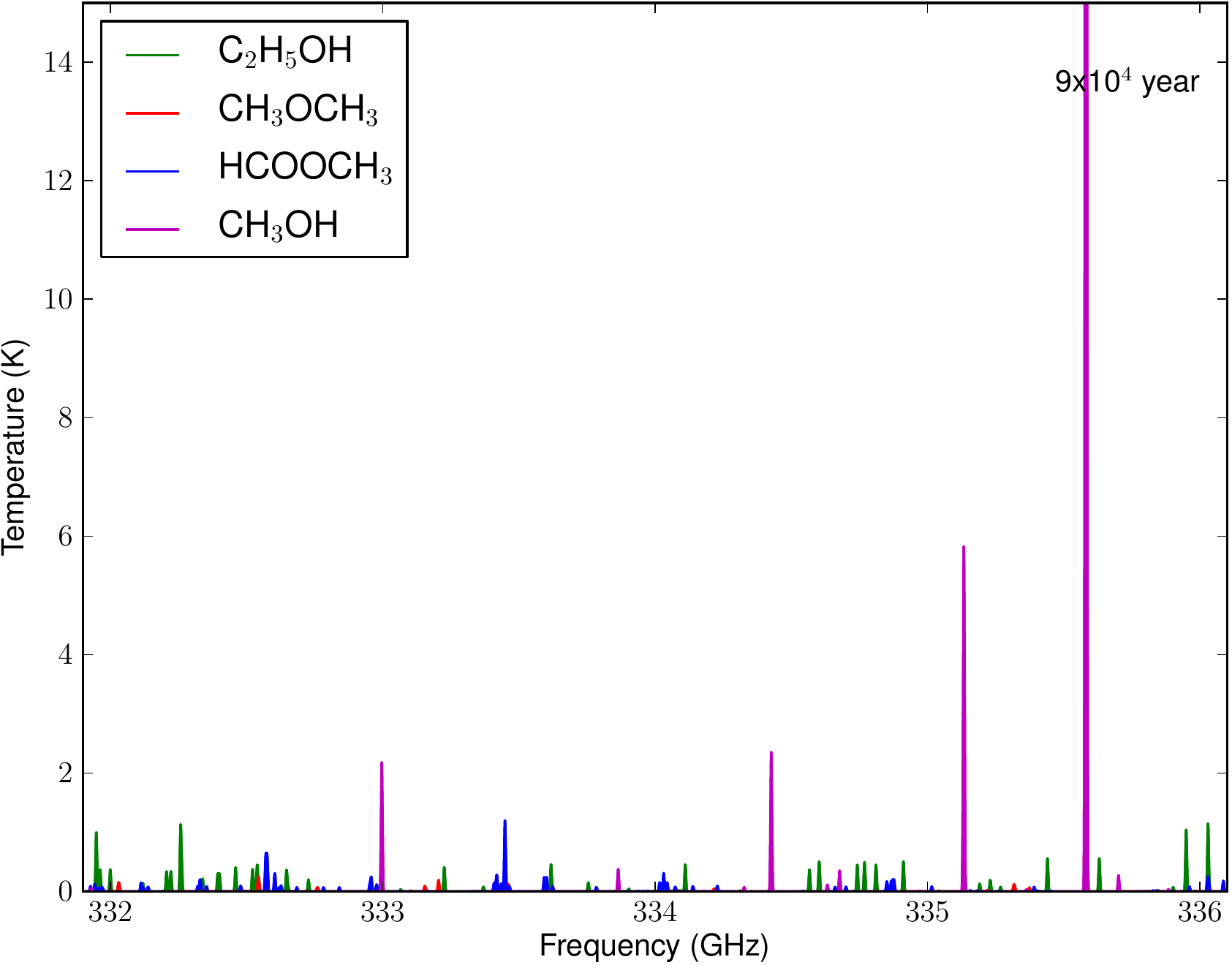}&
\includegraphics[width=8.5cm]{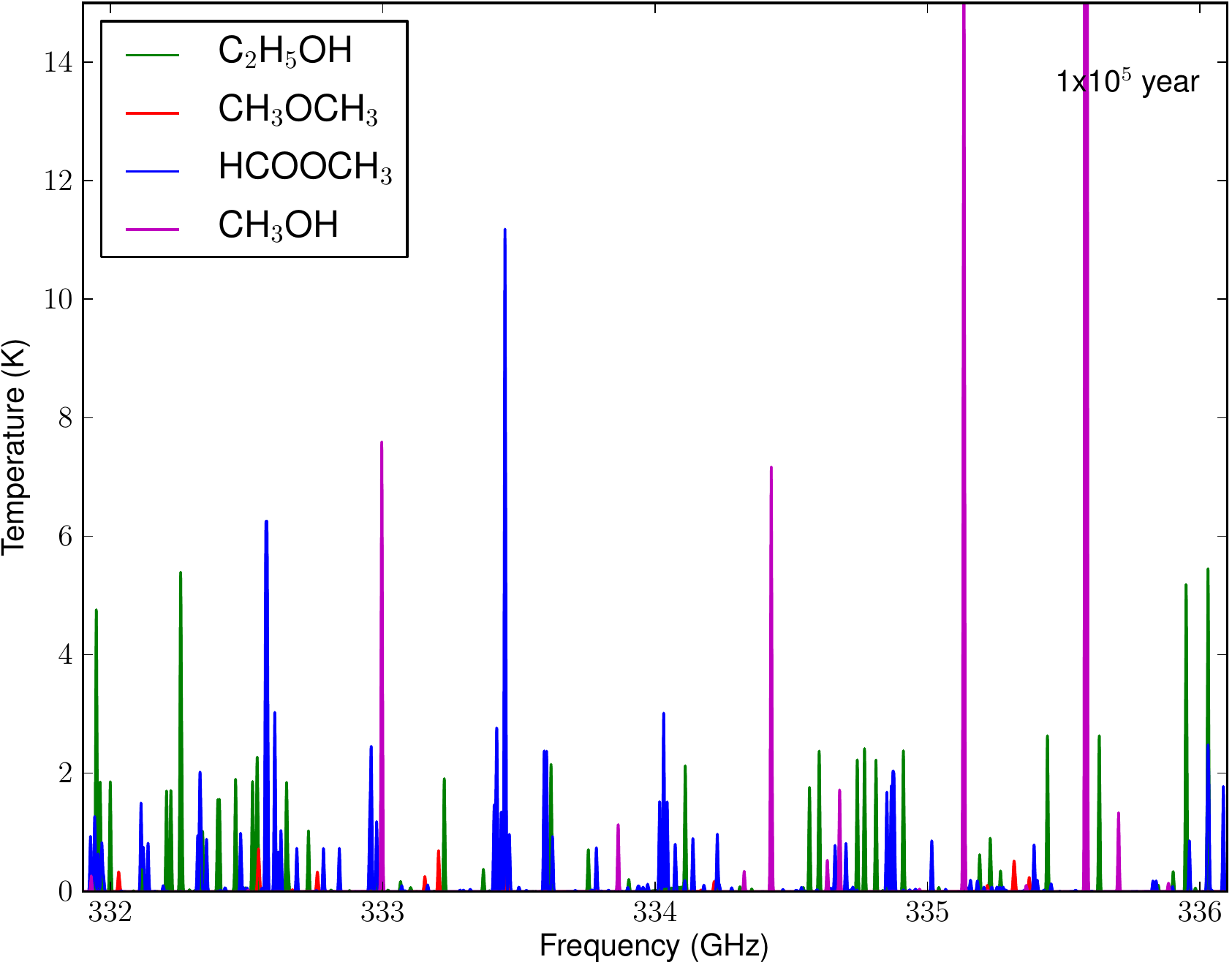}\\
\includegraphics[width=8.5cm]{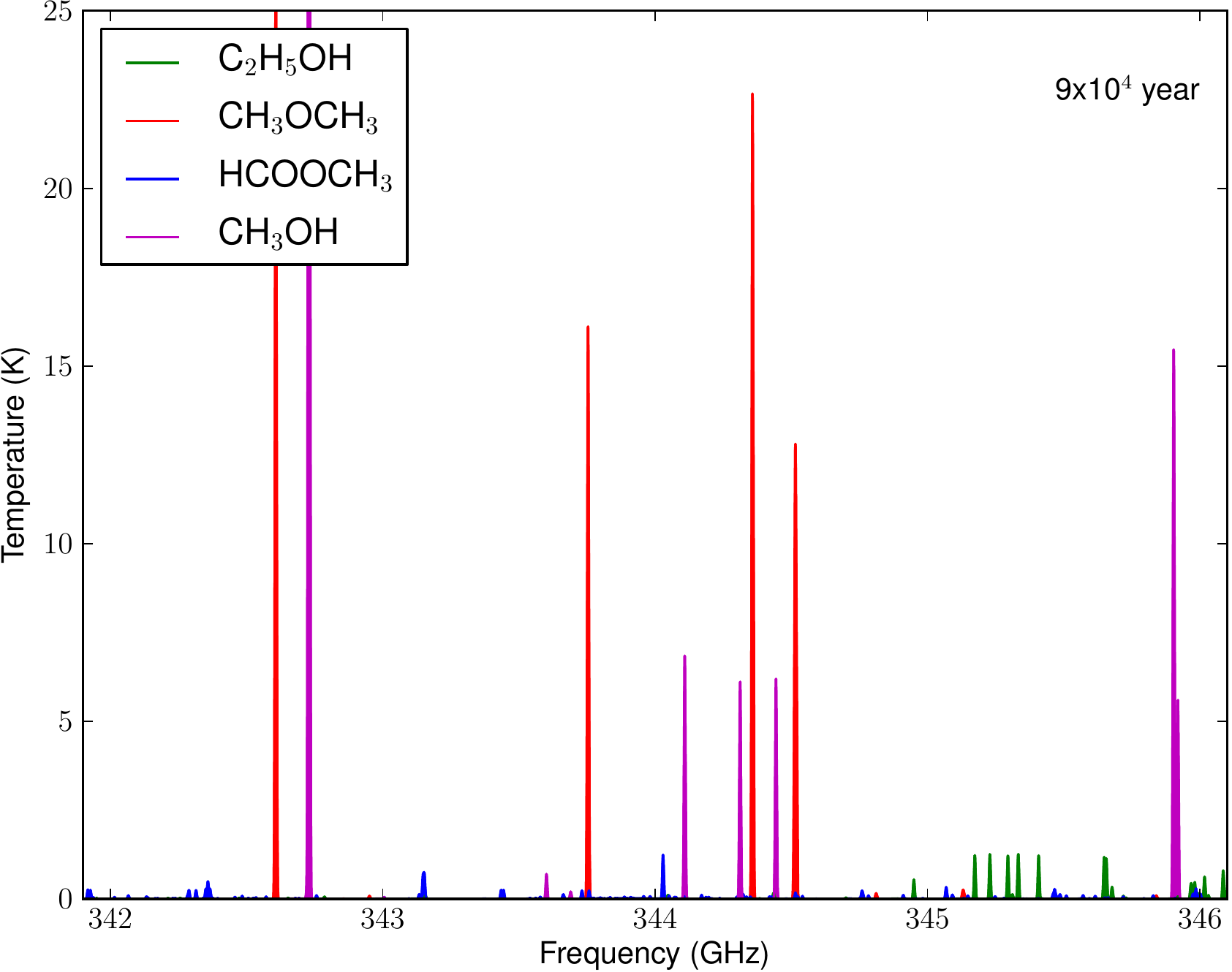}&
\includegraphics[width=8.5cm]{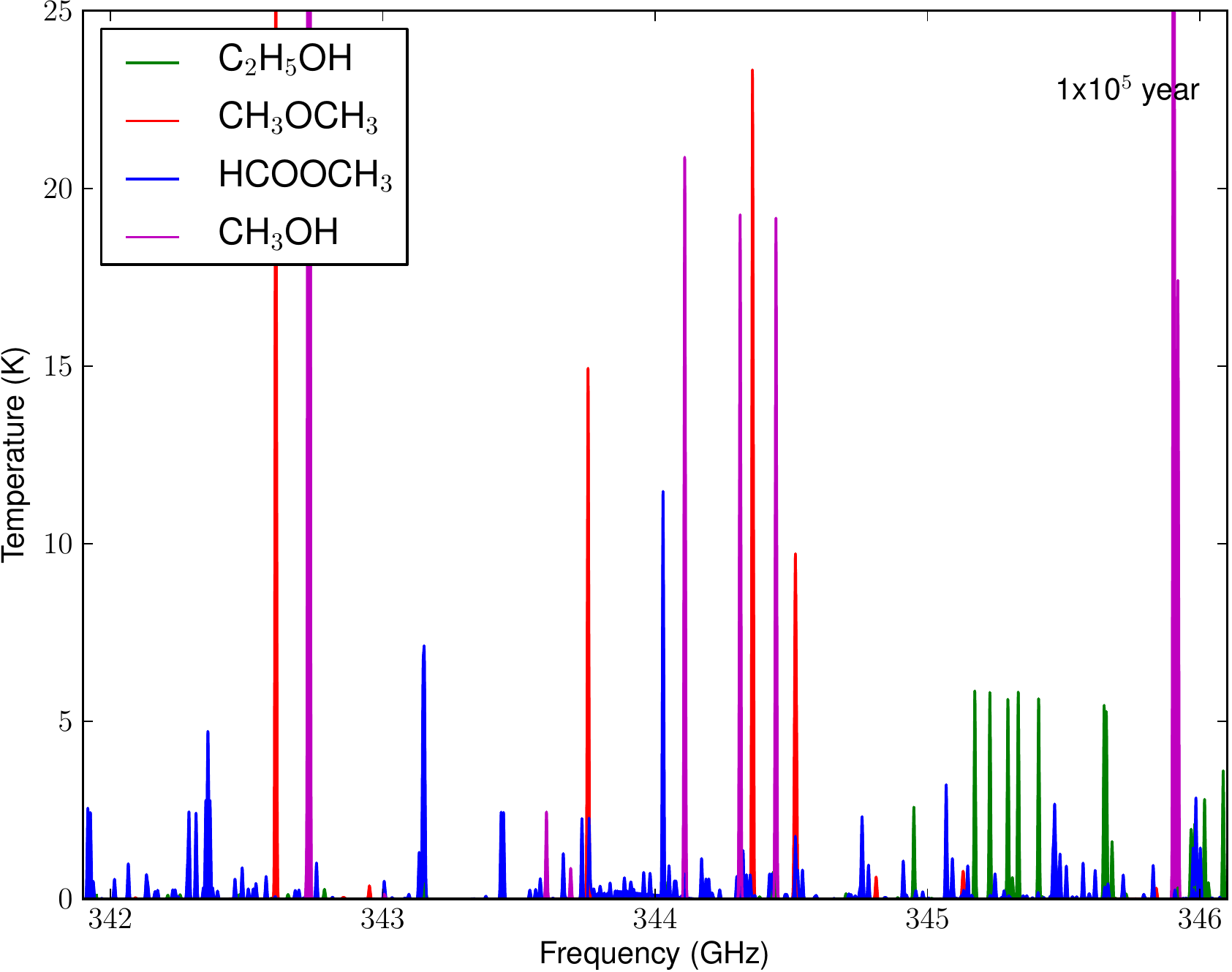}
\end{tabular}
\caption{Simulated spectra of selected COMs for the  hot core model \emph{cr16-r3-l7} at \num{9e4} year (left) and \num{1e5} year (right) at different
wavebands. The adopted distance to the source and the telescope beam size is 2~kpc and 2.2\arcsec.}
\label{fig:spec_coms_332}
\end{figure*}

\begin{figure*}
\centering
\begin{tabular}{@{}cc@{}}
\includegraphics[width=8.5cm]{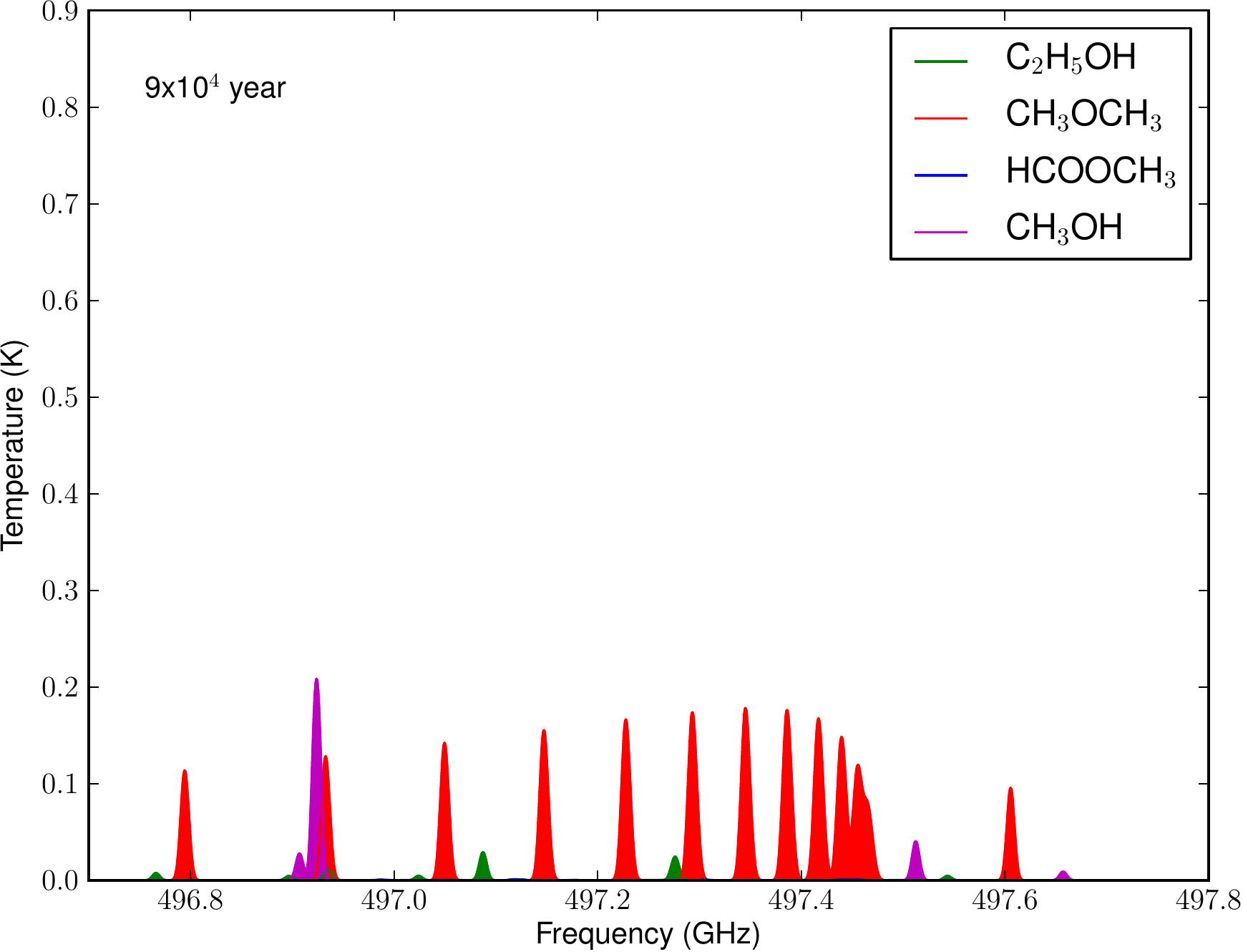}&
\includegraphics[width=8.5cm]{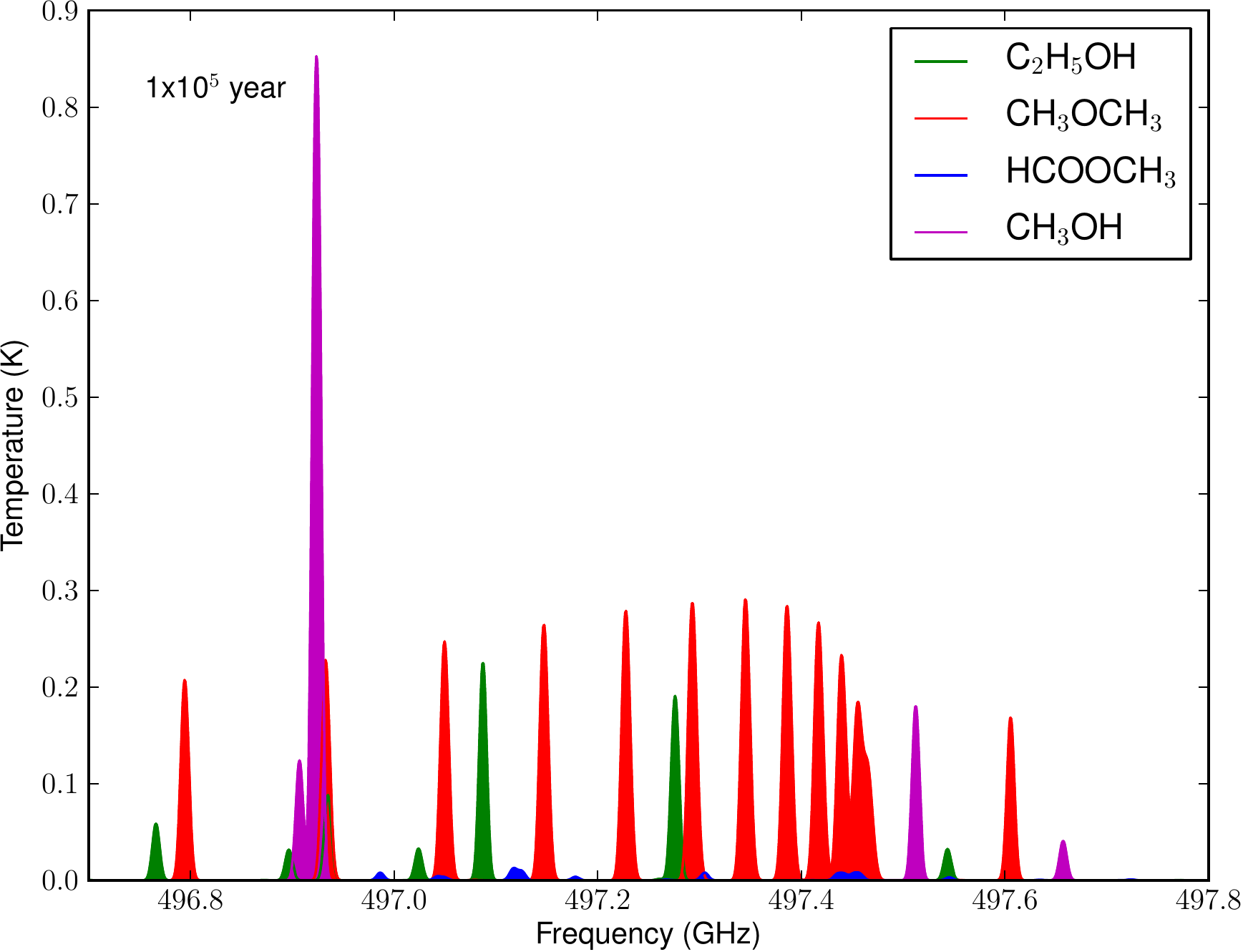} 
\end{tabular}
\caption{Simulated spectra of selected COMs for the hot core model \emph{cr16-r3-l7} at \num{9e4} year (left) and \num{1e5} year (right) at Herschel-HIFI
Band 1. The adopted distance to the source and the telescope beam size is 2~kpc and $\sim$40\arcsec.}
\label{fig:spec_hrsl} 
\end{figure*}

%---------------------------------------------------------------------------------
The  most  noticeable spectral  features  of  high-mass star  forming
regions are  the increase  of the total  number of spectral  lines and
their        intensities        with        evolutionary        stages
\citep{Beuther2009,Ceccarelli2010,Gerner2014}.  Spectral variations are
also   observed  in  different   sources  with   similar  evolutionary
stages. Therefore, a characterization of the spectral evolution based on the
inline physio-chemical models can be used as a unique tool to constrain
the  evolutionary sequence  of the  initial stages  of high-mass star
formation.  Our simplistic model  is unsuitable for any  quantitative
comparisons with the observed spectra, but it is expected that our
model spectra  at least reproduce the general trends of spectral
variation  provided  that  our  input  hot core  models  resemble  the
conditions  of  the  observations.  Comparing the simulated  and
observed spectra is thus also  useful to determine the figure of merit
of the  input models. Motivated by  this aim, we  simulated the spectra
for  various  hot core  models  at  different  evolutionary stages  at
various   wavebands    and   present   the    simulated   spectra   in
Figs.~\ref{fig:spec_four_evol}--\ref{fig:spec_hrsl}.

%--------------------------------------------------------------------------------------------------

\begin{figure*}
\centering
\includegraphics[width=.9\textwidth]{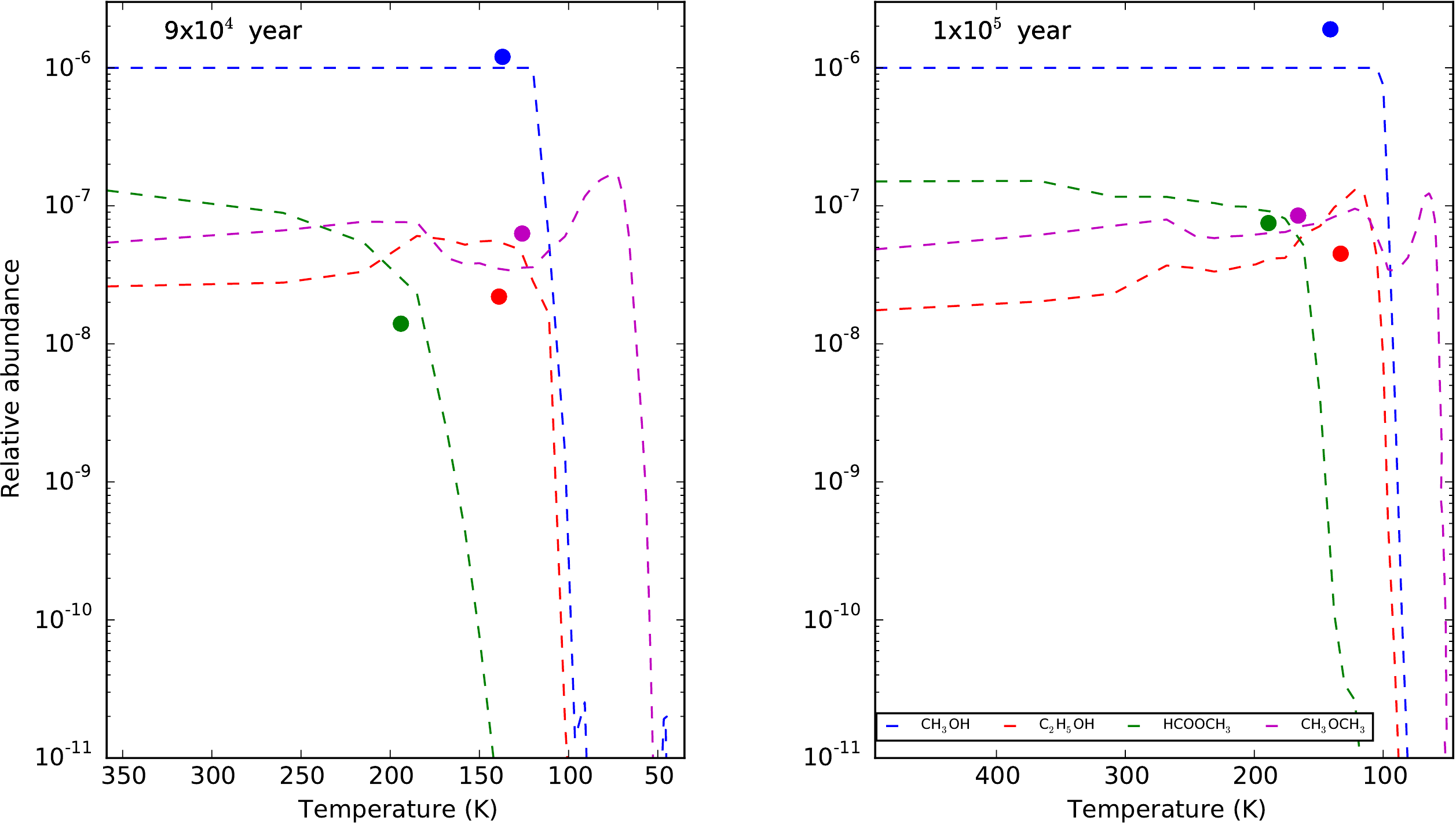} 
\caption{Comparison of the input abundance distribution of the hot core model \emph{cr16-r3-l7} and the single-point physical parameters
obtained by the myXCLASS analysis.}
\label{fig:myxclass_fit}
\end{figure*}

Simulated   spectra   of   \ce{CH3OH}   for   the   hot   core   model
\emph{cr16-r3-\emph{l}2}  at  different  evolutionary  stages  are
shown   in  Fig.~\ref{fig:spec_four_evol};   the  spectral   range  was
chosen because  it is often used in  SMA/ALMA observations. In
the  literature  the term  \emph{line  forest}  is  often used  when
numerous  complex molecules emit  within a  given frequency  range and
thereby create a forest  of lines that  perhaps hide  emission from
weaker  molecules.   In  this  paper  we  refer to  the  densely  spaced
ro-vibrational spectral lines of the COMs over a few GHz  bandwidth
as \emph{line  forest}.  These lines arise  from various transitions
spanning over a range of energy states that trace different layers within
the  hot  core.  The  spectra  shown in  Fig.~\ref{fig:spec_four_evol}
cover   two  different  \emph{line   forests}  of   \ce{CH3OH}  with
relatively   lower  ($\sim$60--300~K   at  $\sim$338~GHz)   to  higher
($\sim$350--650~K  at $\sim$337~GHz)  excitation energies.  It  can be
easily  seen  from Fig.~\ref{fig:spec_four_evol}  that  the number  of
spectral lines and their respective intensities vary with evolutionary
timescales.   Moreover, it  is  also evident  from  the spectra  that
temporal  evolution  of spectral  features  are  not simply a  scaled-up
version  of  the previous  time-steps  but   originated  from  the
combined  variation   of  spatial  distribution   of  temperature  and
abundances             with             evolutionary            stages
(Fig.~\ref{fig:sim_spec_t9_on_t1}).   Simulated spectra  of \ce{CH3OH}
around  337~GHz for  various hot  core models  with  different density
distributions are shown in Fig.~\ref{fig:spec_four_model}. To  avoid the  self absorption
features  in the  simulated spectra  the highest value  of the \ce{CH3OH}
relative    abundance    was     restricted    to    \num{1e-6}    (see
Fig.\ref{fig:myxclass_fit})   compared   with   the  values   shown   in
Fig.\ref{fig:com_evol_a}.    This  discrepancy  also   indicates  that
\ce{CH3OH}  is  overproduced in  our  models  compared  with the  other
COMs. This  trend is also seen  in the recent chemical  models for
hot cores \citep{Garrod2013}. One  possibility to explain this might be that there are
some missing reactions in the chemical network that are useful to form
other complex  molecules using \ce{CH3OH}. Since \ce{CH3OH}  is one of
the parent molecules of other COMs, it would be interesting to compare
the  predicted  relative  abundance  of  other COMs  with  respect  to
\ce{CH3OH}  with  the observations  to  improve  our understanding  of
the chemical  evolution  in  hot  cores \citep{Neill2014}.  However,  this
discrepancy also emphasizes  the  importance of  simulated  spectra  over the  chemical
abundances that are generally quoted by the astrochemical models to provide a
better constraint on physio-chemical model of hot cores.
%----------------------------------------------------------------------------------------------------------------------------------------------------
%---------------------------------------------------------------------------------------------------------------------------------------------------------
%-----------------------------------------------------------------------------------------------------------------------------------------------------
\begin{landscape}
  \begin{figure}
  \centering
  \begin{tabular}{@{}cccc@{}}
  \includegraphics[width=6.1cm,height=6.0cm]{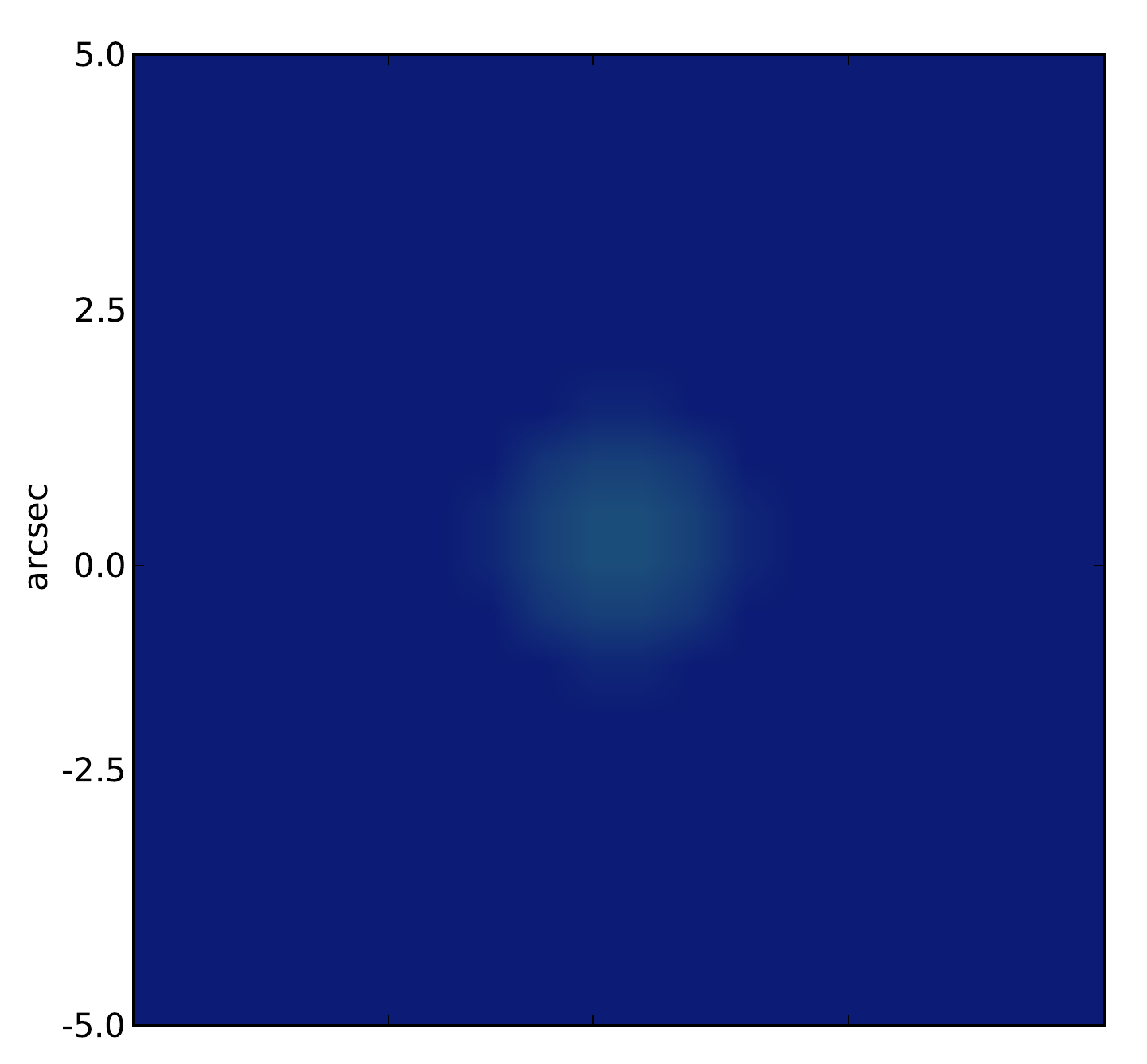}&
  \includegraphics[width=6.0cm,height=6.0cm]{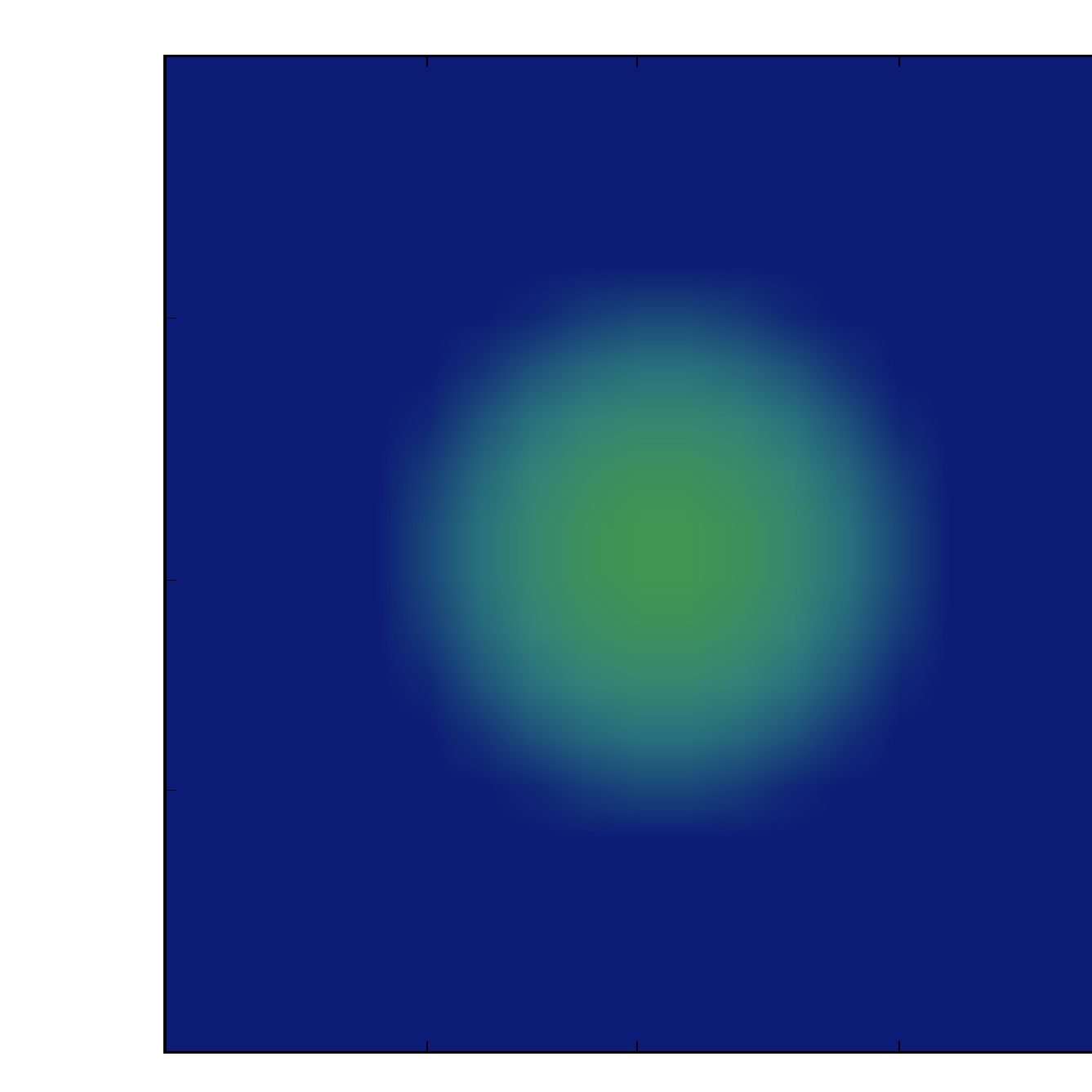}&
  \includegraphics[width=6.0cm,height=6.0cm]{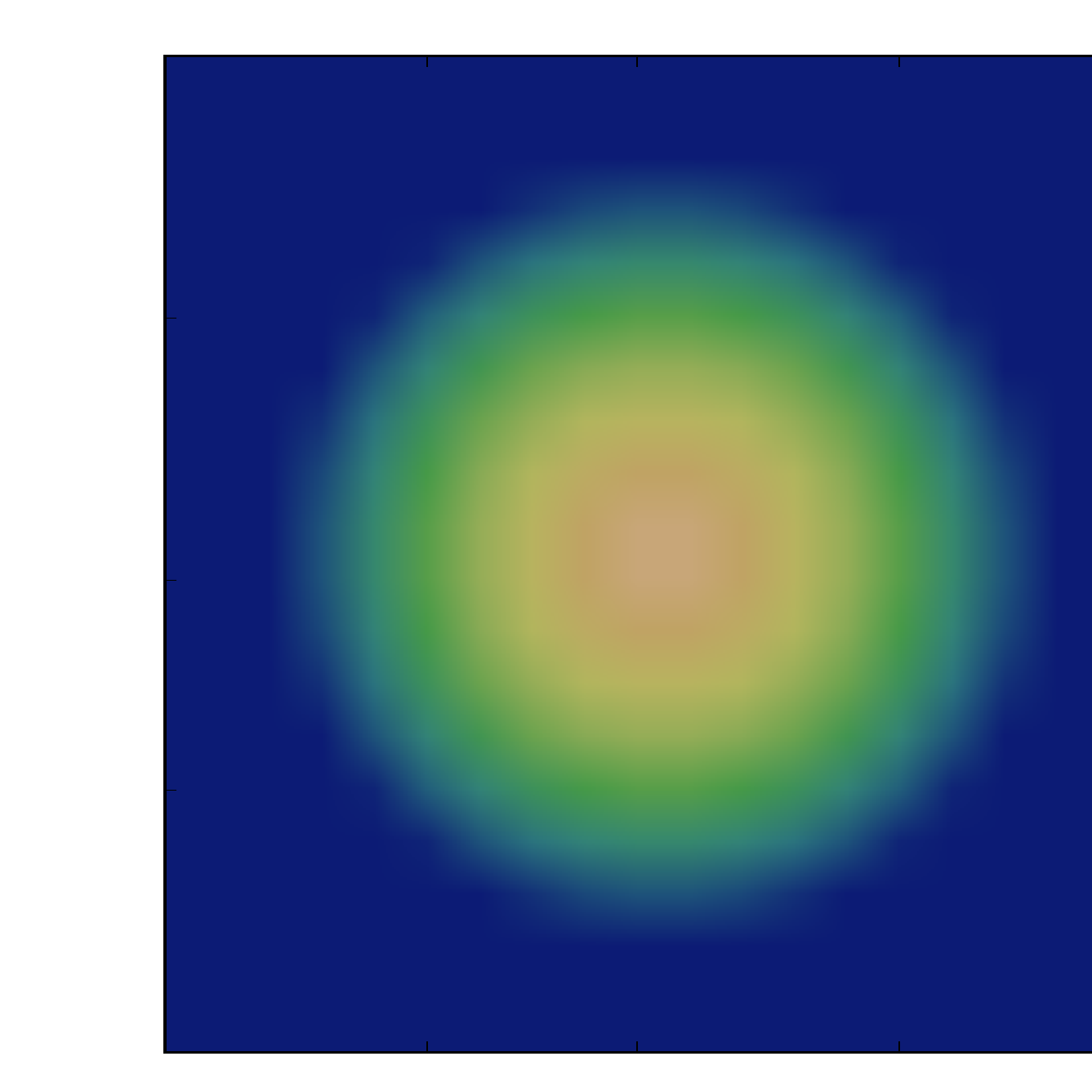}&
  \includegraphics[width=6.5cm,height=6.0cm]{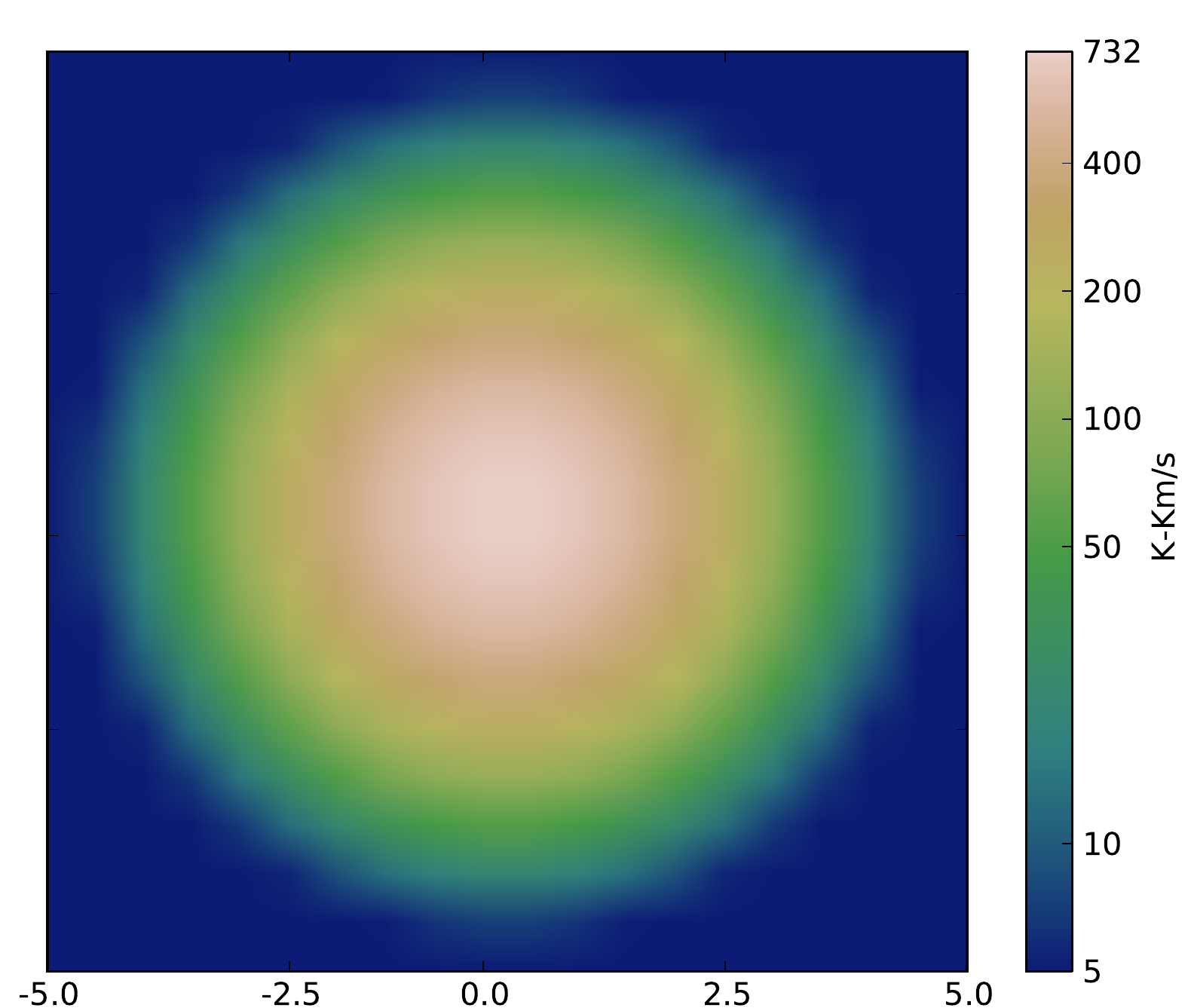}
  \end{tabular}
  \caption{Integrated line intensity maps of the \ce{CH3OH} (v$_{t}$= 0) [7(0,7)--6(0,6)] transition at 338.1245~GHz for the hot core model
\emph{cr16-r3-\emph{l}2} at \num{7.0e4}, \num{8.0e4}, \num{9.0e4} and \num{1.0e5} year (left to right). The adopted distance to the source 
and the telescope beam size is 2~kpc and 2.2\arcsec.}
  \label{fig:chan_maps_time}
 \end{figure}
  \begin{figure}
  \centering
  \begin{tabular}{@{}cccc@{}}
  \includegraphics[width=6.1cm,height=6.0cm]{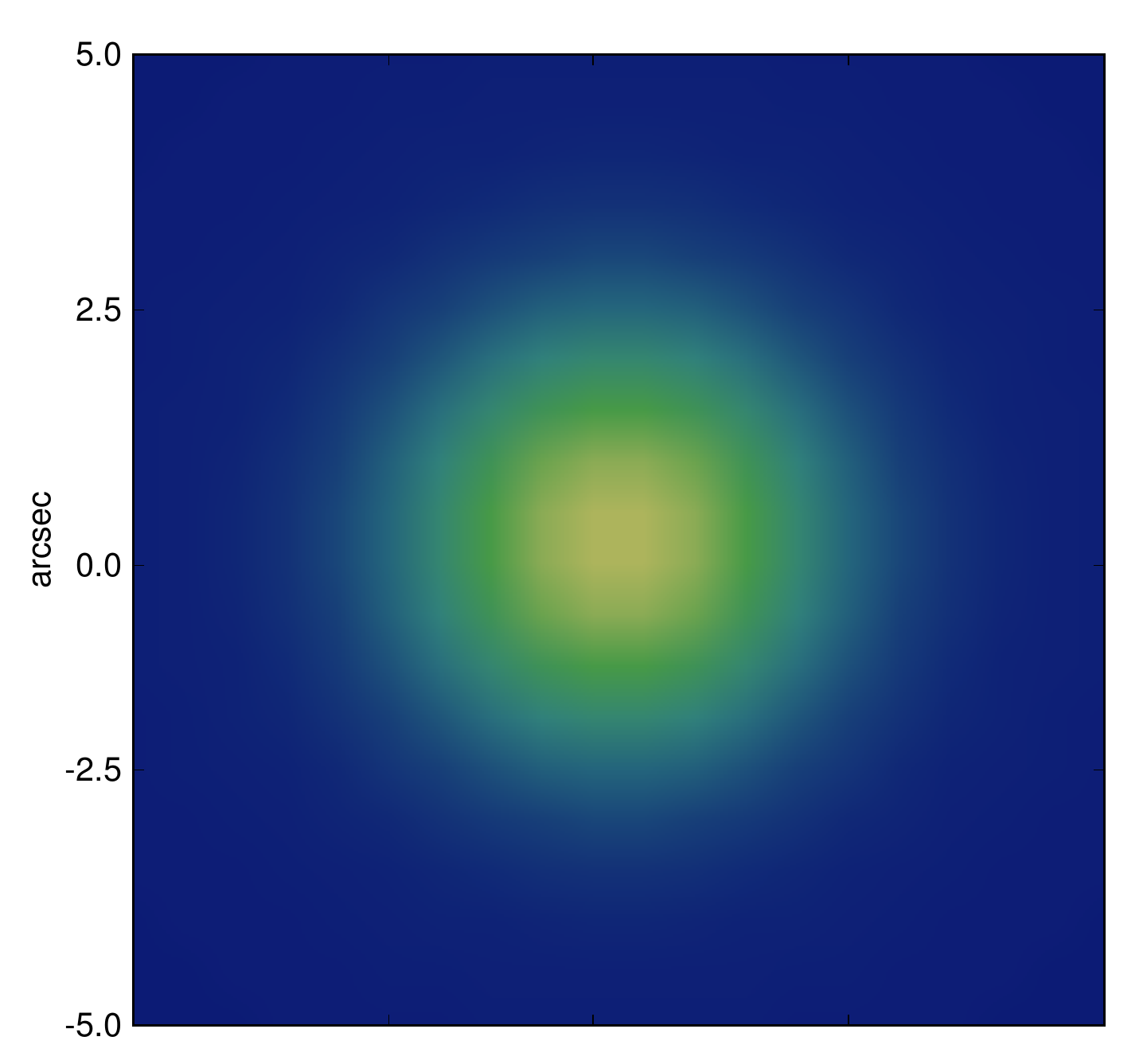}&
  \includegraphics[width=6.0cm,height=6.0cm]{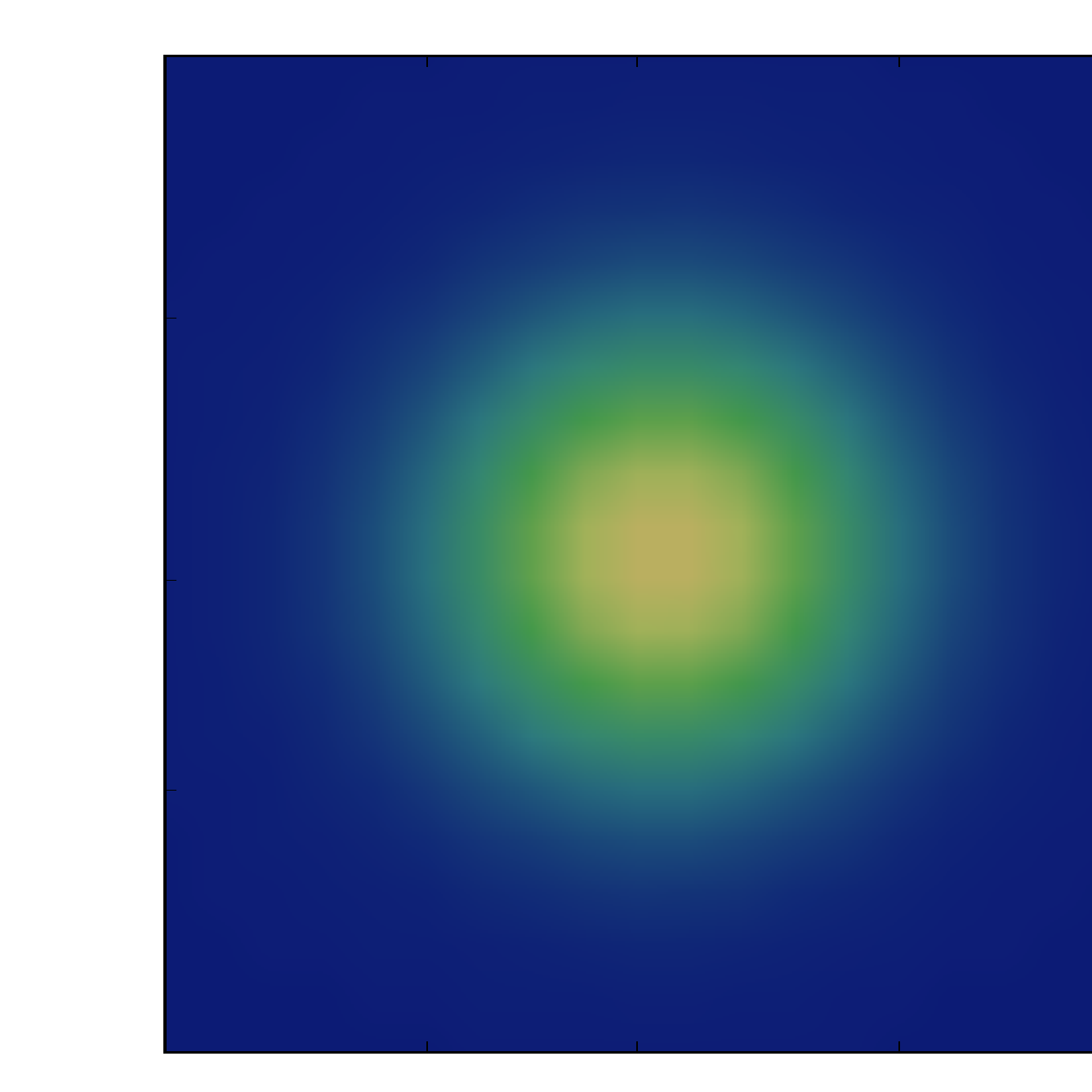}&
  \includegraphics[width=6.0cm,height=6.0cm]{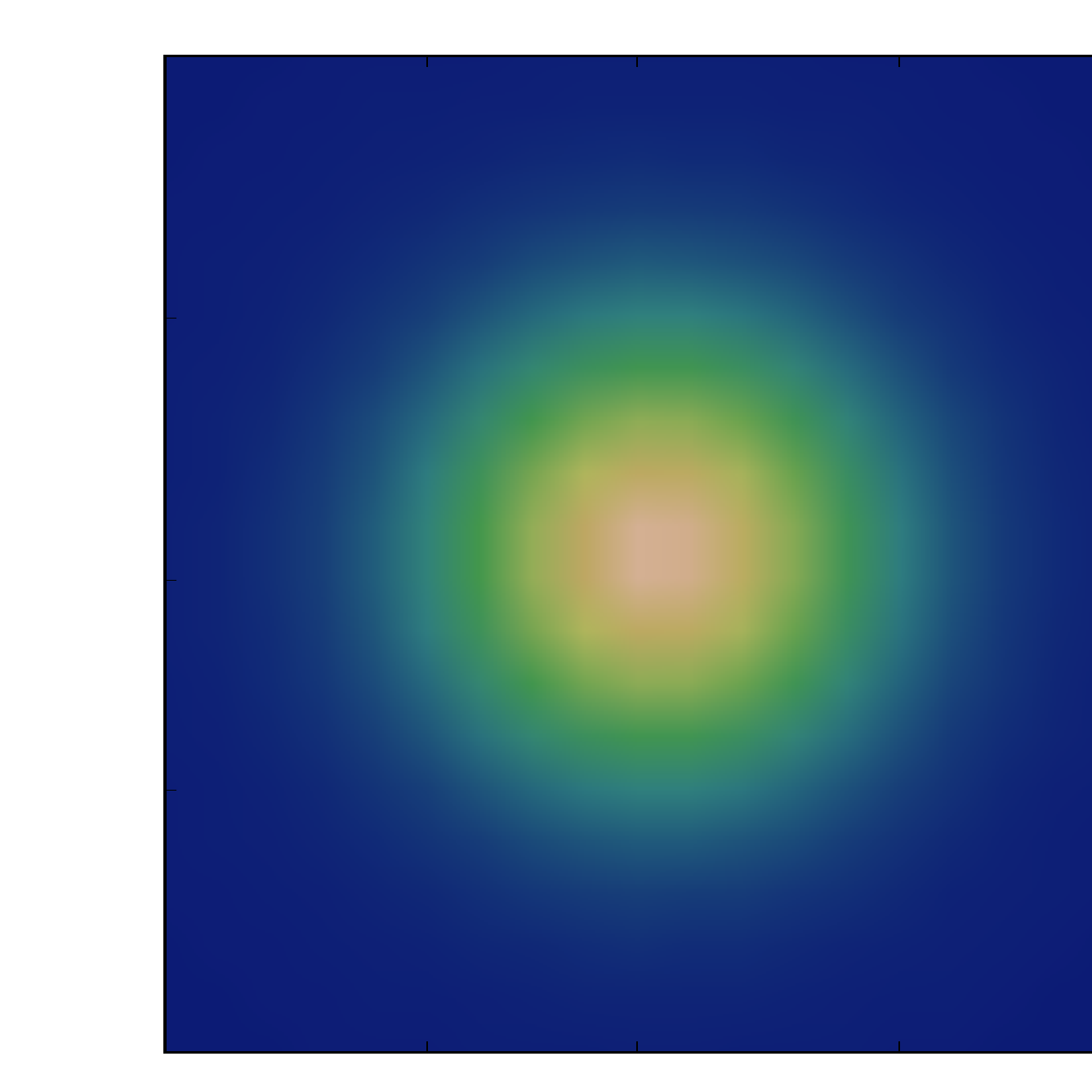}&
  \includegraphics[width=6.5cm,height=6.0cm]{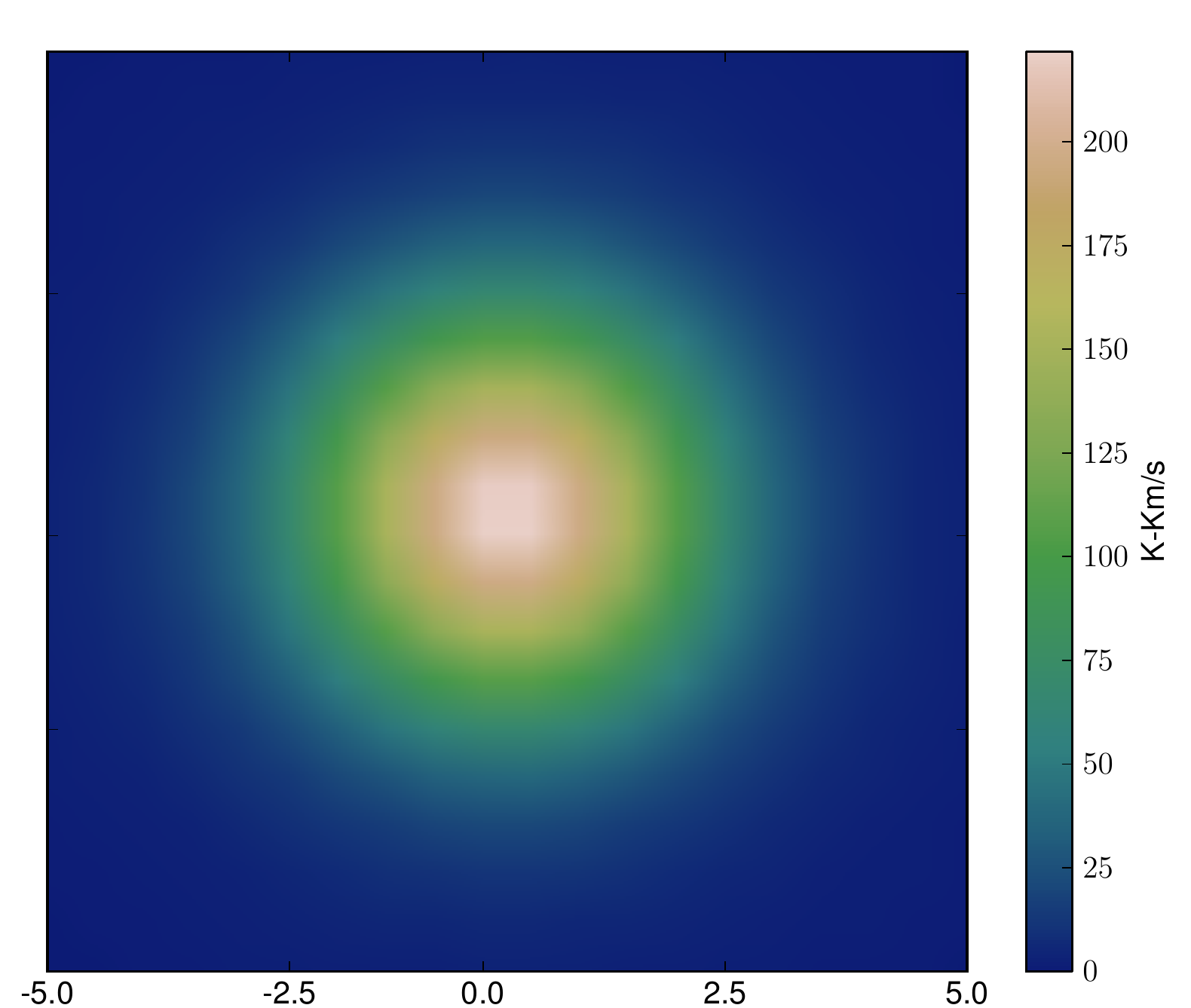}
  \end{tabular}
  \caption{Integrated line intensity maps of the \ce{CH3OH} (v$_{t}$= 1) [7(4,4)--6(4,3)] and [7(4,3)--6(4,2)] transitions at 337.68559~GHz for the hot core models
\emph{cr16-r3-\emph{l}2}, \emph{cr16-r3.5-\emph{l}2}, \emph{cr16-r4-\emph{l}2} and \emph{cr16-r4.5-\emph{l}2} (left to right) at \num{1.0e5} year. The adopted 
distance to the source and the telescope beam size is 2~kpc and 2.2\arcsec.}
  \label{fig:chan_maps_dens}
 \end{figure}
\end{landscape}
\clearpage
%--------------------------------------------------------------------------------------------------------------------------------------
\begin{figure}
\centering
\begin{tabular}{@{}c@{}}
\includegraphics[width=.45\textwidth]{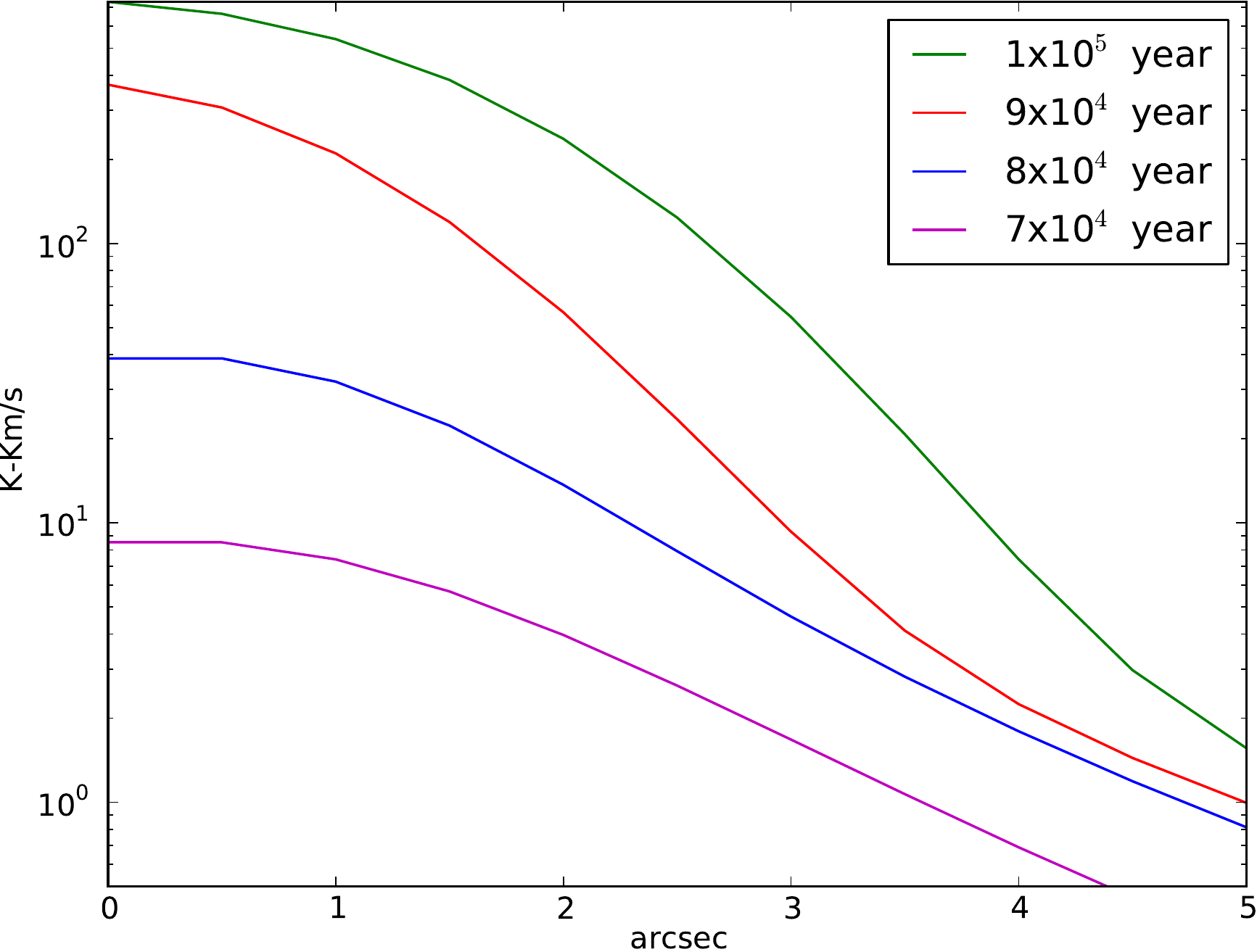} \\
\includegraphics[width=.45\textwidth]{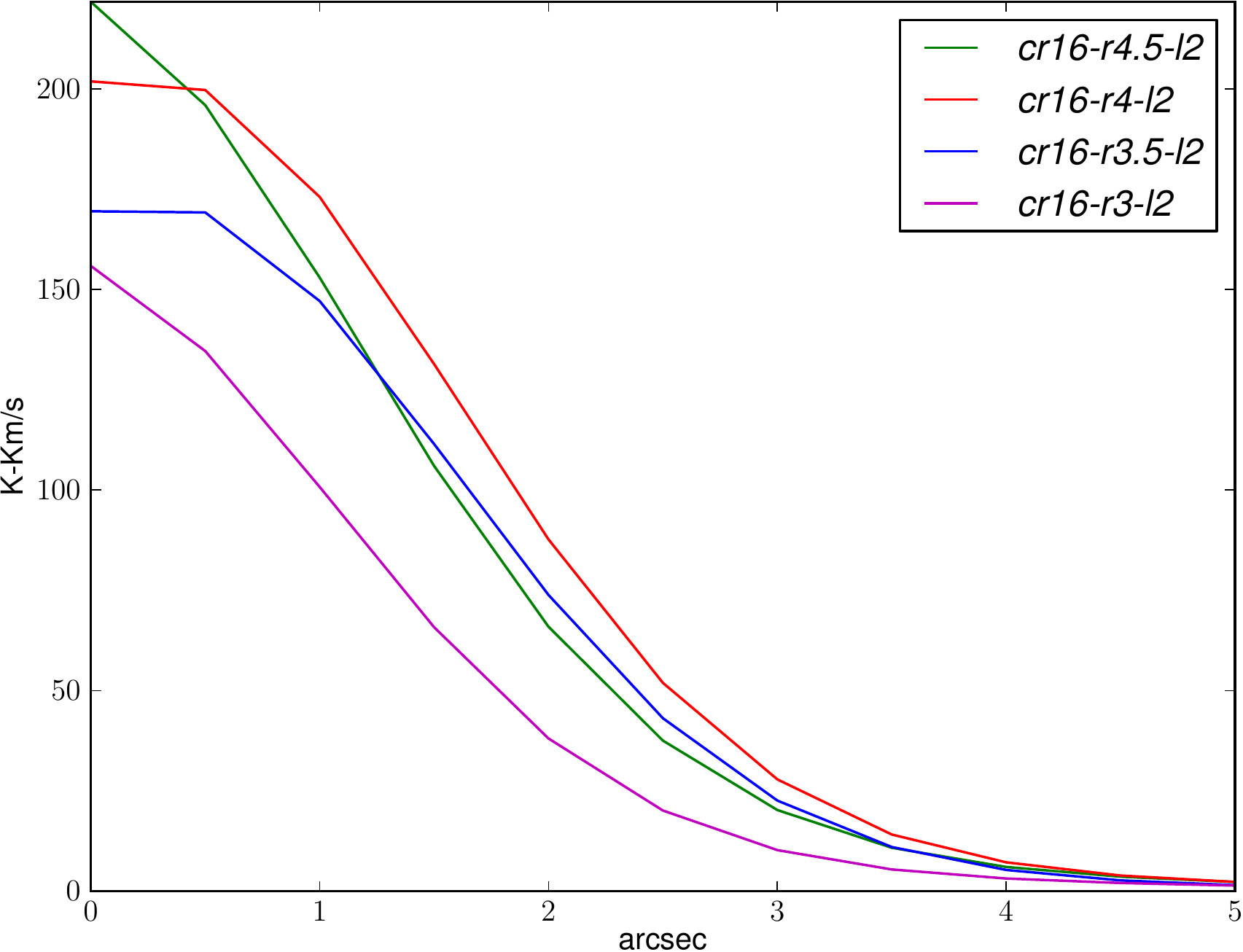}  
\end{tabular}
\caption{\emph{top}: Radial profiles of the \ce{CH3OH} (v$_{t}$= 0) [7(0,7)--6(0,6)] integrated intensity maps at 338.1245~GHz for the hot core model
\emph{cr16-r3-\emph{l}2} at various timescales. \emph{Bottom}: Radial profile of the \ce{CH3OH} (v$_{t}$= 1) [7(4,4)--6(4,3)] and [7(4,3)--6(4,2)] integrated intensity
maps at 337.68559~GHz for various hot core models. The adopted distance to the source and the telescope beam size is 2~kpc and 2.2\arcsec.}
\label{fig:rad_profs}
\end{figure}

\begin{figure}
\centering
\begin{tabular}{@{}c@{}}
\includegraphics[width=.45\textwidth]{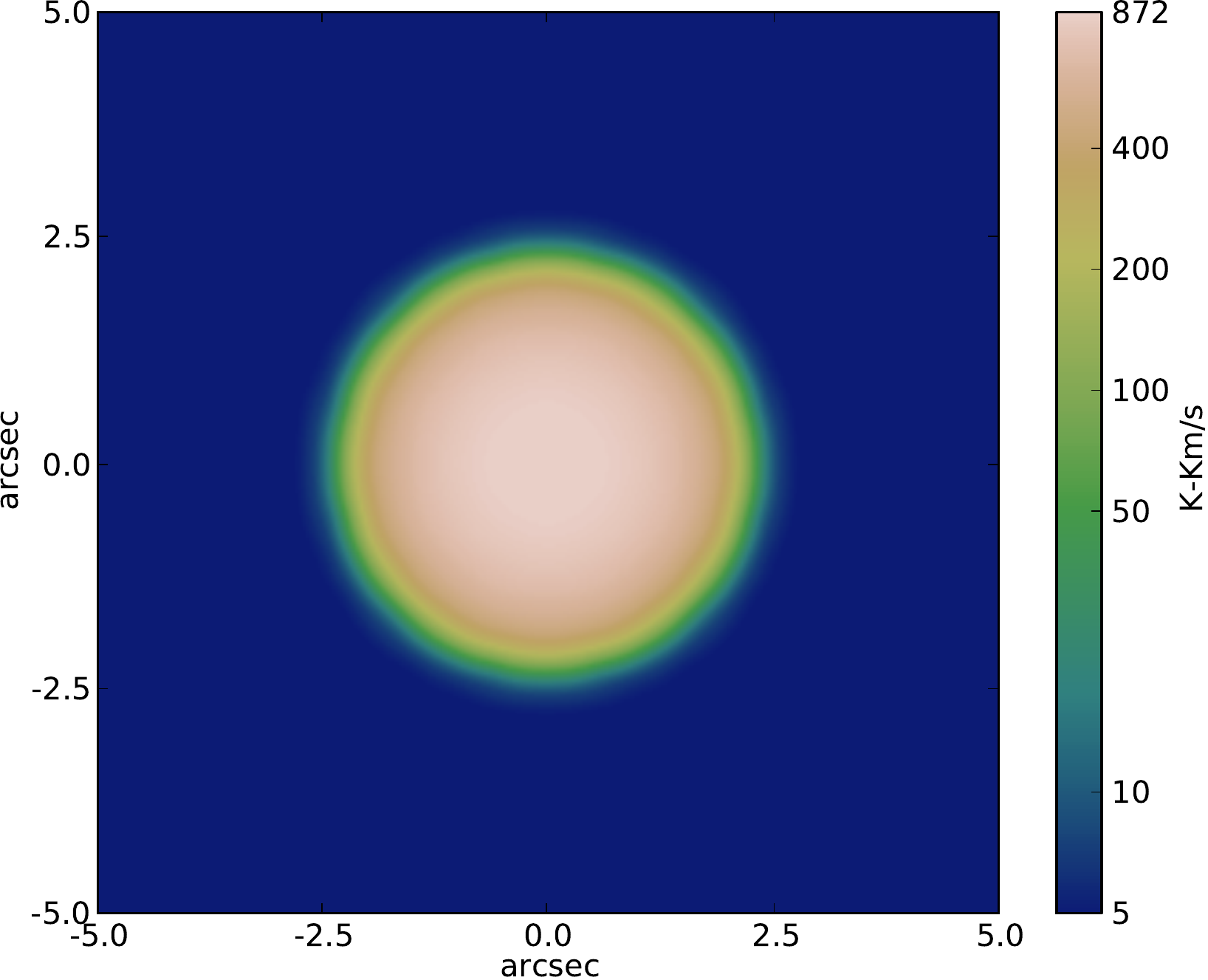} \\
\includegraphics[width=.45\textwidth]{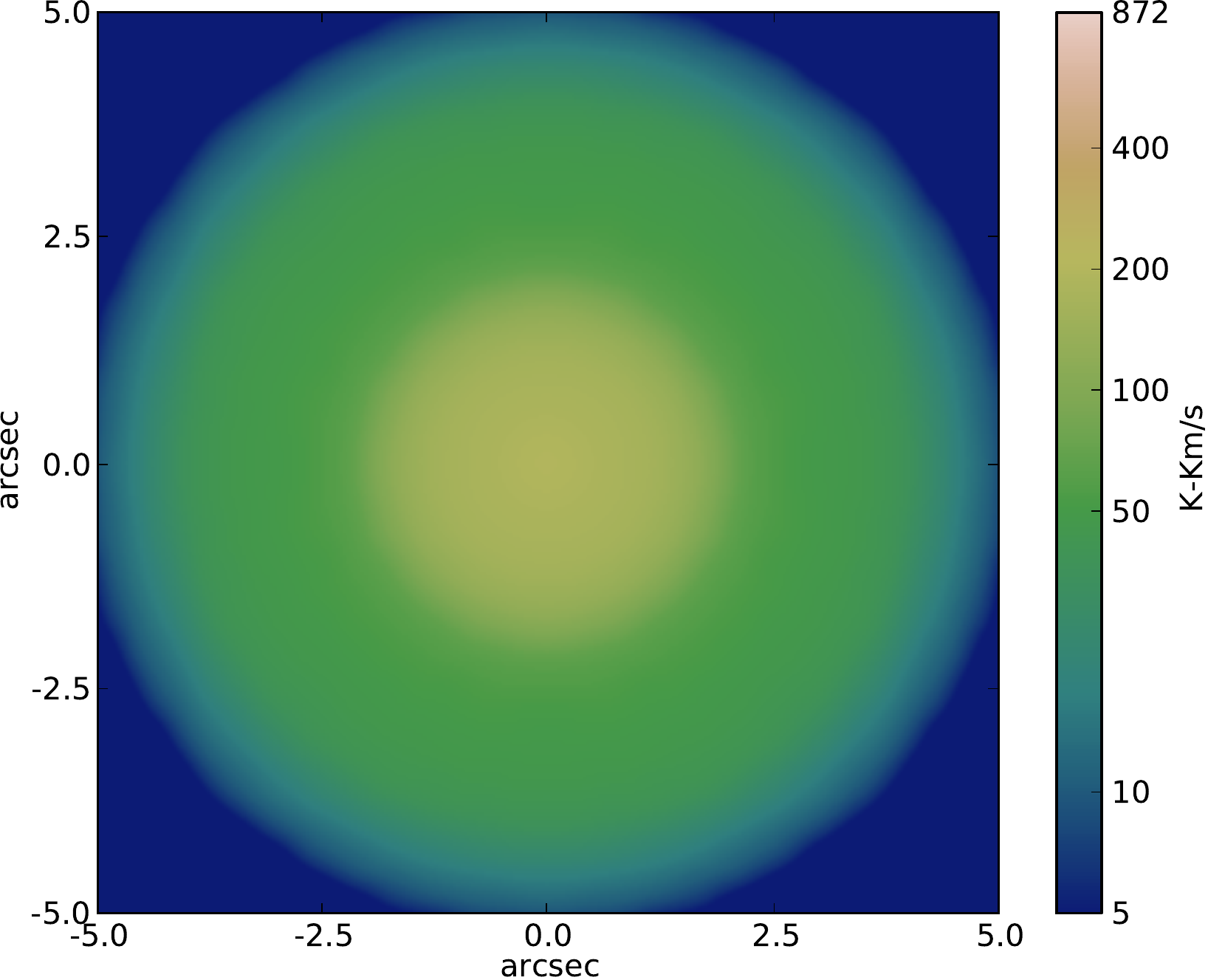}  
\end{tabular}
\caption{\emph{top}: \ce{CH3OH} 241.8877~GHz (E$_{u}$= 72.5~K), and \emph{bottom}: \ce{CH3OCH3} 241.9465~GHz (E$_{u}$= 81.1~K) integrated line intensity
maps of the hot core model \emph{cr16-r3-l7} at \num{1e5} year. The adopted distance to the source and the telescope beam size are 2~kpc and 0.3\arcsec.}
\label{fig:chan_meth_di}
\end{figure}
%----------------------------------------------------------------------------------
  
Simulated spectra  of different COMs such  as \ce{CH3OH}, \ce{C2H5OH},
\ce{CH3OCH3}    and    \ce{HCOOCH3}   of    the    hot   core    model
\emph{cr16-r3-l7} for different frequency bands, evolutionary stages
and        angular        resolution        are        shown        in
Figs.~\ref{fig:spec_coms_332}--~\ref{fig:spec_hrsl}. These spectra also
follow  the  same trend:  the  specific  line intensities  become
stronger  and  the  spectra  become more  densely  populated  at  later
evolutionary    stages.     The    frequency    range     shown    in
Fig.~\ref{fig:spec_hrsl}  is also covered  by the  Herschel-HIFI line
surveys \citep{Zernickel2012,  Crockett2014}. A qualitative comparison
of the simulated  and observed spectra reveals that  the relative line
intensities   of  \ce{CH3OCH3}  and   \ce{HCOOCH3}  with   respect  to
\ce{CH3OH}      are     comparable      with      the     observations
(Fig.~\ref{fig:simspec_on_hifi}).

Integrated   line   intensity  maps   of   the \ce{CH3OH}  transition   at
338.1245~GHz    at     different    timescales    are     shown    in
Fig.~\ref{fig:chan_maps_time}.  The  corresponding radial profiles for
the  integrated intensity maps  are shown  in Fig.~\ref{fig:rad_profs}
(top panel).  In both  figures, the  radial profiles  of molecular
emission expand with time.  This is due to the radial expansion of the
gas  phase abundance  of \ce{CH3OH}  with evolutionary  time resulting
from      the     increase      in     the      stellar     luminosity
(Fig.~\ref{fig:lum_and_temp}(a)) and the successively higher mass that
is heated to higher temperatures.  However, as our models use a static
density  distribution and  do not  yet include dynamics,  the spatial
extension  of temperature,  abundance, emission  maps, etc.   should be
considered       as      approximate       values.       In
Fig.~\ref{fig:chan_maps_dens},  integrated   line  intensity  maps  of
\ce{CH3OH} transition at 337.68559~GHz for various hot core models are
shown.    The   corresponding   radial    profiles   are    shown   in
Fig.~\ref{fig:rad_profs} (bottom panel). The variation in the emission
maps can be attributed to the different density distributions of these
models  since  the  other  input  parameters  are  the same  for  all  the
models.  These  maps  demonstrate  the importance  of  the  underlying
physical structure on chemical  evolution and spectra of hot molecular
cores. It  also vouches for the  potential of this study  to achieve a better
understanding of the  hot cores and thus the  high-mass star formation
scenario.

%-----------------------------------------------------------------------------------------------------------------------------------------------------------
%-----------------------------------------------------------------------------------------------------------------

One  of the  important parameters  of  astrochemical models  is the
desorption  energy  of  the  respective molecules\footnote{A  list  of
  binding   energies   for  various   molecules   can   be  found   at
  \url{http://www.astro.cornell.edu/~rgarrod/wp-content/uploads/2013/03/GH06_binding.txt}};
it also controls the gas-phase abundance of COMs. Recent observational
studies also started to compare the spatial distribution of COMs using
high resolution emission line maps \citep{Oberg2013}. The relative gas
phase  abundance  of  \ce{CH3OCH3}  is  more  extended  than  that  of
\ce{CH3OH} in the  hot core models, therefore, it  will be interesting
to verify whether this trend is also seen in observations. We show the
integrated  intensity  maps   of  the \ce{CH3OH}:  241.8877~GHz  (E$_{u}$=
72.5~K) and  \ce{CH3OCH3}: 241.9465~GHz (E$_{u}$=  81.1~K) transitions
for  the  model hot  core  \emph{cr16-r3-l7}  at  \num{1e5} year  in
Fig.~\ref{fig:chan_meth_di}.  This figure  demonstrates that  the bond
strength of  the molecule to the  grain can influence the  size of the
emissive region (i.e., a lower desorption energy places the molecule in
the gas phase over a larger surface area). We can compare the observed
emission  maps from  a range  of parent  molecular species,  each with
different desorption energies, but sampling transitions with comparable
excitation  conditions. Then these  emission profiles  can be  used to
constrain   the  relative  desorption   energies  of   the  associated
molecules.  Moreover,   with  grounding  to  laboratory   work  for  a
particular molecule,  we might  be able to  place this on  an absolute
scale.

It has been found that  the abundances of various COMs in \emph{cr17}
models are typically lower than that of \emph{cr16} models, and these
values are  unsuitable  to reproduce the  observed line  strength of
different COMs. Since our current  approach is based on static density
distribution,  it is  not  possible  to distinguish  the  effect  of
dynamics vs. cosmic-ray ionization  rates on the initial ice formation
and  hence on  the  final gas  phase  abundance of  COMs. More  detailed
studies are needed to constrain  the cosmic-ray ionization rate at the
sites of high-mass star formation.

\subsection{myXCLASS fit to the simulated spectra}
\label{subsec:myxclass_fit}
\begin{table}
\caption{Physical parameters of the hot core model \emph{cr16-r3-l7}: myXCLASS analysis}              
\label{tab:myxclass}     
\centering  
\resizebox{8.5cm}{!}{%
\begin{tabular}{llllll}          
\hline\hline   
Time    & Molecule  &   \multicolumn{4}{c}{myXCLASS analysis} \\
(year)  &           &   Frequency & Source        &Tempera-           & Relative \\
        &           &   (GHz)     & size (\arcsec)&-ture(K)           & abundance \\
\hline 
\multirow{4}{*}{\num{9e4}} & \ce{CH3OH}  &   248--251                  & 2.6   & 139 & \num{1.2e-6} \\
                           &             &   337--339                  & 1.86  & 137 & \num{1.2e-6} \\
                           & \ce{C2H5OH} &   \multirow{3}{*}{332--336} & 1.42  & 139 & \num{2.2e-8} \\
		           & \ce{CH3OCH3}&	                       & 1.75  & 126 & \num{6.3e-8} \\
		           & \ce{HCOOCH3}&	                       & 1.16  & 194 & \num{1.4e-8} \\
\multirow{4}{*}{\num{1e5}} & \ce{CH3OH}  &   248--251                  & $\textgreater$2.2 & 144 & \num{1.8e-6}\\ 
                           &             &   337--339                  & $\textgreater$2.2 & 141 & \num{1.9e-6}\\
                           & \ce{C2H5OH} &   \multirow{3}{*}{332--336} & $\textgreater$2.2 & 133 & \num{4.5e-8} \\
		           & \ce{CH3OCH3}&                             & 1.85  & 166 & \num{8.5e-8} \\
		           & \ce{HCOOCH3}&	                       & 1.75  & 189 & \num{7.5e-8} \\
\hline		           
\end{tabular}%
}
\end{table}
Many  of  the  line-survey  studies  used  XCLASS  \citep{Schilke1997,
  Schilke2001,  Zernickel2012,  Crockett2014,  Neill2014}  or  similar
programs to estimate the physical parameters from the observed spectra
and derived  the relative abundances  using \ce{H2} or  \ce{CO} column
densities at  constant temperature. We used the  total hydrogen density
to estimate  the relative abundance  of the respective  molecules (see
Tables~\ref{table:cr16_mol_abun}   and  \ref{table:cr17_mol_abun}).
There is  another notable difference  between the XCLASS  analysis and
this work: the  hot core models predict the  temperature and abundance
profiles  for hot cores,  whereas  XCLASS and  other methods  can only
derive  beam-averaged values.  Therefore, it  would be  interesting to
compare  these single-point  physical parameters  with the
spatial distribution predicted by our  hot core models. We carried out
an analysis to re-estimate  the physical parameters from the simulated
spectra                using                the               myXCLASS
interface\footnote{\url{https://www.astro.uni-koeln.de/projects/schilke/myXCLASSInterface}}
(M\"{o}ller     et      al.     2014,     in      preparation)     for
CASA\footnote{\url{http://casa.nrao.edu/}};    the    interface   also
includes  the  model optimizer  package  \emph{MAGIX} (Modeling  and
Analysis   Generic    Interface   for   eXternal    numerical   codes)
\citep{Moeller2013}.  We fitted the  simulated spectra  of \ce{CH3OH},
\ce{C2H5OH}, \ce{CH3OCH3},  and  \ce{HCOOCH3} convolved with a telescope
beam  size of  2.2\arcsec at  two different  evolutionary  stages over
selected frequency ranges.  The  fitted spectra are overplotted on the
simulated   spectra   in   Fig.~\ref{fig:myxclass_fit_overplot}.    We
list  the physical  parameters  obtained from the  myXCLASS analysis  in
Table.~\ref{tab:myxclass}; the  relative abundances were  derived using
the hydrogen column  density,\num{1.05e24}~cm$^{-2}$, as estimated
for the hot core  models \emph{hmc-r3}. We also added white
noise  to our  simulated spectra  and  fitted the  noisy spectra  with
myXCLASS.  We do  not find  any significant  difference in  the fitted
parameters, indicating that the myXCLASS  fitting values are robust against
noise.  We  also  plot  the  relative  abundances  as  a  function  of
temperature   for  different   molecules  at   different  evolutionary
timescales and overplot the single-point physical parameters
estimated  by  myXCLASS  analysis in  Fig.~\ref{fig:myxclass_fit}.  The myXCLASS estimates  are  comparable  with the  physical
parameters at  the jump radius of these  molecules (where the
gaseous      abundances      sharply      decrease)     (see      also
Fig.~\ref{fig:100k_radius});  these  regions  have  the  largest  area-filling  factor   in  the  beam,   therefore,  the  
results   are expected.  However, myXCLASS underestimates  the temperature  of the
hot cores. Our analysis indicates  that the typical temperatures of the
hot cores  are a few times  higher than the  canonical value of
100~K,   which   is    also   consistent   with   recent   observations
\citep{Sridharan2014}. These  results also suggest  that a comparison of
the simulated spectra  of the 3D hot core models  with the high resolution
observations (e.g., ALMA) will  reveal a more realistic physio-chemical
structure  of  the hot  cores  than the  simple fitting methods.

\section{Summary and discussion}
\label{sec:dis}

Molecular  line  observations  are  important  probes  to  reveal  the
chemical  reservoir   of  hot  cores  \citep{Schilke1997,Schilke2001}.
These observations are also  useful to explore the physical conditions
and   chemical   evolution  of   high-mass   star  forming   regions
\citep{Beuther2009,Zernickel2012,Oberg2013}. Recent  studies also used
single  dish observations  of a  large  number of  sources at  various
evolutionary stages to investigate the chemical evolution of high-mass
star  forming  regions  \citep{Hoq2013,Gerner2014}.  Selected  studies
also explored the radial distribution of density and temperature along
with  chemical  evolutionary   models  to  understand  the  underlying
physical  structure,  abundance  evolution,  evolutionary  stages,  etc.
\citep{Doty2002,  Oberg2013,  Gerner2014}.  In general,  these  studies
analyzed the observations by assuming a density distribution guided by
empirical studies and an estimate  of the thermal structure that would
result  from the  density  profile  and a  source  with a  particular
luminosity.   \cite{Doty2006}  determined  the  temperature  evolution
including the  luminosity evolution  for high-mass protostars,  but did
not include any grain surface diffusion reactions,  which are crucial
for COM formation on grain surface.

In this work we have used different temporal evolution profiles of the
luminosity for a forming high-mass star.  The density distribution of
the   hot  core   models  were   setup  using   recent  dust-continuum
observations  assuming  a gas to dust  mass  ratio  of  100. Within  this
framework, the  thermal evolution was determined with the radiative
transfer  code  \emph{RADMC3D}  based  on the  adopted  protostellar
luminosity  evolution. This then  resulted in  a time-variable thermal
profile  that  has  a   profound  effect  on  the  chemistry  through
sublimation  and   grain  surface  diffusion,   subsequent  gas  phase
processing,   etc.    We   used   the    chemical   evolutionary   code
\emph{Saptarsy} along  with a gas-grain chemical  network to explore
the spatio-temporal  evolution of  molecular abundances in  hot cores.
We then  explored the evolution  of abundance distribution  of selected
oxygen-bearing  COMs such as  \ce{CH3OH},  \ce{C2H5OH}, \ce{HCOOCH3},  and
\ce{CH3OCH3} in  detail under varying physical  conditions. We showed
that  the spatial  distribution of  gas phase  abundances of  the COMs
expand with  the increasing temperature of hot  cores.  These profiles
provide  a  detailed  variation   in  abundances  against  the  physical
structure of the hot cores. The average behavior of these profiles is
similar to the so-called jump models (two different abundances are  used  for the  upper  and lower  temperature
regime relative to a  characteristic temperature, typically 100~K) for
species that primarily  form on the grain surface and  are transformed to
the gas phase mainly via thermal desorption. In the evolutionary model the
jump radius  or the evaporation font (where the temperature
is  equal   to  the  characteristic   temperature of the
jump  abundance) also  expands with  time. We  also  showed the
variation of  the evaporation front associated  with different temperatures
for     the     hot     core     model     \emph{hmc-r3-l7}     (see
Fig.~\ref{fig:100k_radius}),   which  will   be  useful   to   set  the
jump    abundances    for    any   particular    evolutionary
stages.  Moreover, the  predicted abundance  profiles can  be  used as
templates  to model  the observed  spectra of  hot cores,  but one
should  also  keep in  mind  that  these  profiles vary  significantly
depending on the adopted physical structure and luminosity evolution.

Emission lines of  COMs (often referred to as weeds)  are one of
the salient  observational features of hot cores.  These molecules are
highly  abundant ($>$10$^{-7}$ with  respect to  hydrogen) and  have 
rich ro-vibrational emission  spectra.  A significant improvement over
previous studies  was achieved by simulating  the spectra for
hot core models  that can be directly compared  with observations. The
simulated spectra for different hot core models were generated by using
\emph{RADMC-3D}  at  different  spectral  wavebands to  explore  the
spectral  evolution  in  hot  cores.  The qualitative  comparison  of  the
observed  and  simulated  spectra  showed  that  the  hot  core  models
successfully reproduce the observed  trends, that is the increase of the total
number  of  emission  lines  and  their  associated  intensities  with
evolutionary  timescales.  The  adopted  timescales for  the  chemical
simulation of the  cold core phase  ($\sim$10$^{4}$ year) or the hot  core lifetime
($\sim$10$^{5}$ year)  are consistent  with  recent  independent studies
\citep{Tackenberg2012, Gerner2014}.   These results indicate  that the
self-consistent models present  a reasonable evolutionary scenario for
hot  cores.  Based  on these evolutionary  models, we  discuss some
additional features that can be compared with observations.

The hot  core models  predict that the  emission line maps  of various
transitions of different COMs also expand with time. With increasing
temperature,  the material around  the central protostars  heats up,
and consequently, the  gas phase abundance of various  COMs also spreads
out,  depending on the  binding energies  of the  respective molecules.
These  processes  collectively  contribute  to the  expansion  of  the
emitting regions  of various  transitions. We argue  that for  a given
molecule the  \emph{line forest} (see Sec.~\ref{subsec:sim_compare})
provides detailed  information about the excitation conditions, 
which can  be used  to extract key  information regarding  the binding
energies  of the molecules  to the  grain surface,  abundance profiles, 
etc.  This will provide further constraint to the chemical models
that  sometimes suffer  from  lack of  laboratory  measurements of the binding energies. A satisfactory match  between the observed and simulated spectra
along with  the emission  line maps thus  provides a  well-constrained
physio-chemical  structure of  the hot  cores, which  is  a significant
improvement over  existing fitting methods that  estimate the beam-averaged
single-point   physical   parameters.    High   resolution
observations  (such  as from  ALMA)  for  a statistically  significant
samples of high-mass prestellar cores will  be available soon.  The
temperature structure  of these cores  can be re-constructed  by using
radiative  transfer  modeling  and   can  be  further  constrained  by
comparing the simulated emission  line maps with the observations. The
fitting process  also place some  constrain  on the luminosity of  the central
stars, gas phase  abundance profile and the  desorption energy of  the associated molecules.  These models
will be particularly  useful for the molecules that  primarily form on
the grain  surface  and  are transformed   to  gas  phase  mainly  by  thermal
desorption. Moreover, modeling the high resolution observations of the
high-mass protostellar  cores with  similar masses  but  at different
evolutionary  stages  will  be  useful to  estimate  the  protostellar
luminosity  function  for  high-mass  protostars,  which  is  an  input
parameter of  these models.  In summary, the  full potential  of these
models can be explored by using the observations that can probe scales
of few  thousands AU. Currently, it  is also possible  to estimate the
evolutionary  stages using  these models, but the  estimated age
should be considered with caution  because a static density distribution is
used.  With  a temporal variation  of density distribution, it  may be
possible to  diagnose the spectral  features or abundance  ratios that
can be used  as a proxy for the evolutionary stages,  similar to the so-called chemical clocks.

Along with  the variation in  chemical abundances, the  differences in
formation and destruction pathways  of various molecules as a function
of  physical conditions  can  be  explored using  our  model. This  is
specifically  helpful  to verify  the  proposed chemical  evolutionary
scenarios. \ce{CH3OH} acts  as the parent molecule for  the other COMs
that  are considered in  this work.  The spatio-temporal  variation of the
abundance ratio  of these COMs with  \ce{CH3OH} can be  used to verify
the  reaction pathways  discussed  in Sect.~\ref{subsec:com_form}.  The
abundance  and   spectra  of  intermediate   species  such as  \ce{CH3O} and
\ce{CH2OH} will also be useful  to constrain the formation scenario of
COMs.

\section{Conclusion}
\label{sec:conc}
We  investigated  the  formation  and  evolution of  COMs  in  HMCs  by
constructing 3D physio-chemical models  guided by recent empirical and
modeling  studies of high-mass star  formation. We  combined radiative
transfer   calculation  with   chemical  evolution   to   explore  the
spatio-temporal  evolution  of the  COMs  in  hot  cores and  also  to
generate the  synthetic spectra at  selective wavebands. The grain surface 
abundances  of COMs are converted into gas phase,
where  the  temperature  is  higher  than the  water  ice  evaporation
temperature (100~K).  The
gas-phase abundance profile of various COMs will  furthermore be useful for
modeling  the  spectra  of  individual  sources.  We  also  find  that
jump abundance profiles are  still a good approximation for the
COMs   that mainly  form on  the grain  surface and
return to  the gas phase  by thermal desorption (e.g., \ce{C2H5OH}).  We predicted  the temporal
variation of the  jump radius that can  be used as  a template to
set the jump abundances according to the evolutionary stages. 
We successfully reproduced the trends in spectral
variation observed  in hot cores  within the typical lifetime  of HMCs
of  \num{1e5} year.   These results  indicate that  the adopted physical
parameters  and the  chemical model  \citep{Garrod2008}  are realistic
estimates of the observed hot cores.  We also discussed that a satisfactory
match between the simulated and observed spectra obtained by iterating
over  the  input  parameter  space  will be  useful  to  decipher  the
physio-chemical structure  of the hot cores. This technique when
applied to  model the observation of  a handful of hot  cores can also
constrain the luminosity evolution  of the central protostars and 
the  binding energies  of  the  molecules to  the  grain surface.  
Our  approach is  more effective  in extracting  the  physical parameters
(such  as temperature)  using  observations than  the simple  spectral
fitting methods  that estimate  single-point values  for the
whole source.  The modeling approach  discussed in this paper  will be
particularly  helpful to  reconstruct the  physical structures  of hot
cores  and  to  constrain the timescales and  chemical
evolution during the initial stages of high-mass star formation.

%------------------------------------------------------------------
\begin{acknowledgements}
This work is carried out within the Collaborative Research Centre 956,
sub-project    Astrochemistry   [C3],    funded   by    the   Deutsche
Forschungsgemeinschaft (DFG).  The authors would like to thank the referee and the editor, M. Walmsley, 
for a careful reading of the manuscript. RC would  like to thank C.P. Dullemond,
C. Endres, R.  Garrod, N. Harada, G. Hassel,  E. Herbst, A. Hindmarsh,
\'{A}. S\'{a}nchez-Monge, D. Semenov,  S. Viti, and M. Walmsley for useful
discussions. This  research has made  use of NASA's  Astrophysics Data
System.
\end{acknowledgements}
%--------------------------------------------------------------------
%---------------------------------------------------------------------------------------------------
\bibliographystyle{aa}

% Online Material
\Online
\begin{appendix} %First online appendix
\section{Spatio-temporal variation of selected COMs in various hot core models}
\begin{figure*}
\centering
\includegraphics[width=\textwidth,height=0.49\textheight]{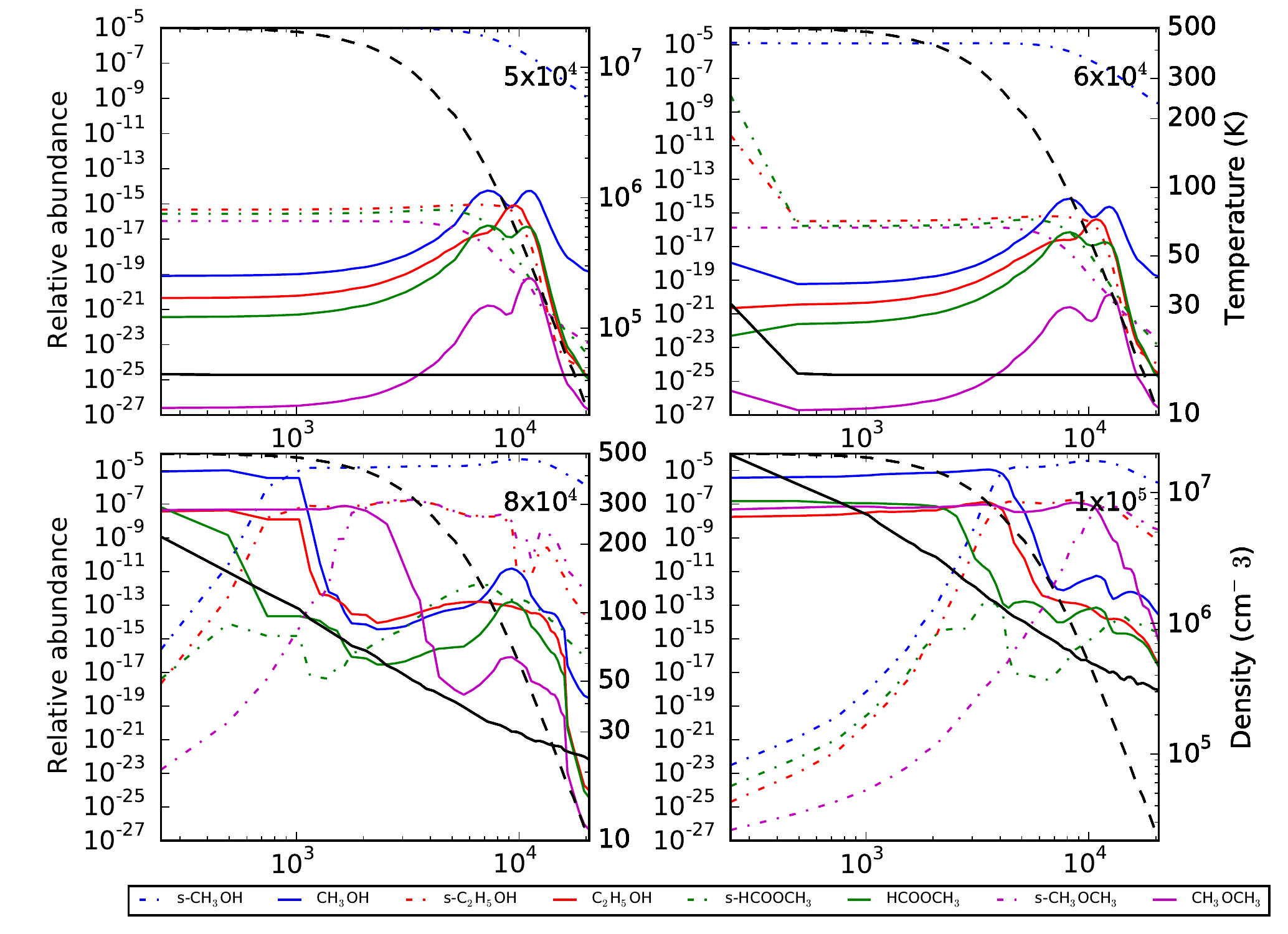}
\includegraphics[width=\textwidth,height=0.49\textheight]{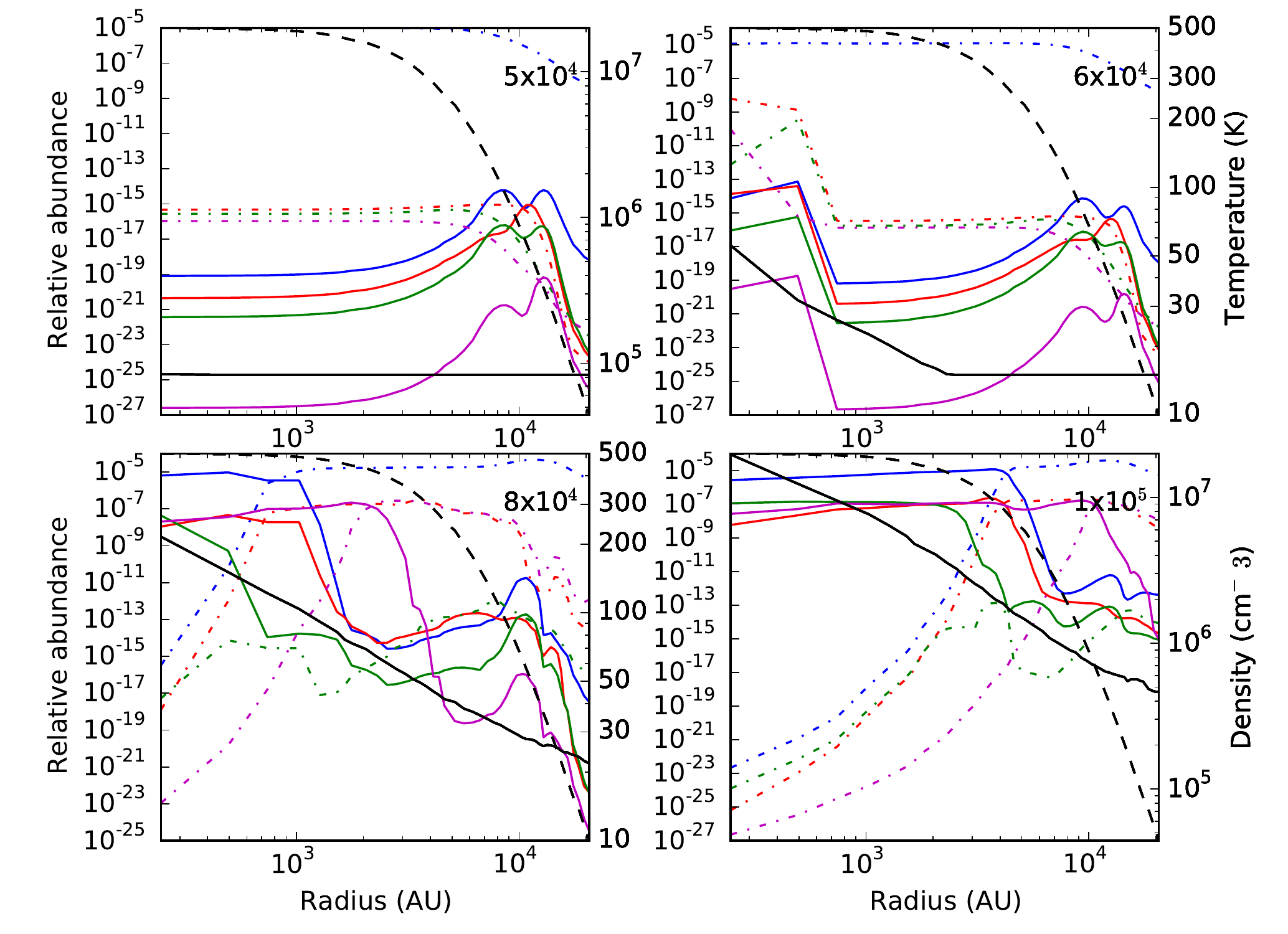}
\caption{Same as Fig.~\ref{fig:com_evol_a}, but for hot core models \emph{cr16-r3-l7} (upper panel) and \emph{cr16-r3.5-l4} (lower panel). Note the
alternative labeling
of the \emph{density} and \emph{temperature} axes on the right side of the respective plots.}
\label{fig:com_evol_b}
\end{figure*}
\begin{figure*}
\centering
\includegraphics[width=\textwidth,height=0.49\textheight]{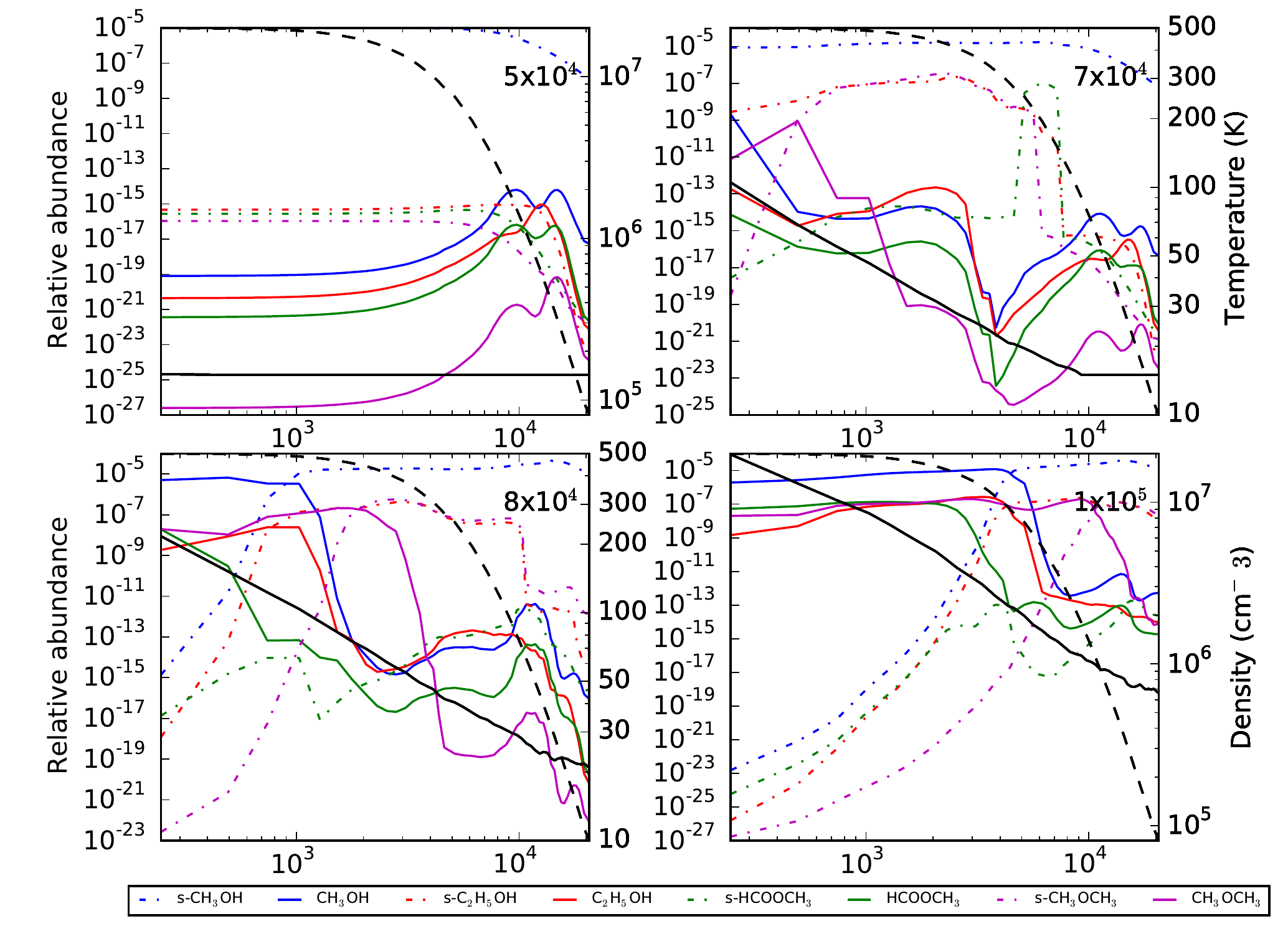}
\includegraphics[width=\textwidth,height=0.49\textheight]{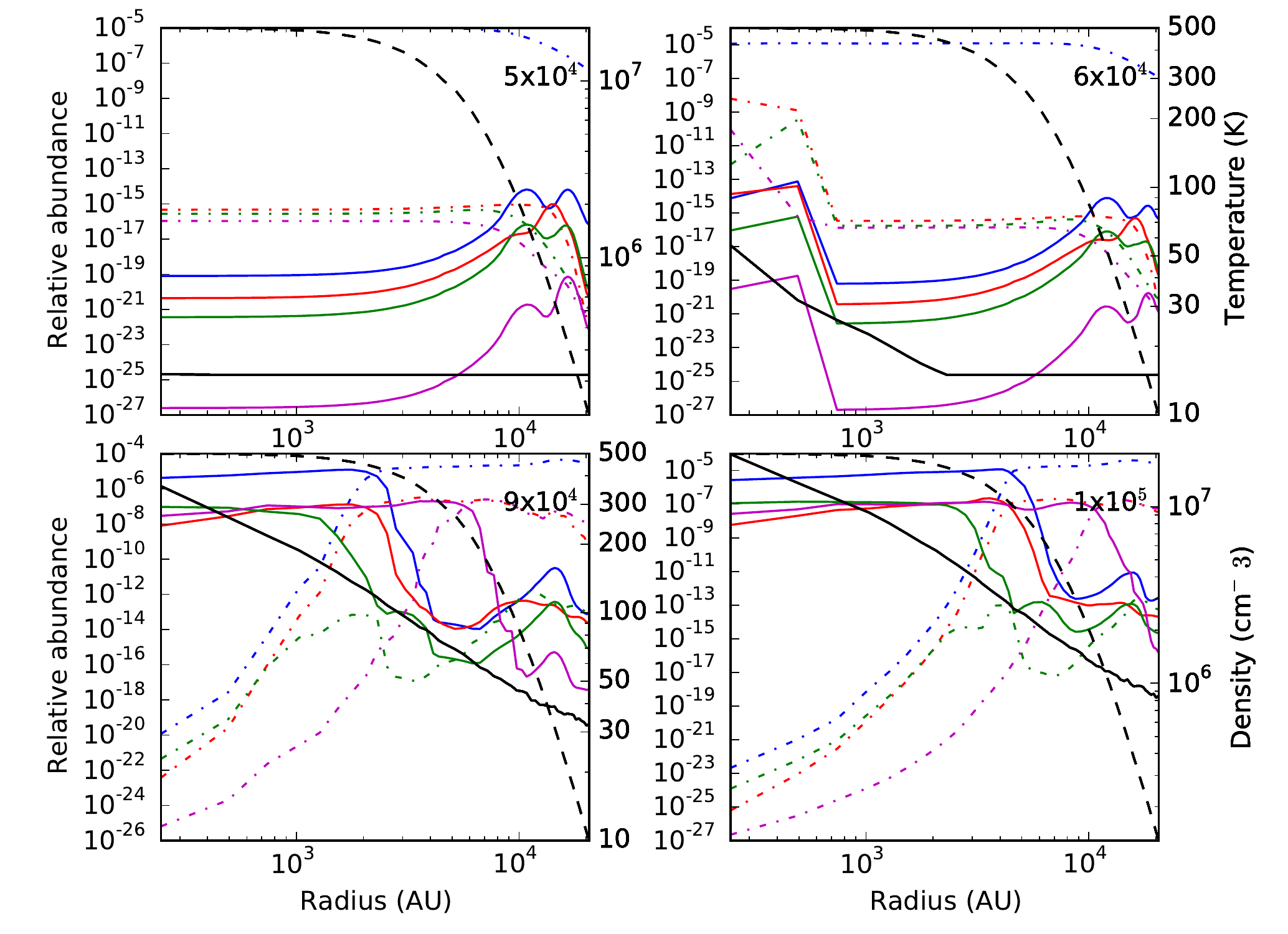}
\caption{Same as Fig.~\ref{fig:com_evol_a}, but for  hot core models \emph{cr16-r4-l2} (upper panel) and \emph{cr16-r4.5-l4} (lower panel). Note the
alternative labeling of the \emph{density} and \emph{temperature} axes on the right side of the respective plots.}
\label{fig:com_evol_c}
\end{figure*}
\begin{figure*}
\centering
\includegraphics[width=\textwidth,height=0.49\textheight]{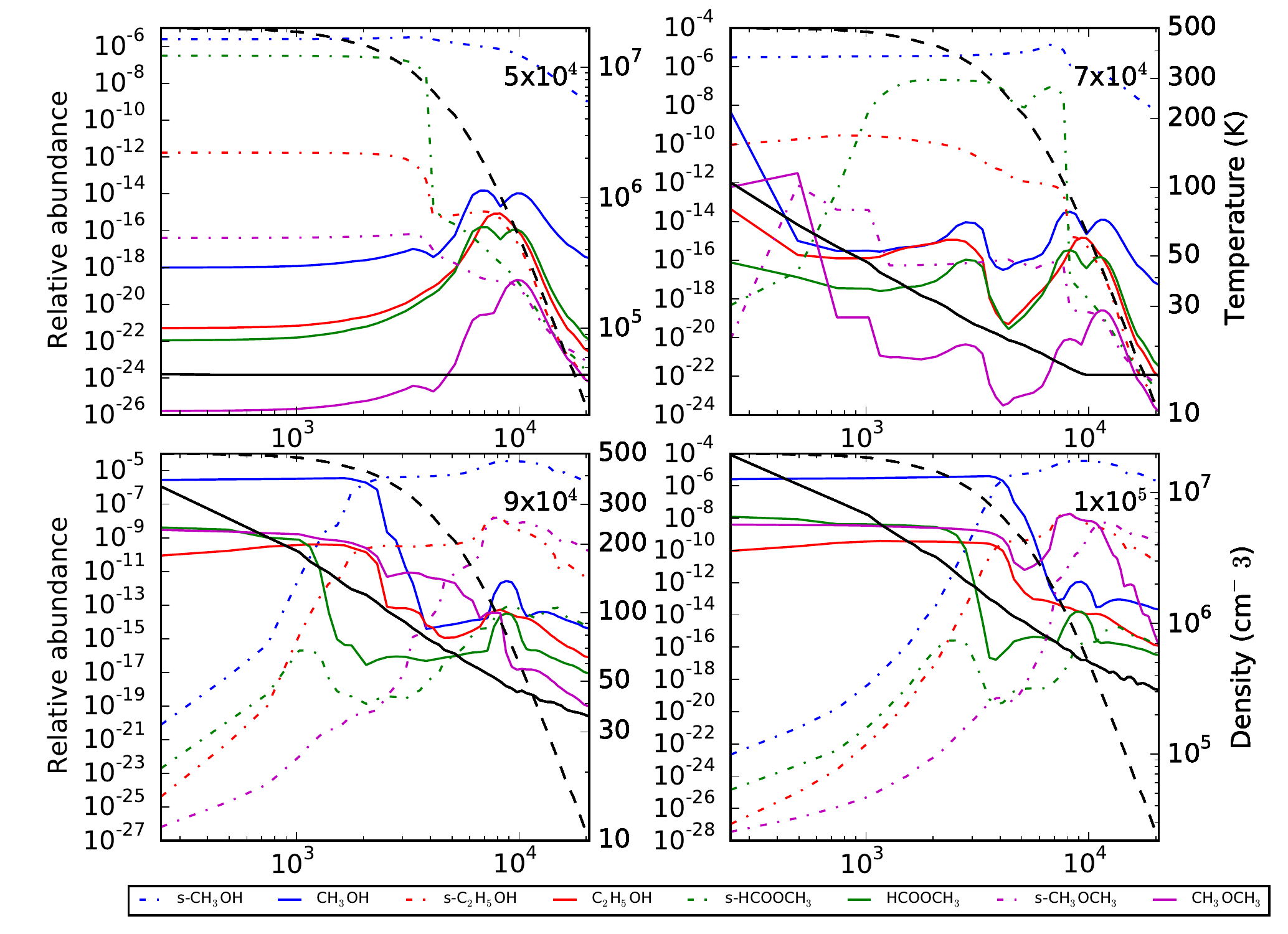}
\includegraphics[width=\textwidth,height=0.49\textheight]{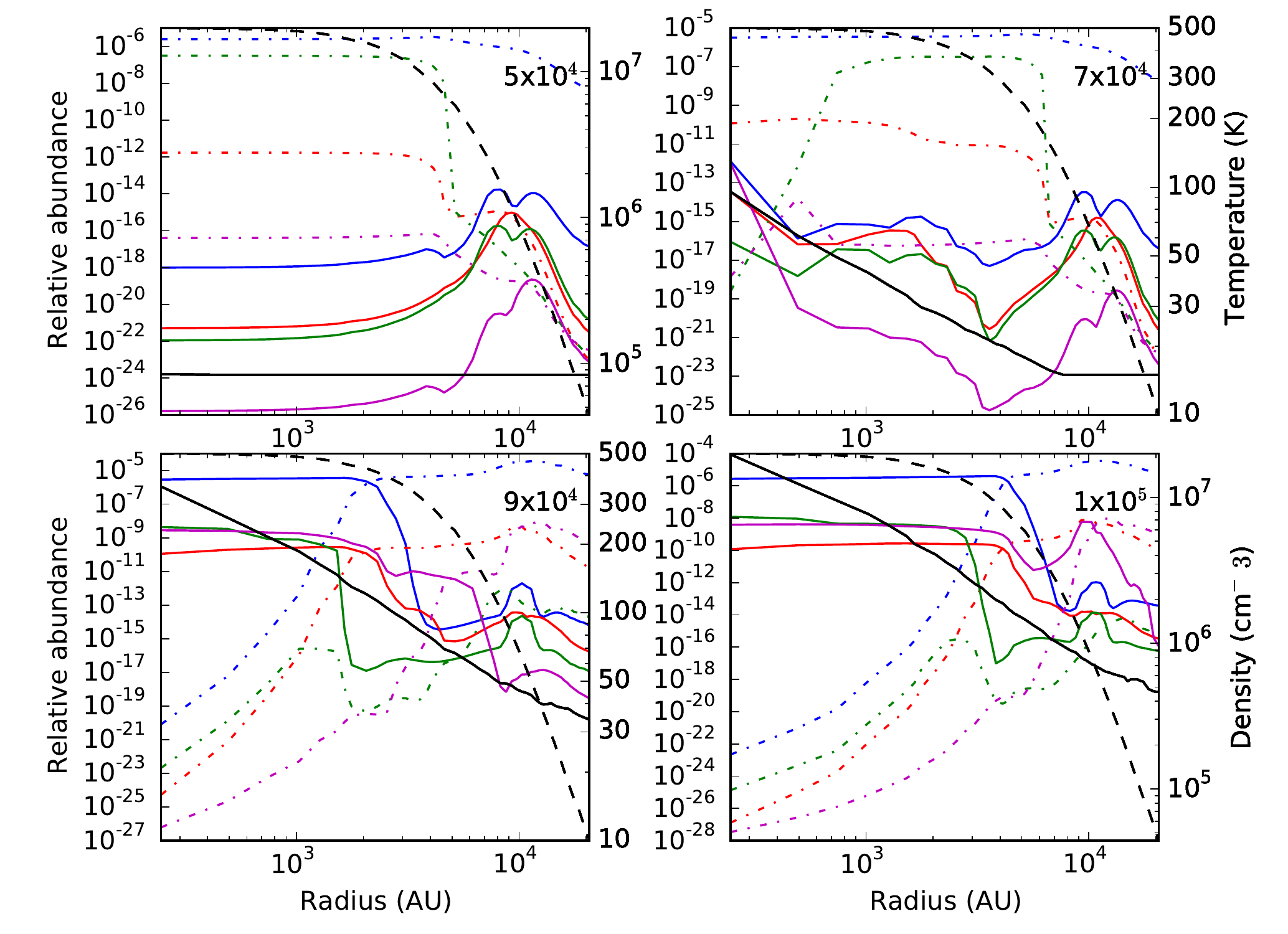}
\caption{Same as Fig.~\ref{fig:com_evol_a}, but for hot core models \emph{cr17-r3-l2} (upper panel) and \emph{cr17-r3.5-l4} (lower panel). Note the
alternative labeling of the \emph{density} and \emph{temperature} axes on the right side of the respective plots.}
\label{fig:com_evol_d}
\end{figure*}
\begin{figure*}
\centering
\includegraphics[width=\textwidth,height=0.49\textheight]{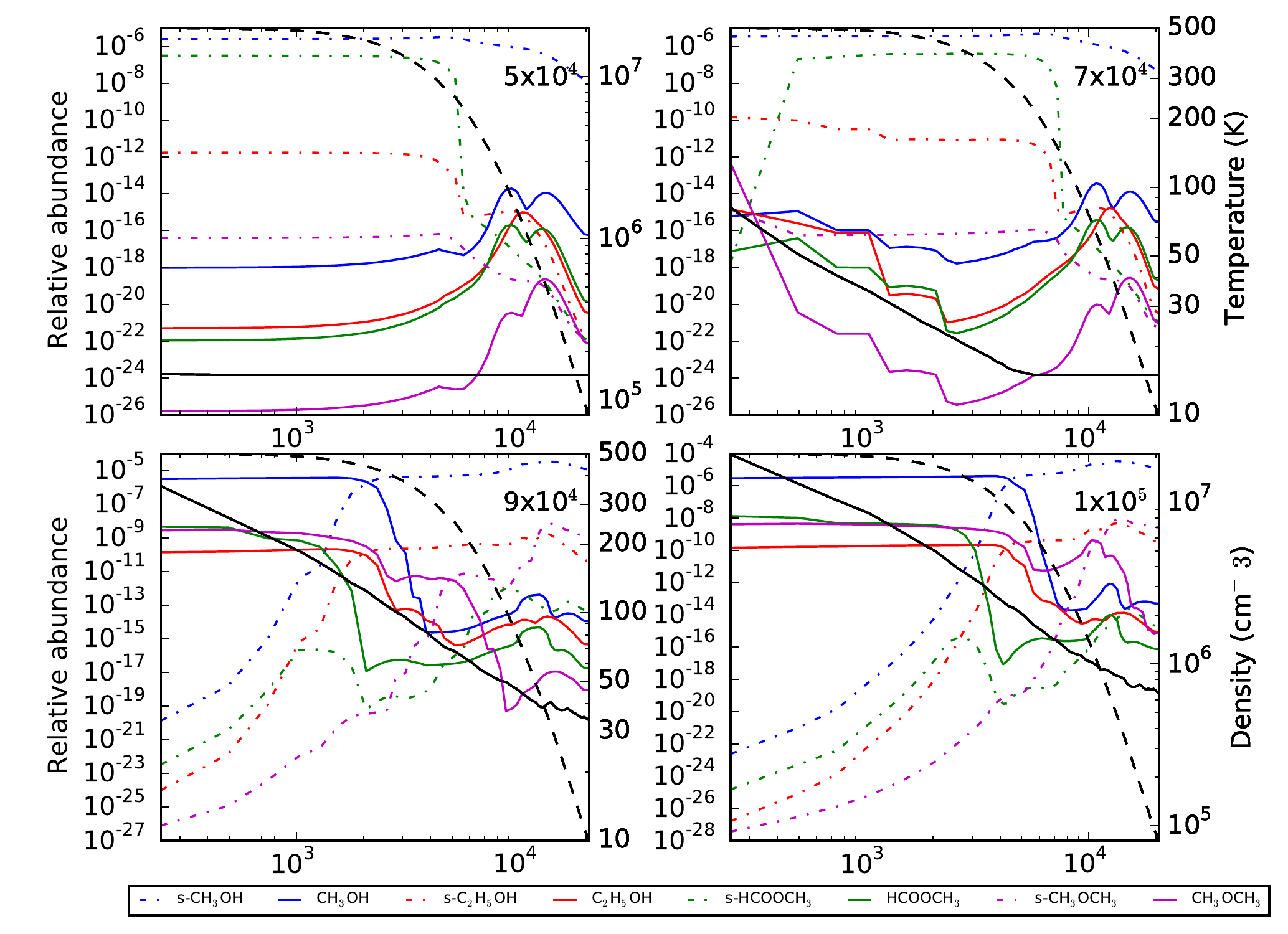}
\includegraphics[width=\textwidth,height=0.49\textheight]{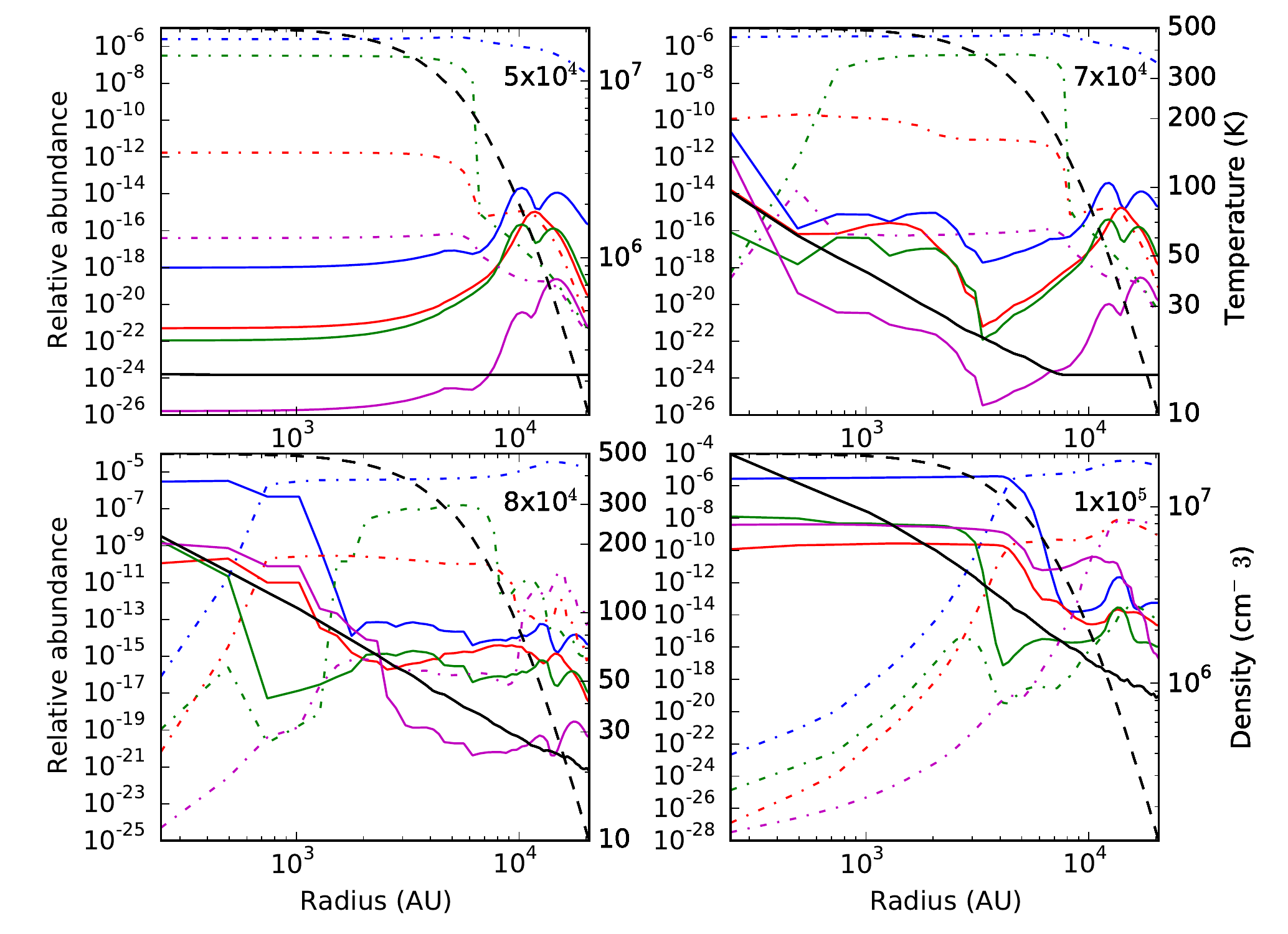}
\caption{Same as Fig.~\ref{fig:com_evol_a}, but for hot core models \emph{cr17-r4-l7} (upper panel) and \emph{cr17-r4.5-l4} (lower panel). Note the
alternative labeling of the \emph{density} and \emph{temperature} axes on the right side of the respective plots.}
\label{fig:com_evol_e}
\end{figure*}

%----------------------------------------------------------------------------------------------------------------------------------
\section{Temporal variation of the radii with temperature above 100~K, 150~K, and 200~K}
\begin{figure*}
\centering
\includegraphics[width=0.4\textwidth]{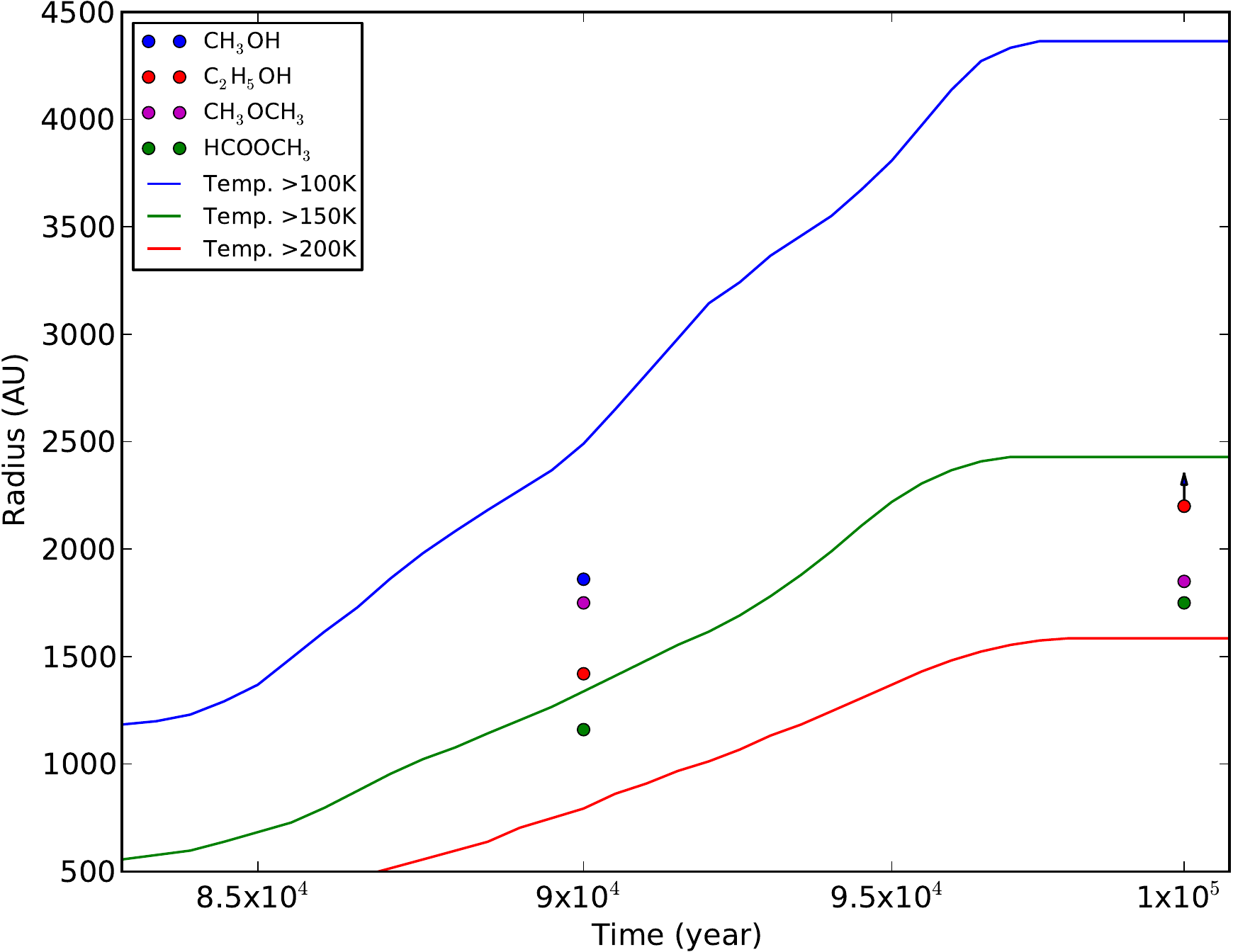} 
\caption{Temporal variation of the radii with temperature above 100~K, 150~K, and 200~K. Radii of the spatial distribution of various molecules as derived
from the myXCLASS analysis (see Table~\ref{tab:myxclass}) are overplotted with different colors. The arrow indicates that the radii of \ce{CH3OH} and \ce{C2H5OH} at
\num{1e5} year are only lower limits.}
\label{fig:100k_radius}
\end{figure*}
%------------------------------------------------------------------------------------------------------------------------------
\section{Comparison of simulated spectra for different evolutionary timescales}
\begin{figure*}
\centering
\includegraphics[width=0.5\textwidth]{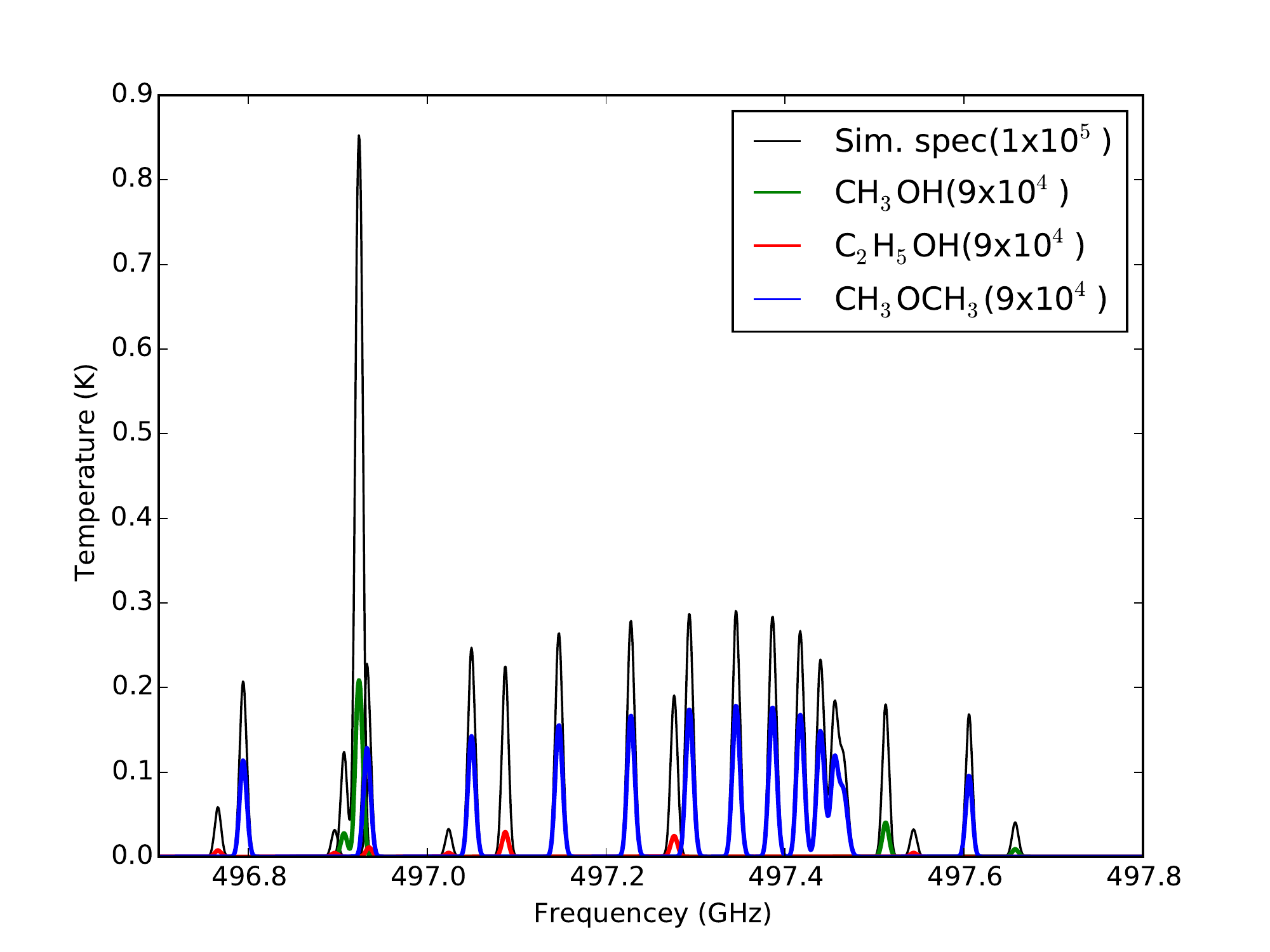} 
\caption{Simulated spectra for the hot core model \emph{cr16-r3-l7} at \num{9e4} year, overplotted on the spectra at \num{1e5} year. The intensity of various spectral lines
at two different epochs cannot be matched using a scaling factor. Spectral changes originate from the variation of temperature and abundances at different timescales.}
\label{fig:sim_spec_t9_on_t1}
\end{figure*}
%-------------------------------------------------------------------------------------------------------------------------------------
\section{Comparison of simulated and Herschel-HIFI spectra}
\begin{figure*}
\centering
\includegraphics[width=0.5\textwidth]{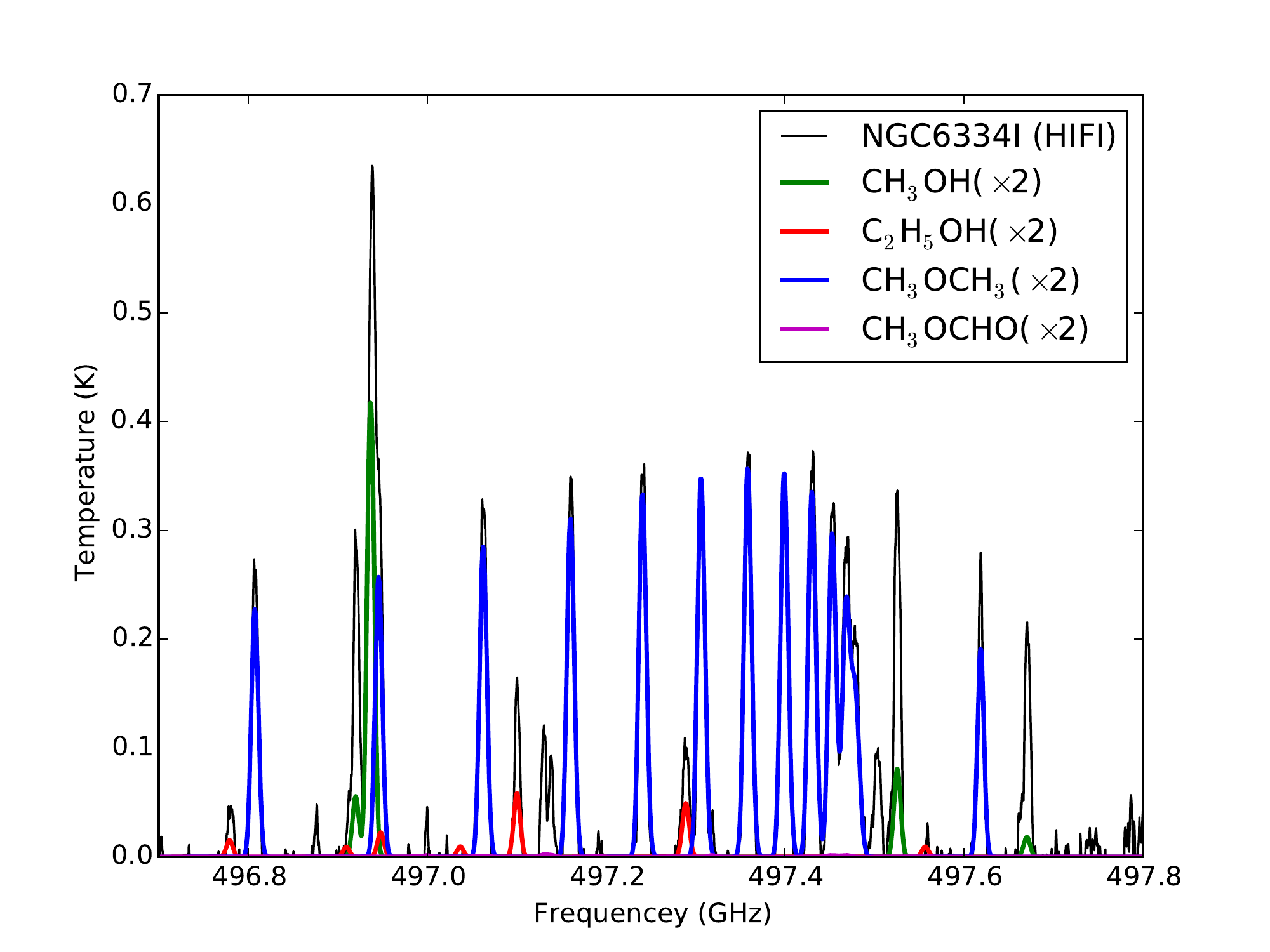} 
\caption{Scaled-up version (multiplied by a factor of 2) of simulated spectra for the hot core model \emph{cr16-r3-l7} at \num{9e4} year, overplotted on the Herschel-HIFI spectra of NGC6334I.}
\label{fig:simspec_on_hifi}
\end{figure*}
%------------------------------------------------------------------------------------------------------------------------
\section{myXCLASS fit to the simulated spectra}
\begin{figure*}
\centering
\includegraphics[width=0.5\textwidth]{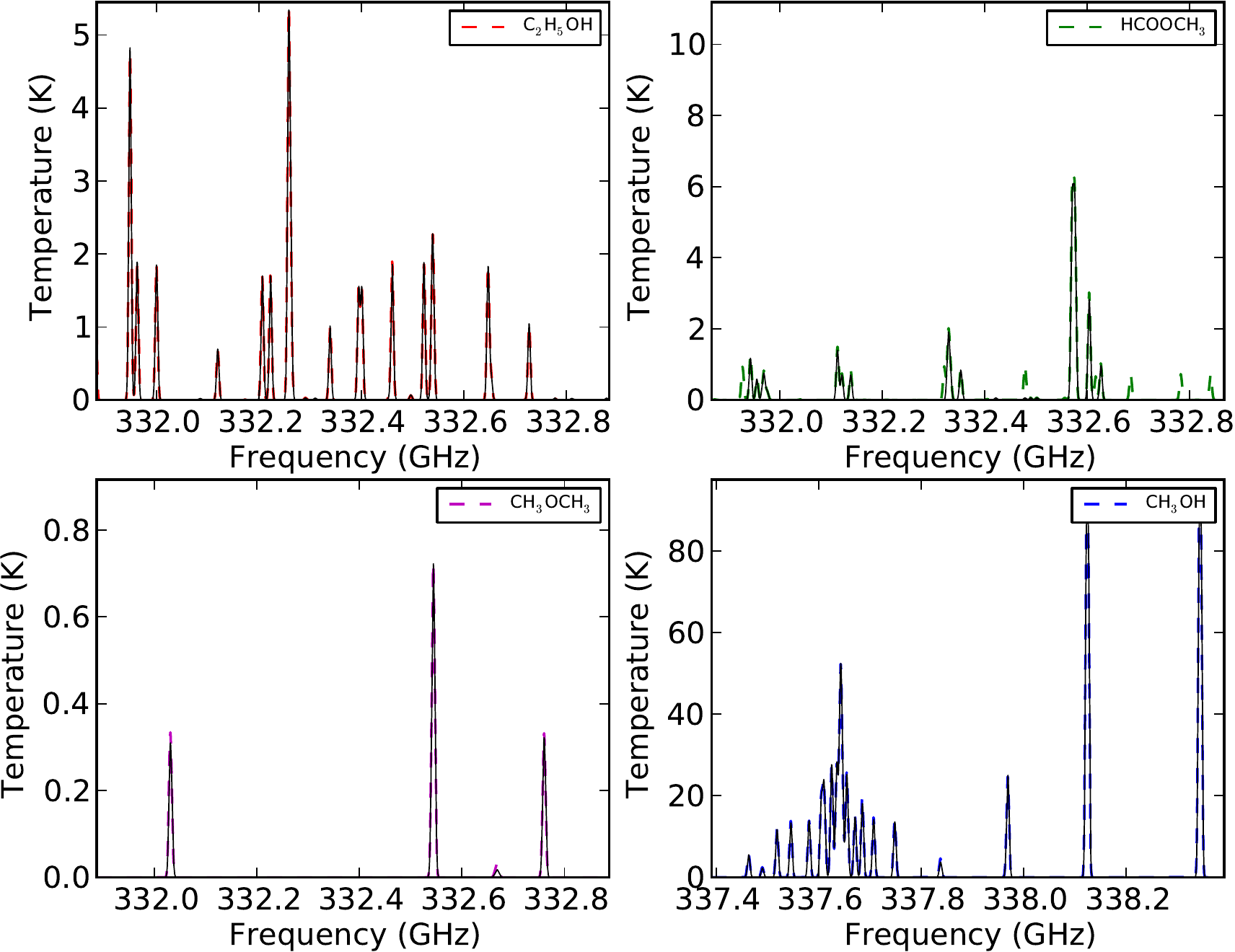} 
\caption{Fitted spectra obtained from the myXCLASS analysis (\emph{solid black line}), overplotted on the simulated spectra (\emph{colored dashed lines})  of
different COMs.}
\label{fig:myxclass_fit_overplot}
\end{figure*}
\end{appendix}
\end{document}